\documentclass[preprint,3p]{elsarticle}
\usepackage{epstopdf}
\usepackage{pdfpages}
\usepackage{amssymb}
\usepackage{empheq}
\usepackage{cases}
\usepackage{amsthm,amsmath}
\usepackage{caption,lipsum}
\usepackage{stmaryrd}
\usepackage{graphicx,epsfig}
\usepackage{tabularx}
\usepackage{color}
\usepackage{hyperref}
\usepackage{empheq}
\usepackage{enumitem}
\usepackage{algorithm, algorithmic}

\usepackage{float}
\usepackage{caption}
\usepackage{subfigure}
\usepackage{graphicx}
\usepackage{comment}
\usepackage{multirow}
\usepackage{mathrsfs}
\usepackage{amsmath,amssymb} 
\usepackage{color}
\usepackage{epsfig}
\usepackage{diagbox}
\usepackage{afterpage}
\usepackage{makecell}
\usepackage{empheq}
\usepackage{lineno}
\usepackage{amsthm}   

\DeclareMathOperator*{\argmin}{arg\,min}
\usepackage{bm}
\usepackage{algorithm}
\usepackage{algorithmic}
\usepackage{booktabs}       
\usepackage{hyperref}

\numberwithin{equation}{section}
\newcommand{\fe}{\mathrm{e}}

\newcommand{\bR}{{\mathbb R}}

\newcommand{\bN}{{\mathbb N}}

\newcommand{\ro}[1]{\uppercase\expandafter{\romannumeral#1}}

\newcommand{\bx}{\mathbf{x}}
\newcommand{\by}{{\mathbf y}}

\newtheorem{theorem}{Theorem}[section]
\theoremstyle{remark}
\newtheorem{remark}[theorem]{Remark}
\newtheorem{example}[theorem]{Example}
\journal{Elsevier}

\begin{document}

\begin{frontmatter}

\title{Computing ground states of Bose-Einstein condensation by normalized deep neural network}

\author[SG]{Weizhu Bao}\ead{matbaowz@nus.edu.sg}
\address[SG]{Department of Mathematics, National University of Singapore, Singapore 119076}

\author[Whu]{Zhipeng Chang}\ead{changzhipeng@whu.edu.cn}
\address[Whu]{School of Mathematics and Statistics, Wuhan University, 430072 Wuhan, China}
\author[Whu,Hubei]{Xiaofei Zhao}\ead{matzhxf@whu.edu.cn}
\address[Hubei]{Computational Sciences Hubei Key Laboratory, Wuhan University, 430072 Wuhan, China}

\begin{abstract}
We propose a normalized deep neural network (norm-DNN) for computing ground states of Bose-Einstein condensation (BEC)
via the minimization of the Gross-Pitaevskii energy functional under unitary mass normalization. Compared
with the traditional deep neural network for solving partial differential equations,
two additional layers are added in training our norm-DNN for solving
this kind of unitary constraint minimization problems: (i) a normalization layer is introduced to enforce the unitary
mass normalization, and (ii) a shift layer is added to guide the training to non-negative ground state.
The proposed norm-DNN gives rise to an efficient unsupervised approach for learning ground states of BEC.
Systematical investigations are first carried out through extensive numerical experiments
for computing ground states of BEC in one dimension. Extensions to high dimensions and multi-component
are then studied in details. The results demonstrate the effectiveness and efficiency of norm-DNN for learning
ground states of BEC. Finally, we extend the norm-DNN for computing the first excited states of BEC and discuss
parameter generalization issues as well as compare with some existing machine learning methods for computing
ground states of BEC in the literature.

\end{abstract}



\end{frontmatter}

\section{Introduction}
{The idea of Bose-Einstein condensation (BEC) originated from A. Einstein's extension of S.N. Bose's work on quantum statistics of photons \cite{Bose} to the case of  Bose gases. A. Einstein} predicted that particles in a Bose gas below a critical temperature would occupy the same quantum state, which was later experimentally realized \cite{ObserBEC} and considered as a new state of matter. Such state is now known as BEC and it has been widely concerned in atomic, molecule and optical (AMO) physics  \cite{BaoCai,DBEC,Pitaevskii}. For a BEC, the fundamental physics is described by a many-body Schr\"odinger equation which is extremely difficult to solve due to the high dimensionality, {and then researchers have been seeking for effective simplifications.} For the dilute  Bose gas below the critical temperature and confined in an external trap,
by  the mean-field approximation, the following Gross-Pitaevskii equation (GPE) (in dimensionless form) \cite{BaoCai} is widely concerned in applications for modelling BEC:
\begin{equation}
	\label{eq:GPE}
	i\frac{\partial \psi(\mathbf{x},t)}{\partial t} = -\frac{1}{2} \nabla^{2}\psi(\mathbf{x},t) + V(\mathbf{x})\psi(\mathbf{x},t)+\beta |\psi(\mathbf{x},t)|^{2}\psi(\mathbf{x},t), \quad \mathbf{x}=(x_1,x_2,\cdots,x_d)^\top\in \mathbb{R}^{d},
\end{equation}
where $\psi:=\psi(\mathbf{x},t)$ represents the unknown complex-valued wave function and $d=1,2,3$. Here $V(\mathbf{x})$ is a given real-valued function denoting the external trapping potential that in applications satisfies the confining condition
\begin{equation}\label{confinement}
	\lim_{|\bx|\to\infty}V(\bx)=\infty.
\end{equation}
We would consider the following two types of potentials which are commonly used in experiments \cite{BaoCai,DBEC,Pitaevskii}:
\begin{align}\label{ho potential}
	&\mbox{harmonic oscillator:}\quad V(\bx)=\sum_{j=1}^d\frac{1}{2}\gamma_j^2x_j^2;\quad
	\mbox{optical lattice:}\quad V(\bx)=\sum_{j=1}^d(\frac{1}{2}\gamma_j^2x_j^2+a_j^2{\sin^2(\alpha_j x_j)});
\end{align}
{with} ${\gamma_j,a_j,\alpha_j\in\bR^+}$.
{The parameter $\beta\in\bR$ in (\ref{eq:GPE}) is a given constant with $\beta>0$ denoting the repulsive interaction and $\beta<0$ denoting the attractive case.}

Though GPE is originated for BEC, as a mathematical model it is of interest in arbitrary dimension \cite{LiebLoss} and can be used to describe general nonlinear wave phenomenon.
For the past few decades, extensive efforts have been devoted to solving the GPE for its dynamics or stationary solutions \cite{Besse,BaoCai}. Along the dynamics in the model (\ref{eq:GPE}), the total mass  and energy of the system are conserved, i.e.,
\begin{equation}
	\label{eq:nc}
	M(\psi(\cdot, t)) := \int_{\mathbb{R}^{d}} |\psi(\mathbf{x},t)|^2 d\mathbf{x} \equiv  M(\psi(\cdot, 0))=1, \quad t\geq 0,
\end{equation}
and
\begin{equation}\label{energy}
	E(\psi(\cdot, t)) := \int_{\mathbb{R}^d} \left[\frac{1}{2}|\nabla \psi(\mathbf{x},t)|^2+V(\bx)|\psi(\mathbf{x},t)|^2+\frac{\beta}{2} |\psi(\mathbf{x},t)|^4 \right] d\mathbf{x} \equiv E(\psi(\cdot, 0)), \quad t\geq 0.
\end{equation}
The stationary solutions, which are a particular class of interesting solutions for \eqref{eq:GPE},  are given by the ansatz
$\psi(\mathbf{x},t) = \phi(\mathbf{x})\fe^{-i\mu t}$,
 where $(\mu,\phi)$ satisfies the following nonlinear eigenvalue problem:
\begin{equation}
	\label{eq:eigen}
	\mu \phi(\mathbf{x}) = -\frac{1}{2} \Delta \phi(\mathbf{x}) + V(\mathbf{x})\phi(\mathbf{x}) + \beta|\phi(\mathbf{x})|^2 \phi(\mathbf{x}), \quad \mathbf{x}\in \mathbb{R}^{d},
\end{equation}
under a unitary mass normalization condition
\begin{equation}\label{mass-norm}
||\phi||_{2}^2:=\int_{\mathbb{R}^d}|\phi(\mathbf{x})|^2\,d\mathbf{x}=1.
\end{equation}
Here $\mu$ is known as the chemical potential of the condensate given as
\begin{equation*}
\mu = \mu(\phi) = \int_{\mathbb{R}^d} \left[\frac{1}{2}|\nabla \phi(\bx)|^2+V(\mathbf{x})|\phi(\bx)|^2+\beta |\phi(\bx)|^4 \right]
	d\mathbf{x}=E(\phi) + \int_{\mathbb{R}^d}\frac{\beta}{2} |\phi(\bx)|^4 d\mathbf{x}.
\end{equation*}
Among all the {eigenstates}, the most important one is the so-called \emph{ground state (GS)} \cite{BaoCai,Du,mathofbec}, which  minimizes the energy functional under the unitary mass constraint:
\begin{equation}
	\label{eq:min}
	\phi_g:=\argmin_{\Vert \phi \Vert_2 = 1} E(\phi),\qquad   E_g := E(\phi_g).
\end{equation}
{We remark here that \eqref{eq:eigen} is the Euler-Lagrange equation for minimizing the energy \eqref{energy} under the constraint of mass conservation  \eqref{mass-norm}, where the chemical potential $\mu$ is viewed as a Lagrange multiplier for the mass conservation.}
In this paper, we mainly focus on how to efficiently solve the GS solution by a deep learning method.

To numerically compute the GS of BEC, there are already quite many powerful methods. The most traditional method is {the gradient flow with discrete normalization method or known as the imaginary-time evolution method} \cite{Du,Tosi,Liu}, and later on many more methods like the constrained optimization techniques \cite{ALT2017,Danaila1,Danaila2} and the nonlinear eigenvalue solvers \cite{Altmann,Maday} have also been developed. These methods though mathematically elegant, are all based on traditional spatial discretizations of the wave function in the computational domain. Therefore, they are basically limited to rather low-dimensional setup. For the original many-body Schr\"odinger equation or {high-dimensional} GPE, the curse-of-dimensionality will preclude their availability for applications.

In recent years, the deep neural network (DNN) has made enormous success in a wide scientific computing area. It is {hopefully} to overcome the curse-of-dimensionality, and this gives it great potential to be considered for solving the quantum many-body system. Although the final and meaningful target might be the many-body Schr\"odinger model, we need to begin with the low-dimensional problem to develop the detailed method and understand its performance. Thus in this work, we would like to begin with the development of the DNN for computing GS of the GPE. This kind of trend in fact has already begun among physicists recently. For instance, the work \cite{Liang} first considers the parameter generation of GS by using deep convolution neural networks for 2-dimensional (2D) GPE and 1-dimensional (1D) two-component case. The Gaussian process is used in \cite{bayesian} later to reduce the required data for training by incorporating Bayesian optimization. The work \cite{Bai}  {investigates} the theory-guided neural network for 1D GPE with spin-orbit coupling. These existing DNN approaches are {to some extent} based on the supervised learning which requires data of the unknown for label, and the data is obtained from the traditional numerical methods like the imaginary-time evolution. Such requirement of data seems not possible or very expensive again for high-dimensional or multi-component problems.

In this work, we propose to compute the GS via an unsupervised learning approach.
We shall propose a suitable DNN which we call \emph{normalized DNN (norm-DNN)}, to approximate the GS and directly solve {the optimization problem \eqref{eq:min}} without calling data from traditional numerical methods. In fact with norm-DNN, we can convert the original constraint optimization in functional space into an unconstrained minimization in finite dimensional parameter space, where the unitary mass normalization \eqref{mass-norm}  is naturally provided by the network by introducing a normalization layer. By utilizing some mathematical features of the solution,  some other key techniques including a shift layer in the network, Gaussian initialization and a regularization in the loss are introduced to provide efficient training process and accurate approximations of GS. Investigations through extensive numerical experiments will firstly be made in a systematical way for 1D GPE and then extended to the high-dimensional case and the two-component case. {The numerical results will also show that the application of standard neural {network to \eqref{eq:min}} could fail to produce GS, which illustrates the importance of norm-DNN.} The proposed norm-DNN is also extended to compute the first excited states of BEC. Furthermore, parameter generations of the proposed norm-DNN will be given in the end and comparisons will be made with traditional supervised learning DNN.

The rest of the paper is organised as follows. In Section \ref{sec:2}, we shall review some useful features of the GS and define the  structure of norm-DNN. Section \ref{sec:3} investigates the norm-DNN for 1D GS problem, and Section \ref{sec:4} considers extensions to solve GS of BEC in high-dimensions and two-component case and
to compute the first excited states. We shall present the parameter generation and comparisons with existing machine learning methods in Section \ref{sec:5}. Some conclusions are summarized  in Section \ref{sec:6}.

\section{Normalized deep neural network for ground state}\label{sec:2}
In this section, we shall first go through some preliminary for the ground state (GS) which will be helpful for computations later. Then, we shall propose our main neural network and present the computational setup for the training.

\subsection{Feature of GS and traditional numerical approach}\label{sec:pre_gs}
{Under (\ref{confinement}) and  the case $\beta\geq0$, GS always exists for (\ref{eq:min}). Moreover, we have the following key mathematical features for it \cite{BaoCai,LiebLoss,mathofbec}:
\begin{enumerate}[label=(\roman*)]
	\item GS is unique up to a phase shift, i.e., $\fe^{i\xi}\phi_g$ for any $\xi\in\bR$ is also a GS; so $\phi_g$ can be chosen nonnegative;
	\item If the potential $V$ is smooth, then $\phi_g\in C^\infty$ and it is exponentially decaying at far field;
	\item If $V$ has radial symmetry, then $\phi_g$ is also radially symmetric (so it is an even function in 1D).
\end{enumerate}
The design of our neural network and computations later will benefit from these features.}
	
{Before we introduce the network, we first review the traditional numerical method.}
The most traditional and popular way to compute GS is the gradient flow with discrete normalization (GFDN) method or known as the imaginary-time evolution method: set a time series $0=t_0<t_1<t_2<\cdots<t_n<\cdots$ with an initial guess $\phi(\mathbf{x}, 0)=\phi_0(\mathbf{x})$ and evaluate
\begin{equation}
    \label{eq:gfdn}
    \phi_t=-\frac{1}{2}\frac{\delta E(\phi)}{\delta {\phi}} = \frac{1}{2} \nabla^2 \phi - V\phi - \beta|\phi|^2\phi, \quad   t_n<t<t_{n+1},
    \quad \phi(\mathbf{x},t_{n+1}) = \frac{\phi(\mathbf{x}, t_{n+1}^{-})}{||\phi(\cdot, t_{n+1}^{-})||_{2}},\quad  n\geq0,
\end{equation}
till a steady state is reached. To numerically implement (\ref{eq:gfdn}),
one can truncate the whole space to a bounded domain  and impose zero boundary conditions. Then a basic scheme for discretizating  the partial differential equation (PDE) in (\ref{eq:gfdn}) is to apply the backward Euler finite difference in time and the sine pseudospectral method in space. We refer the readers to \cite{BaoCai,Du} for the detailed scheme,  and we shall use this scheme as the benchmark for reference GS solutions in our numerical experiments later. For more advanced traditional schemes, we refer to \cite{Altmann,ALT2017,Maday,Danaila1,Danaila2} and references therein.

The traditional approaches contain two drawbacks. (i) The first drawback is of course the {curse-of-dimensionality} for discretizing the PDE in high dimensions. The storage of mesh grids and computational costs can be  expensive. (ii) The second drawback comes from the fact that the GS solution $\phi_g$ depends on the free parameters in the model, e.g., the value of $\beta$ and the values of the parameters in the potential (\ref{ho potential}), and for each single group of chosen parameters one will have to run the GFDN (\ref{eq:gfdn}) for the corresponding $\phi_g$. This is also quite costly and tedious in applications.

These drawbacks motivate us to study the deep neural network approach for the GS problem. We shall propose a normalized deep neural network (norm-DNN) for the effective approximation of GS, and the norm-DNN can be trained to directly provide any GS within a prescribed parameter plane.

\subsection{Normalized deep neural network}

The concept of neural network exists for a long time. By including multiple hidden layers to make it deep, the deep neural network (DNN) mimics the function of a human brain. It can be trained to learn complex relations with substantial amounts of data, which has found great success in computer vision and natural language processing \cite{RepreL, ImagnetCNN, DLnature, DLNN}.
Here let us first provide a brief introduction to the network architecture of DNN.
In general, given an input $\mathbf{y}_0\in \mathbb{R}^{n_0}$ with $n_0\in\bN^+$, a feedforward DNN with $L\in\bN$ hidden layers reads as a composite function of the form:
\begin{equation}
	f_{\theta}(\mathbf{y}_0) = {\bf F}_{L+1} \circ \sigma \circ {\bf F}_{L} \circ \sigma \circ\cdots \circ {\bf F}_{2} \circ \sigma \circ {\bf F}_{1}(\mathbf{y}_0),\quad \theta:=\{W_l,b_l\}, \label{eq:DNN}
\end{equation}
where all the ${\bf F}_{l}:\bR^{n_{l-1}}\to\bR^{n_l}$ are some affine transforms:
\begin{equation*}
	{\bf F}_{l}(\mathbf{y}_{l-1})=W_l \mathbf{y}_{l-1}+b_l,\quad \by_{l-1}\in\bR^{n_{l-1}},\  1\leq l \leq L+1,
\end{equation*}
with $W_l \in \mathbb{R}^{n_{l} \times n_{l-1}},\ b_l \in \mathbb{R}^{n_{l}}$ for $n_l\in\bN^+$. The last one ${\bf F}_{L+1}$ is called as the output layer and ${\bf F}_{l}$ for $1\leq l \leq L$ are called as the hidden layers, with $n_l$ interpreted as the number of neurons in the $l$-th layer. In particular,  $n_0$ and $n_{L+1}$ denote the dimensions of the input and output, respectively. Moreover, $\sigma:\bR\to\bR$ is known as the activation function that acts on each component of ${\bf F}_l$, i.e., $\sigma \circ {\bf F}_{l}\in \bR^{n_l}$.
For simplicity, we shall restrict to consider the fully connected neural network (FCNN) and assume that all the hidden layers use the same number of neurons, i.e.,  $W=n_1=\ldots=n_{L}$. We refer the total number of hidden layers $L$ as `depth', and we refer the number of neurons $W$ in the hidden layers as `width'.

The power of DNN for approximating a general nonlinear function has received many theoretical supports. The early universal approximation theorem  \cite{GCybenko} already states that for any continuous function $h: [0,1]^d \rightarrow \mathbb{R}$ and any $\varepsilon > 0$, there exists a neural network $f_{\theta}(\bf{x})$ with at least one hidden layer and a sigmoidal activation function  such that
\begin{equation}\label{dnn err}
    \|h(\bx) - f_{\theta}(\bx)\|_{L^\infty} < \varepsilon.
\end{equation}
Later, the theorem has been extended \cite{hornikappro, horniksome,hornikuniversal} to cover cases of other types of target functions and activation functions. For smooth target functions, more delicate error bounds have been established to reveal how the error depends on the depth and width of DNN \cite{barromuniversal, hornikdegree, provable,shenzuowei}.
For further recent analytical achievements, we refer to \cite{E sci,Snna,Lu,Shen}.
The mathematical features of the GS from Section \ref{sec:pre_gs} shows that $\phi_g$ certainly belongs to the case that can be covered by the theoretical results, so an effective DNN approximation for our GS problem does exist, and then the primary focus of this paper is to address the practical way for computing it.

On the other hand, DNN in recent years has been extensively applied and developed to solve PDEs \cite{MLCM, DLBM, DRM, SHD,PINNs, DGM, WAN}. To tackle the GS problem (\ref{eq:min}), one could think of using these PDE approaches for instance, the physics-informed neural networks (PINNs) \cite{PINNs} to solve the GFDN \eqref{eq:gfdn}. However, we find  this may not be a good option. To apply methods like PINNs to \eqref{eq:gfdn}, besides the special efforts needed to treat the piece-wisely defined PDE and the normalization in \eqref{eq:gfdn}, one will need to fix the time interval $t\in[0,T]$ to solve the PDE for the solution. Unfortunately, the terminal time $T>0$ for the GFDN \eqref{eq:gfdn} to reach a steady state is not known in advance. Improperly prescribed $T$ may either lead to an inaccurate final state or a waste of computational resource.
Thus, in this work we propose to consider DNNs directly for the minimization (\ref{eq:min}), and practical ways will be presented to find the one that theoretically exists in (\ref{dnn err}).

\begin{figure}[t!]
	\centering
	\includegraphics[width=0.75\textwidth]{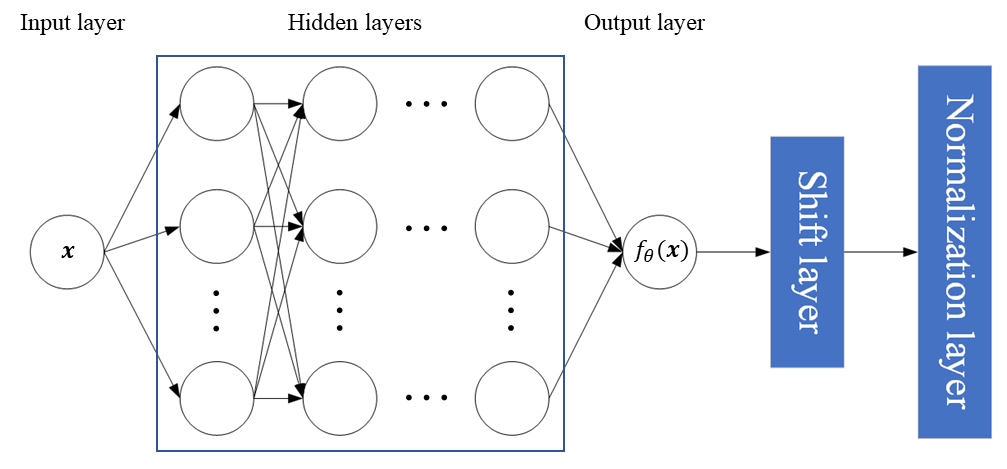}
	\caption{Illustration of the architecture of norm-DNN.}
    \label{fig:archi}
\end{figure}

To construct an effective DNN to approximate the GS $\phi_g$, the basic DNN that we introduce in this paper is the following \textbf{normalized DNN (norm-DNN)}:
\begin{equation}\label{NDNN}
    \phi_\theta(\mathbf{x}) = \mathcal{N} \circ \mathcal{T} \circ {\bf F}_{L+1} \circ \sigma \circ {\bf F}_{L} \circ \sigma \circ\cdots \circ {\bf F}_{2} \circ \sigma \circ {\bf F}_{1}(\mathbf{x}),\quad \bx\in\bR^d,
\end{equation}
where $\mathcal{N}(\by) = \by / ||\by||_2$ with $||\by||_2 = \sqrt{\int_{\Omega} |\by|^2 d\bx}$ and $\by=\by(\bx):\bR^d\to\bR^{n_{L+1}}$ is the \emph{normalization layer} to ensure that the output of norm-DNN always meets the mass constraint in (\ref{eq:min}).
Moreover in (\ref{NDNN}), we have introduced another special layer that we call
the \emph{shift layer}: $\mathcal{T}(\by) = \by  - \min_{\bx}(\by)$, $\by=\by(\bx):\bR^d\to\bR^{n_{L+1}}$. 
Here the minimization is acting on each component of $\by$.
It provides a non-negativity restriction and
helps the norm-DNN (\ref{NDNN}) in a minimization process to turn to the GS instead of an excited state. 
For the scalar/single-component BEC case (\ref{eq:min}), we would clearly take $n_{L+1}=1$.
Figure \ref{fig:archi} illustrates the basic network architecture of norm-DNN.
For the activation function, a classical choice in machine learning would be ReLU, but DNN with ReLU is non-differentiable. Concerning the fact that $\phi_g$ is smooth and the energy \eqref{eq:min} contains a gradient, we choose to consider smooth activation functions such as $\sigma=\tanh$. The detailed strategy to choose $\sigma$ can be problem-dependent, and a better choice for some special occasion like the optical lattice potential case can significantly increase the efficiency of norm-DNN. This will be clarified later.

{We illustrate here that the norm-DNN \eqref{NDNN} with the proposed two additional layers will not break the automatic differentiation feature of DNN.  
The norm-DNN (\ref{NDNN}) can be denoted as $\phi_\theta(\bx)= \mathcal{N} (\by(\bx;\theta))$ with $\by(\bx;\theta)=\mathcal{T} \circ f_\theta(\bf{x}) = \textit f_\theta(\bf{x}) - \min_\bx \textit f_\theta(\bx)$ and $f_\theta(\bf{x})$ 
the standard DNN \eqref{eq:DNN}.}
{It is well-known that the derivatives of $f_\theta$ such as  $\partial_{\bf{x}}f_\theta$ and $\partial_\theta f_\theta$ can be computed by the automatic differentiation in Python for DNN. Now we consider the computation of $\partial_{\bf{x}} \phi_\theta$. Since $\|\by\|_{2}$ and $\min_\bx f_\theta(\bf{x})$ are independent of $\bx$, we can find}
$$
{\partial_{\bf{x}} \phi_\theta= \frac{1}{\left\| \by \right\|_{2}}  \partial_\bx f_\theta.}
$$
{For $\partial_\theta  \phi_\theta$, we have }
\begin{equation*}
	{\partial_\theta  \phi_\theta  = \frac{\partial_\theta \by}{\|\by\|_{2}} - \frac{\by\int_{\Omega}  \partial_\theta \by\cdot\by d\bf{x}}{\|\by\|_{2}^3},}
\end{equation*} 
{with 
	$\partial_\theta \by= \partial_\theta f_\theta(\bf{x}) -{\min_\bx} \partial_\theta \textit f_\theta(\bx)$.}
{Then, $\partial_{\bf{x}} \phi_\theta$ and $\partial_\theta  \phi_\theta$ can be found out with $\partial_{\bf{x}}f_\theta$ and $\partial_\theta f_\theta$ available.  Thus, the derivatives of the norm-DNN (\ref{NDNN}) can still be computed via the automatic 
differentiation.}

To apply the deep learning algorithm, we need to define the optimization objective, also known as the loss function for norm-DNN. For the GS problem \eqref{eq:min}, it is now natural to consider the energy functional $E$ as the loss, i.e.,
$\text{Loss}:=E(\phi_\theta),$ and then $\phi_\theta\approx\phi_g$ would be the target norm-DNN (\ref{NDNN}) that we look for.
For a practical implementation, the whole space integration  in Loss/energy (\ref{energy}) has to be truncated to a finite domain $\Omega\subset\bR^d$ and
the integral has to be discretized. Therefore, the practical loss that we will use is defined as
\begin{equation}
\label{eq:loss}
    \begin{split}
    \text{Loss}=\text{Loss}(\theta)&:=\frac{|\Omega|}{N} \sum_{j=1}^{N} \left[ \frac{1}{2} |\nabla \phi_\theta(\mathbf{x}_j)|^2 + V(\mathbf{x}_j)|\phi_\theta(\mathbf{x}_j)|^2 + \frac{ \beta}{2} |\phi_\theta(\mathbf{x}_j)|^4 \right],
    \end{split}
\end{equation}
where  $|\Omega|$ denotes the measure of the domain and
$\{\mathbf{x}_{j}\}_{j=1}^{N}\subset\Omega$ are the quadrature points for approximating the integral with $N>0$ the total number of points. Since $\phi_g$ is exponentially localized, as long as $\Omega$ is large enough the truncation error of the domain would be negligible. Note that no boundary constraint is imposed in the loss. The quadrature points here are interpreted as the training set for DNN. For a high-dimensional problem, they are often generated by a Monte Carlo method to overcome the curse-of-dimensionality \cite{DRM}. The price is of course the low accuracy from the  random sampling and the result might be polluted by noise. In this work, let us first consider $\{\mathbf{x}_{j}\}_{j=1}^{N}$ as fixed equally distributed grid points in $\Omega$, then (\ref{eq:loss}) is the Riemann sum and so the quadrature error of LOSS can be very small. This will allow us to focus on studying the performance of the proposed norm-DNN on GS computing without the interference from the sampling noise. For the high-dimensional examples, we will test the Monte Carlo sampling. More sampling issues  would be investigated and improved in future by incorporating recent developments like the adaptive sampling \cite{ADNNs} or the quasi-Monte Carlo sampling \cite{QMC}.
\begin{remark}
    {Here we briefly discuss the computational complexity of norm-DNN \eqref{NDNN} compared with the traditional method like GFDN \eqref{eq:gfdn}. The implementation of GFDN needs a mesh grid, which means with $N$ grid points in each direction, the total number of unknowns is then $N^d$. Therefore, its storage and computational costs grow exponentially with respect to the dimensionality $d$. This is not the case for norm-DNN \eqref{NDNN}, whose unknowns are the matrices and vectors in the layers. Note that only the size of  $W_1$ and $b_1$ in the first layer $F_1$ of \eqref{NDNN}  depends linearly on $d$, while the others can be totally independent of $d$.   Additionally, the total error of traditional methods like GFDN is an accumulation of discretization  error in each dimension, so usually the error constant increases along with $d$. For the DNN approach, such approximation error is believed to be independent or weakly dependent on $d$ \cite{shenzuowei}.} 
\end{remark}

Now with (\ref{NDNN}) and (\ref{eq:loss}), we can turn the original GS problem (\ref{eq:min}) which is a constraint optimization in  functional space, into an unconstrained minimization in finite dimensional parameter space, i.e.,
\begin{equation}\label{min dnn}
\min_\theta\{\text{Loss}(\theta)\}.
\end{equation}
To get the optimized norm-DNN (\ref{NDNN}) for (\ref{min dnn}),  the basic approach is the gradient descent method: update the parameters  as
$  \theta^{n+1} = \theta^{n} - \tau \nabla \text{Loss}(\theta^n)$ for $n\geq0,$
where $\tau$ denotes the learning rate and $\theta^0$ is an initial guess.
For the high-dimensional case of (\ref{eq:loss}) where the number of training points may be very large, the stochastic gradient descent (SGD) method or the mini-batch gradient descent (m-BGD) method could also be applied to release the computational costs per iteration.
However, SGD and m-BGD may have extra difficulties to provide fast and reliable convergence  for training  \cite{saddlepoint,ASAM}.
To avoid the additional troubles so that we can focus on the development of the network in this paper, the Adam optimizer \cite{adamKingma2015} will be utilized for (\ref{min dnn}), which is a special optimization method widely considered in deep  learning.
Moreover, we will employ another commonly used training strategy in machine learning: gradually reducing the learning rate during the training process \cite{innb,vdcn}. This can make the training process more stable and accelerate convergence.
More precisely, here the initial learning rate $\tau$ is set as $10^{-3}$ and it decays after every $100$ steps with a base of $0.99$.
{We remark that the very recent work \cite{WANCO} could be an alternative DNN way for the GS problem \eqref{eq:min}.}

To initialize the norm-DNN parameters for optimization (\ref{min dnn}), i.e., $\theta^0$, we employ the Xavier method \cite{glorot2010understanding} which generates the initialization in a random way.
The training stops when the mean value of the loss function in adjacent 100 iterations becomes  relatively stable, i.e., with a given threshold $tol>0$,
$$\left|\sum_{j=100(k+1)}^{100(k+2)} \text{Loss}(\theta^{j}) -  \sum_{j=100k}^{100(k+1)} \text{Loss}(\theta^{j})\right| \bigg/ \left|\sum_{j=100k}^{100(k+1)} \text{Loss}(\theta^{j}) \right| < tol, \quad k=0,1,2\cdots.$$
To quantify the accuracy, we shall compute the relative $L^2$ error 
\begin{equation}
    \label{eq:relative_error}
    \text{error}:=\left\|\phi_\theta-\phi_g\right\|_2 / \left\|\phi_g\right\|_2
\end{equation} 
between the exact solution and the approximate solution from norm-DNN. The ${L}^2$-norm in (\ref{eq:relative_error}) will be calculated as the square-root of the Riemann sum over a finer grid (much finer than the grid for training in (\ref{eq:loss})) which can be  interpreted as the test set. For numerical studies throughout the paper, the number of test points is 2048 for 1D problems and is $400\times400$ for 2D problems. The exact solution $\phi_g$ will be obtained from GFDN (\ref{eq:gfdn}), and owning to the fact (i) in Section \ref{sec:pre_gs}, it will be determined up to a sign.
Notations for the parameters are summarized in Table \ref{tab:parameters}, and the value of them will be specified in each  experiment later.

\begin{table*}[htbp]
	\centering
	\caption{Summary of notations and parameters for the network.}
	\label{tab:parameters}
	\begin{tabular*}{9cm}{ll}
		\hline
		Notation  & Meaning \\
		\hline
		$L$  & Depth of the network \\
		$W$  & Width of the network \\
		$\tau$  & Initial value of learning rate \\
		$tol$  & Threshold to stop training  \\
		$N_{x}$  & Number of training points in  $x$-direction \\
            $N_{y}$  & Number of training points in  $y$-direction \\
		\hline
	\end{tabular*}
\end{table*}

All the numerical experiments in this paper are programmed in PyTorch sequentially and conducted on a laptop with an Intel Core i9-12950HX processor and 32GB of memory. The codes are available in https://github.com/1761121438/Norm-DNN-for-computing-the-ground-state-of-BEC.git.

\section{Application to {1-dimensional problems}}\label{sec:3}
Though DNN is meant for high-dimensional problems which we would eventually go for, here we would like to begin with the {1D} case to clarify the basics and investigate the performance of norm-DNN. The GS problem by norm-DNN (\ref{min dnn}) in 1D now reads:  $\mathbf{x}=x\in \mathbb{R}$ and
\begin{equation}
	\label{eq:loss_gs}
	\theta^{*} = \argmin_{\theta}
	\frac{|\Omega|}{N_{x}} \sum_{j=1}^{N_{x}} \left[ \frac{1}{2} |\nabla \phi_\theta(x_j)|^2 + V(x_j)|\phi_\theta(x_j)|^2 + \frac{1}{2} \beta |\phi_\theta(x_j)|^4 \right].
\end{equation}
We shall fix $\Omega=(-12,12)$, $\beta=400$ in this section. With numerical experiments, we shall illustrate in a sequel the  {superiority and necessity of the normalization layer $\mathcal{N}$ and the shift layer $\mathcal{T}$} in norm-DNN, the acceleration of convergence with pre-training and the influence of the network architecture.
{Then, we will present some special strategies for the optical lattice potential case.}

\subsection{{Normalization layer}}\label{sec:norm}
\begin{figure*}[htbp]
	\centering
	\subfigure{
		\begin{minipage}[t]{0.45\textwidth}
			\centering
			\includegraphics[width=1\textwidth]{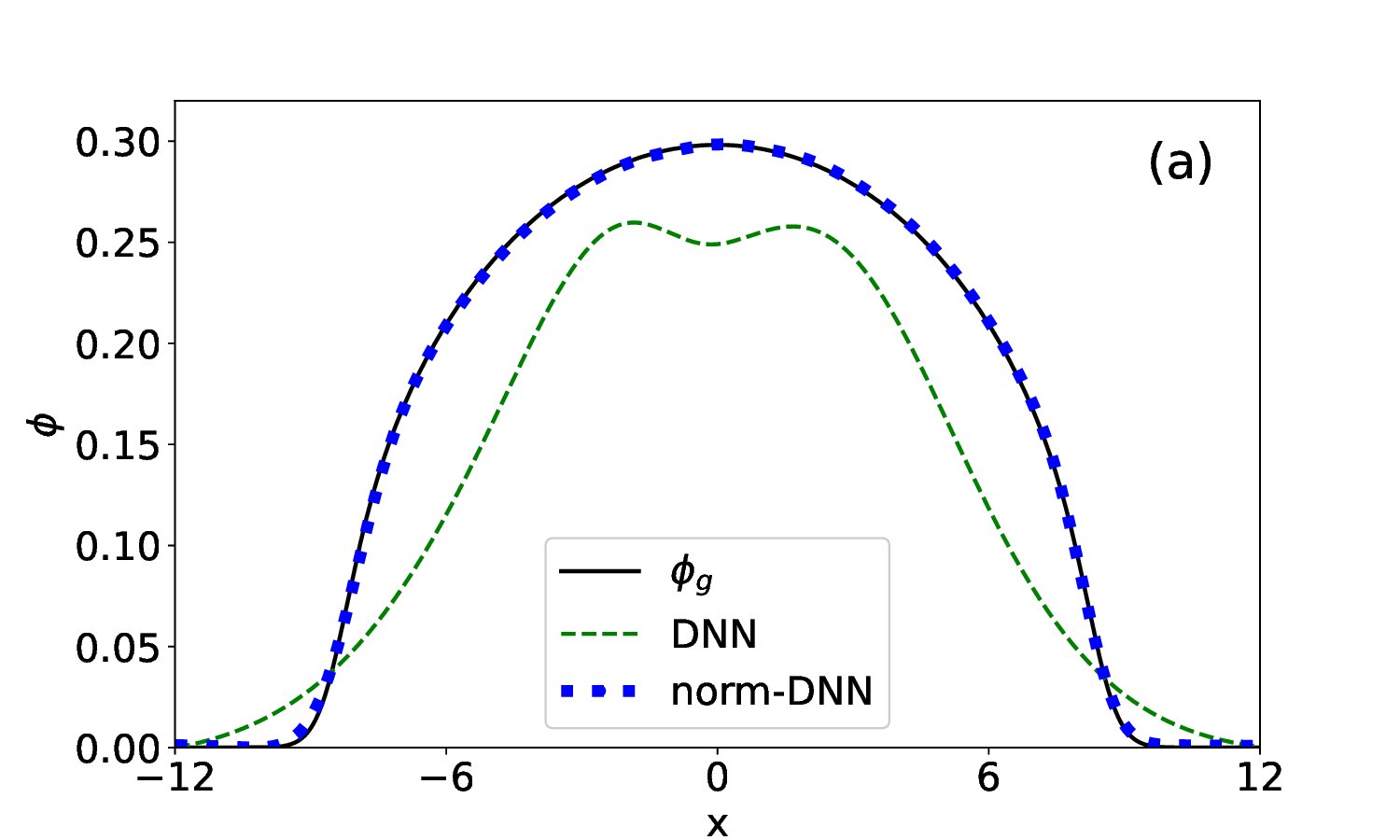}
	\end{minipage}}
	\subfigure{
		\begin{minipage}[t]{0.45\textwidth}
			\centering
			\includegraphics[width=1\textwidth]{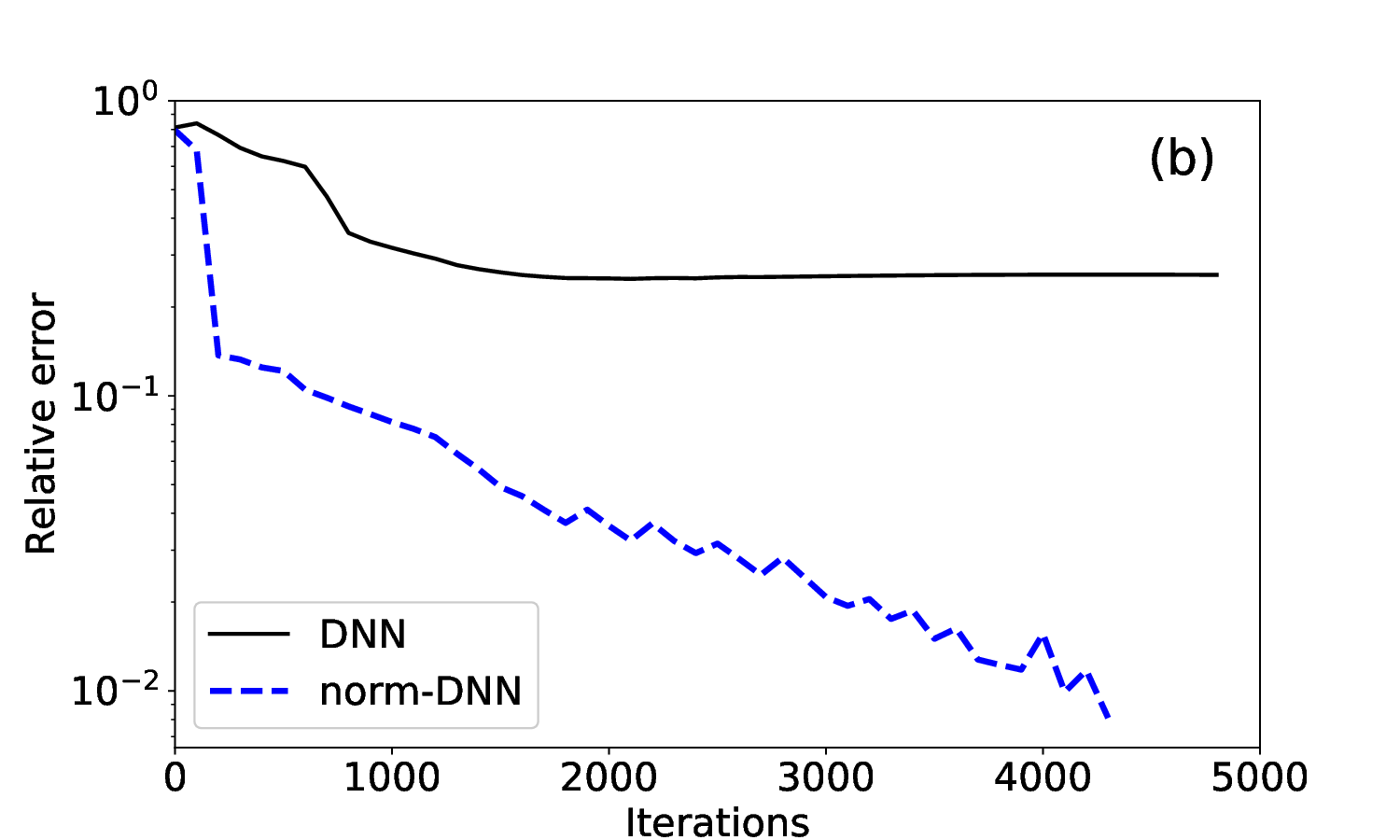}
	\end{minipage}}
        \vspace{-3mm}
	\caption{{Results for GS problem \eqref{eq:min} with norm-DNN \eqref{NDNN} and DNN \eqref{eq:DNN}: (a) exact and numerical solutions ($\left\| f_{\theta} \right\|_2 = 0.7928$ for DNN and $\left\| \phi_{\theta} \right\|_2 = 1$ for norm-DNN); (b) error during the iterations.}}
	\label{fig:norm_dnn}
\end{figure*}

{In norm-DNN \eqref{NDNN}, the  normalization layer $\mathcal{N}$ is proposed to provide the unitary mass in \eqref{eq:min}. To show its superiority, here we  consider the usual soft-constraint way as a comparison.  That is to  consider the standard DNN (\ref{eq:DNN}) to approximate GS and impose the  mass constraint as a penalty term in the loss function \eqref{eq:loss}. The corresponding loss function for the comparison under standard DNN then reads} 
$$
{\text{Loss}(\theta):=E(f_{\theta})+\eta  \left( \left\| f_{\theta}\right\|_2 -1 \right)^2,}
$$
{where $f_{\theta}$ represents the output of  (\ref{eq:DNN}).
Let $V(x)=\frac{1}{2} x^2$, $\sigma=\tanh$, and 
the hyper-parameters are set as $L=4$, $W=70$, $tol=10^{-6}$, $N_x=128$, $\eta=100$, which are not specifically tuned   either for DNN or for norm-DNN. 
The relative error along with iterations, the exact and numerical solutions for norm-DNN and DNN are shown in Figure \ref{fig:norm_dnn}.}

{As it can be seen from Figure \ref{fig:norm_dnn},  the standard DNN \eqref{eq:DNN} with penalty is ineffective under the set of parameters.  It may work by carefully tuning the hyper-parameters especially the weight $\eta$ for penalty, but this could be  relatively involved and tedious. 
On the contrary, the norm-DNN can accurately and efficiently capture the GS.
This example illustrates here that the proposed hard constraint $\mathcal{N}$  goes beyond the typical soft penalty-type constraint, which makes norm-DNN more effective for learning GS.}

\subsection{Shift layer}\label{sec:positive}
From the review  in Section \ref{sec:pre_gs}, it is known that GS is determined up to a phase shift. Since our computational parameters in norm-DNN are all real-valued, a valid $\phi_\theta$ resulting from the training  (\ref{eq:loss_gs}) should be an approximation of either $|\phi_g|$ or $-|\phi_g|$. However, a gradient flow type optimization method in general cannot guarantee to get the global minmizer/GS, because the GS problem (\ref{eq:min}) is a well-known non-convex optimization \cite{BaoCai} and the minimization for norm-DNN in general is far away from being convex as well. In addition, the Xavier initialization \cite{glorot2010understanding} and the SGD (if applied) contain randomness, so it could happen that part of $\phi_\theta$  converges to $|\phi_g|$ and some other part converges to $-|\phi_g|$. This phenomenon will be observed in the coming numerical experiment and the obtained $\phi_\theta$ will only be a local minimum/excited state instead of the correct GS. To overcome this issue, here comes the shift layer $\mathcal{T}$ that we insert between the output layer and the normalization layer $\mathcal{N}$ in the norm-DNN (\ref{NDNN}).
For the 1D GS problem, it is now defined as
\begin{equation}
\label{eq:x_min}
    \mathcal{T}(y)=y - \min_x(y), \quad y=y(x):\bR\to \bR,
\end{equation}
in order to provide a non-negativity restriction on the norm-DNN:
\begin{equation}\label{1d ndnn}
    \phi_{\theta}(x) = \mathcal{N} \circ \mathcal{T} \circ {\bf F}_{L+1} \circ \sigma \circ {\bf F}_{L} \circ \sigma \circ\cdots \circ {\bf F}_{2} \circ \sigma \circ {\bf F}_{1}(x).
\end{equation}
Then the norm-DNN (\ref{1d ndnn}) will always go to a non-negative function through the training (\ref{eq:loss_gs}). Now let us first investigate its performance for computing GS under the harmonic oscillator potential case, and we fix the activation function as $\sigma=\tanh$.

\begin{table*}[htbp] 	
	\centering 	
	\caption{Error (\ref{eq:relative_error}) for 1D GS problem with or without shift layer $\mathcal{T}$.} 	
	\label{tab:positive_error} 	
\begin{tabular}{c|c|ccc}
    \hline
 Error \eqref{eq:relative_error}  & \diagbox{$L$}{$W$}   & 10       & 50       & 70       \\ \hline
    \multirow{3}{*}{with $\mathcal{T}$} & 1 & 5.70E-2 & 5.90E-2 & 6.01E-2 \\
                      & 2 & 6.34E-2 & 4.90E-2 & 1.51E-2 \\
                      & 3 & 4.38E-2 & 6.07E-3 & 1.72E-2 \\
                      & 4 & 4.63E-2 & 7.10E-3 & 1.05E-2 \\ \hline
    \multirow{3}{*}{without $\mathcal{T}$} & 1 & 1.43E+0 & 1.42E+0 & 1.40E+0 \\
                      & 2 & 1.40E+0 & 1.43E+0 & 1.40E+0 \\
                      & 3 & 1.43E+0 & 1.40E+0 & 1.40E+0 \\
                      & 4 & 1.43E+0 & 1.40E+0 & 1.33E+0 \\ \hline

    \end{tabular}
\end{table*}

\begin{table*}[htbp] 	
	\centering 	
	\caption{Numerical energy $E(\phi_{\theta})$ for 1D GS problem with or without shift layer $\mathcal{T}$ (exact $E(\phi_{g})=21.3601$).} 	
	\label{tab:energy} 		
\begin{tabular}{c|c|ccc}
    \hline
 $E(\phi_{\theta})$  & \diagbox{$L$}{$W$}   & 10       & 50       & 70       \\ \hline
    \multirow{3}{*}{with $\mathcal{T}$} & 1 & 21.3874 & 21.3922 & 21.3903 \\
                      & 2 & 21.4010 & 21.3755 & 21.3522 \\
                      & 3 & 21.3715 & 21.3513 & 21.3530 \\
                      & 4 & 21.3734 & 21.3556 & 21.3517 \\ \hline
    \multirow{3}{*}{without $\mathcal{T}$} & 1 & 22.1470 & 22.2626 & 22.1150 \\
                      & 2 & 23.0825 & 23.2284 & 23.1349 \\
                      & 3 & 23.0455 & 23.5141 & 23.6430 \\
                      & 4 & 22.9423 & 23.3252 & 23.4760 \\ \hline

    \end{tabular}
\end{table*}

\begin{figure*}[t!]
	\centering
	\subfigure{
		\begin{minipage}[t]{0.45\textwidth}
			\centering
			\includegraphics[width=1\textwidth]{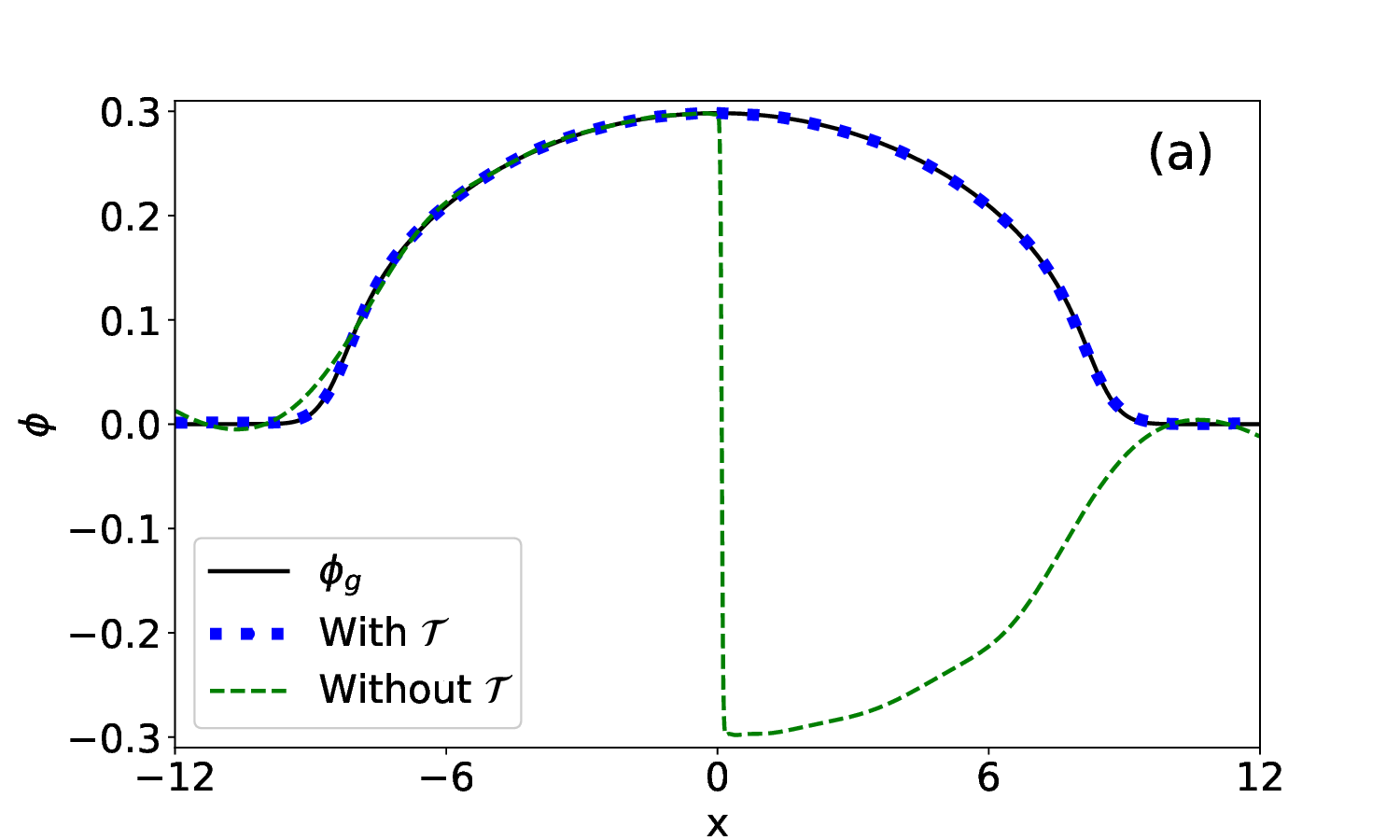}
	\end{minipage}}
	\subfigure{
		\begin{minipage}[t]{0.45\textwidth}
			\centering
			\includegraphics[width=1\textwidth]{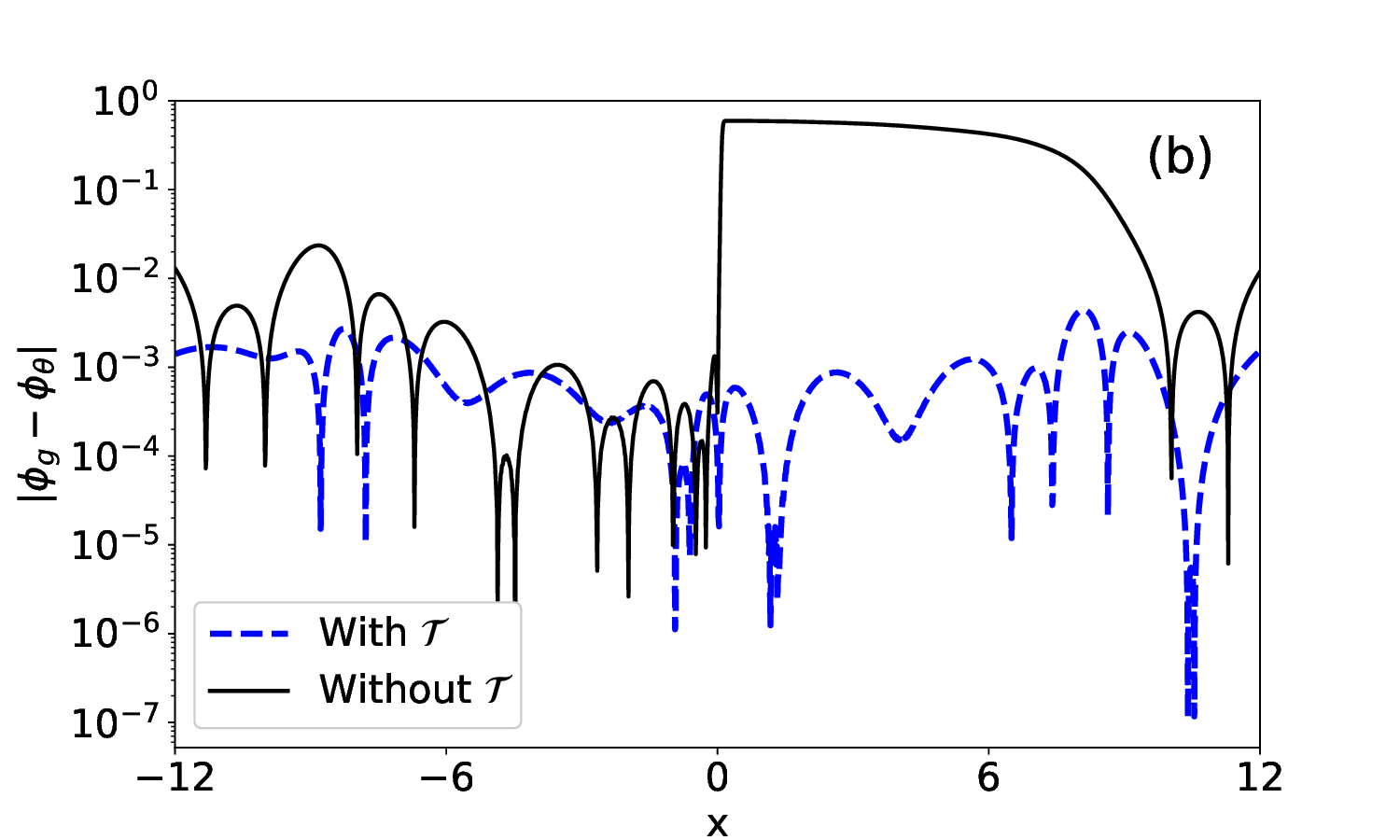}
	\end{minipage}}
        \vspace{-3mm}
	\caption{Results for GS problem under $L=3$, $W=50$: (a) exact solution and numerical solution with or without $\mathcal{T}$; (b) pointwise error of numerical solution.}
	\label{fig:positive}
\end{figure*}

We take $V(x)=\frac{1}{2}x^2$  for the numerical experiment.
Setting the hyper-parameters as $L\in \left\{1,2,3,4 \right\}$, $W\in \left\{10,50,70 \right\}$, $tol=10^{-6}$, $N_{x}=128$, we solve the 1D GS problem as described before. To show the necessity of the shift layer $\mathcal{T}$, we compare the performance of the norm-DNN with $\mathcal{T}$ as defined in  (\ref{1d ndnn}) or without $\mathcal{T}$ (or set $\mathcal{T}=id$).
The relative errors (\ref{eq:relative_error}) under the two cases are presented in Table \ref{tab:positive_error}. The numerical energy $E(\phi_{\theta})$ for norm-DNN with or without $\mathcal{T}$ is shown in Table \ref{tab:energy}. Here the exact GS energy $E(\phi_{g})$ is $21.3601$, which can be obtained from GFDN.

It is clear from the results in Table \ref{tab:positive_error} that in the absence of $\mathcal{T}$, the relative errors always exceed one and so the approximations are completely invalid.
In contrast, when $\mathcal{T}$ (\ref{eq:x_min}) is included, the relative errors of norm-DNN (\ref{1d ndnn}) become much smaller, reaching as low as $6\times 10^{-3}$.
From Table \ref{tab:energy}, we find that $E(\phi_{\theta})$ of norm-DNN with $\mathcal{T}$ is closer to $E(\phi_{g})$, while $E(\phi_{\theta})$ of norm-DNN without $\mathcal{T}$ is larger.
This show that norm-DNN without $\mathcal{T}$ will produce some excited states.
Under $L=3$, $W=50$, we show the exact solution $\phi_g$ and the numerical solution $\phi_\theta$ with or without using $\mathcal{T}$ in Figure \ref{fig:positive}. It can be observed that without $\mathcal{T}$ to fix the sign, the norm-DNN $\phi_\theta$ will partially go to the positive GS and partially go to the negative one, and in the end the training will produce an excited state. This highlights the crucial role of the shift layer $\mathcal{T}$ in enabling norm-DNN to accurately capture the GS.  We remark that though we have used the Adam method for optimization here, using SGD will give the same results and conclusion.

\begin{remark}
Instead of imposing the shift layer $\mathcal{T}$ in the norm-DNN (\ref{1d ndnn}), an alternative approach is to make use of the symmetry feature of GS. We can add a regularization term in the loss to  restrict $\phi_\theta(x)$ being even in space, and then we solve the following minimization problem:
\begin{equation}
	\label{eq:loss_even}
	\min_{\theta}\left\{
	\text{Loss}(\theta) + \text{Even}(\theta)\right\},\quad\mbox{where}\quad \text{Even}(\theta)=\frac{\delta}{N_e} \sum_{j=1}^{N_e} |\phi_\theta(x_j^e) - \phi_\theta(-x^e_j)|^2,
\end{equation}
with a weight $\delta>0$ and some training points $\big \{x^{e}_{j}\big \}_{j=1}^{N_{e}}\subset \Omega\cap\bR^+$. Then, (\ref{1d ndnn}) without $\mathcal{T}$ will also lead to the GS. The detailed results are omitted here for brevity. Note in high dimension when the radial symmetry is absent, the shift layer then becomes the only option.
\end{remark}

 \subsection{Gaussian pre-training}\label{sec:gauss}
 In the linear regime of BEC, i.e., $\beta=0$ in (\ref{eq:GPE}), the GS under a harmonic potential can be found out explicitly as a Gaussian function  \cite{BaoCai}. Therefore, traditional methods like GFDN commonly use Gaussian functions as the initial data in (\ref{eq:gfdn}) so that the flow can  avoid local minimums and can  reach the steady state faster. Here we borrow this idea and implement it into our norm-DNN.

\begin{table*}[h!] 	
	\centering 	
	\caption{Number of iterations, computational time and relative error \eqref{eq:relative_error} for norm-DNN (\ref{1d ndnn}) with or without Gaussian pre-training. }
	\label{tab:gauss} 	
	\begin{tabular}{c|ccc} 		\hline 					
						  & Iterations  & Time  &  Error \eqref{eq:relative_error}     \\ \hline 		
		$\phi_\theta$ 	    & 12100 	& 45s  	& 6.04E-2  \\
		$\phi_\theta^{G}$  & 6800 	& 29s  	& 3.11E-2  \\ \hline			
		
	\end{tabular}
\end{table*}

\begin{figure*}[h!]
    \centering
	\subfigure{
		\begin{minipage}[t]{0.45\textwidth}
			\centering
			\includegraphics[width=1\textwidth]{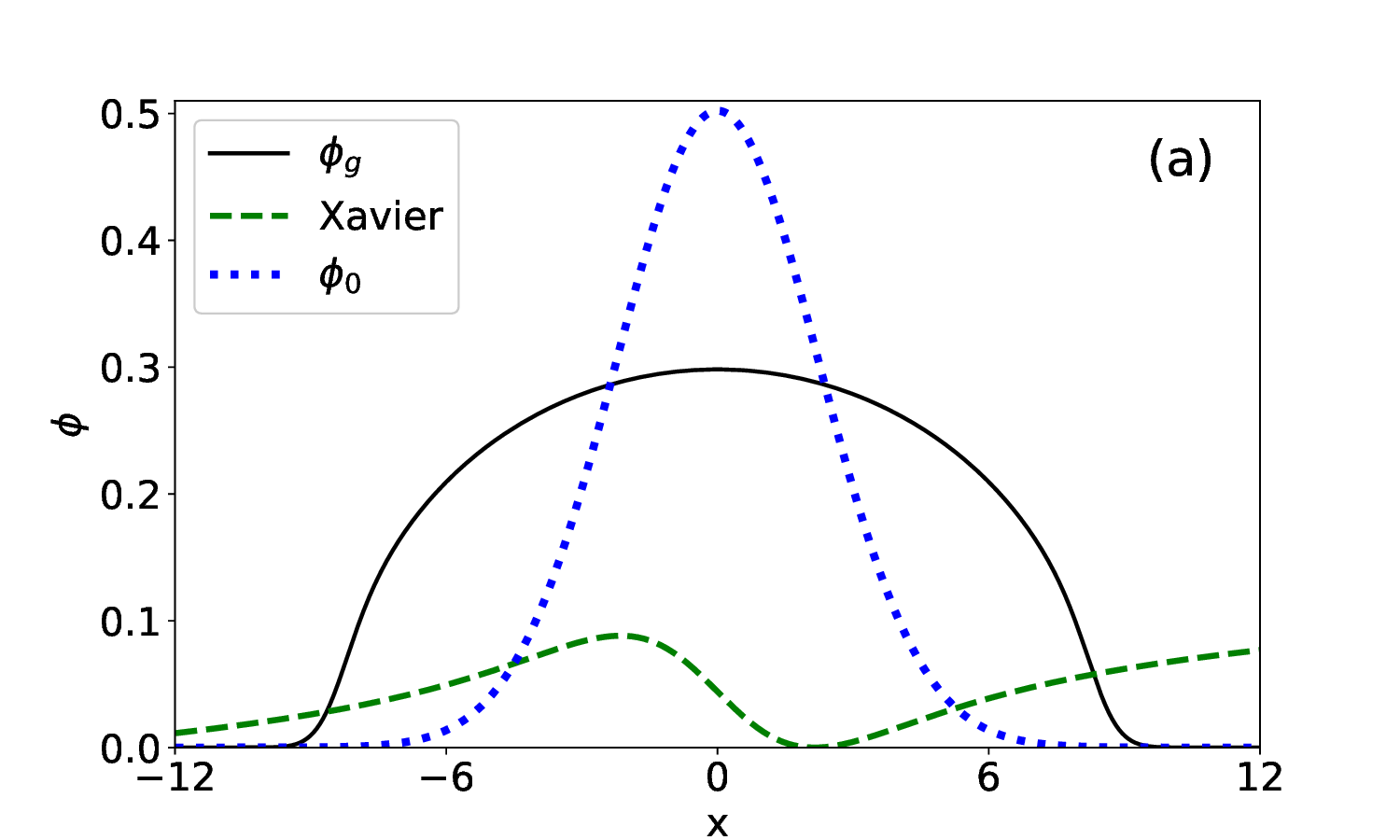}
	\end{minipage}}
 	\hspace{-6mm}
	\subfigure{
		\begin{minipage}[t]{0.45\textwidth}
			\centering
			\includegraphics[width=1\textwidth]{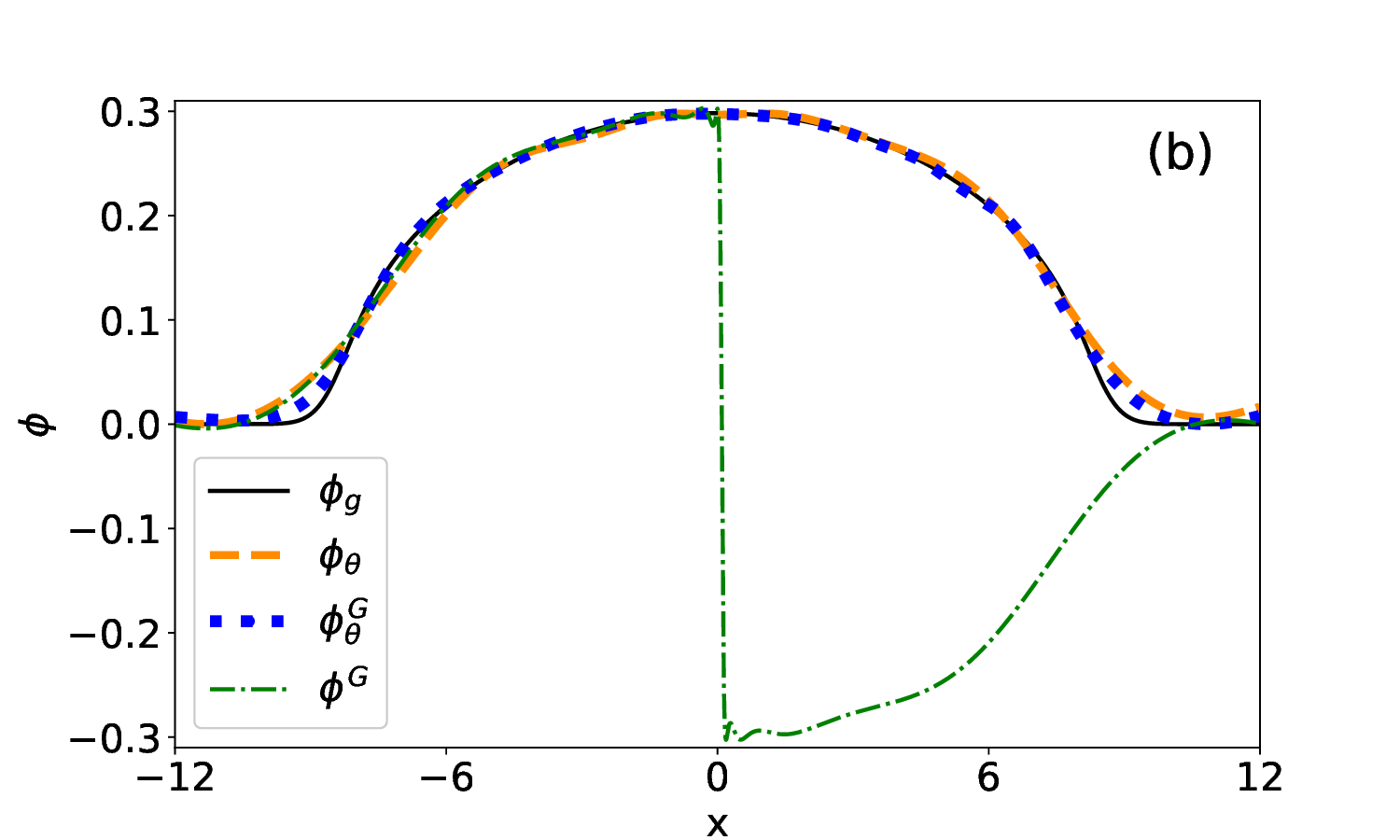}
	\end{minipage}}
	\vspace{-3mm}
	\subfigure{
		\begin{minipage}[t]{0.45\textwidth}
			\centering
			\includegraphics[width=1\textwidth]{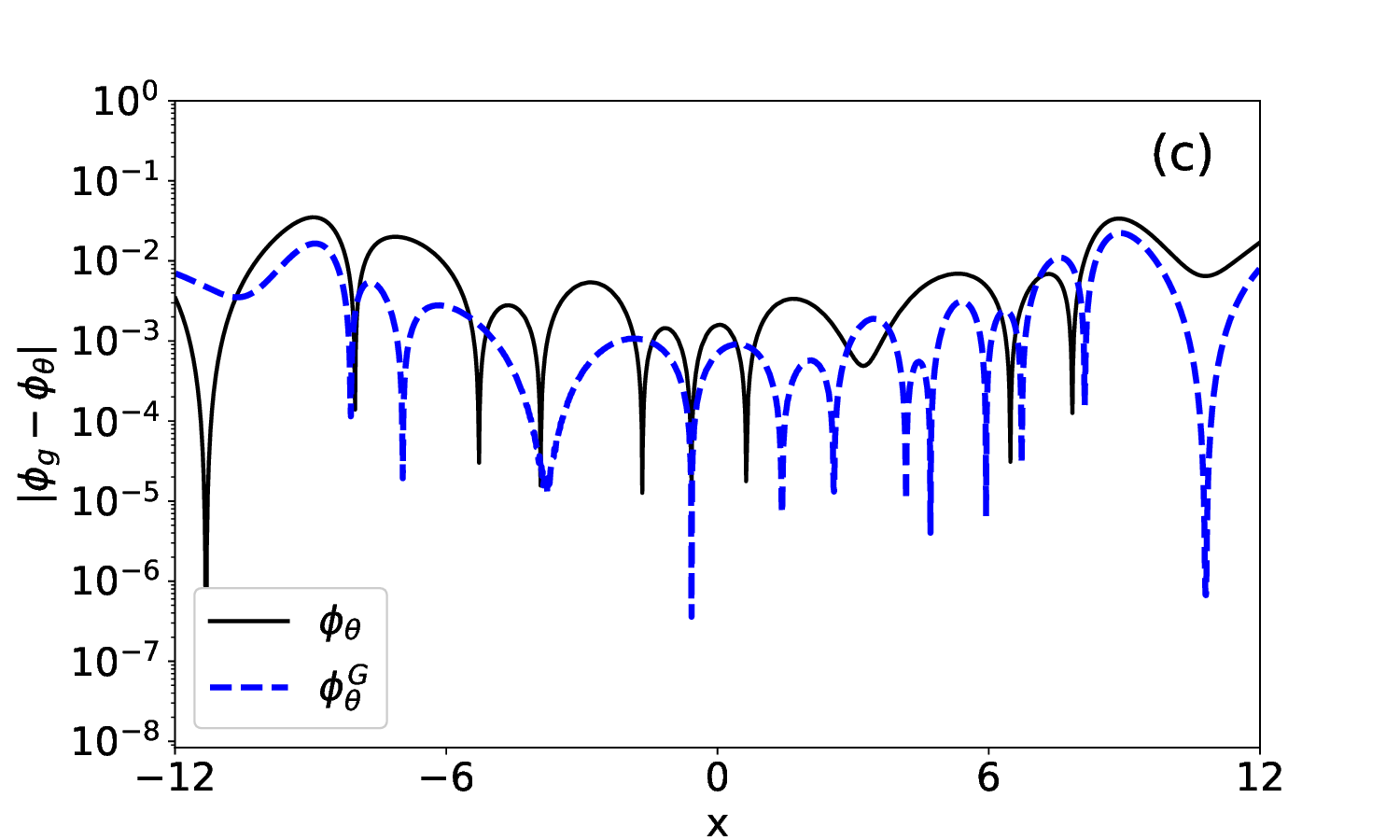}
	\end{minipage}}
	\hspace{-6mm}
	\subfigure{
		\begin{minipage}[t]{0.45\textwidth}
			\centering
			\includegraphics[width=1\textwidth]{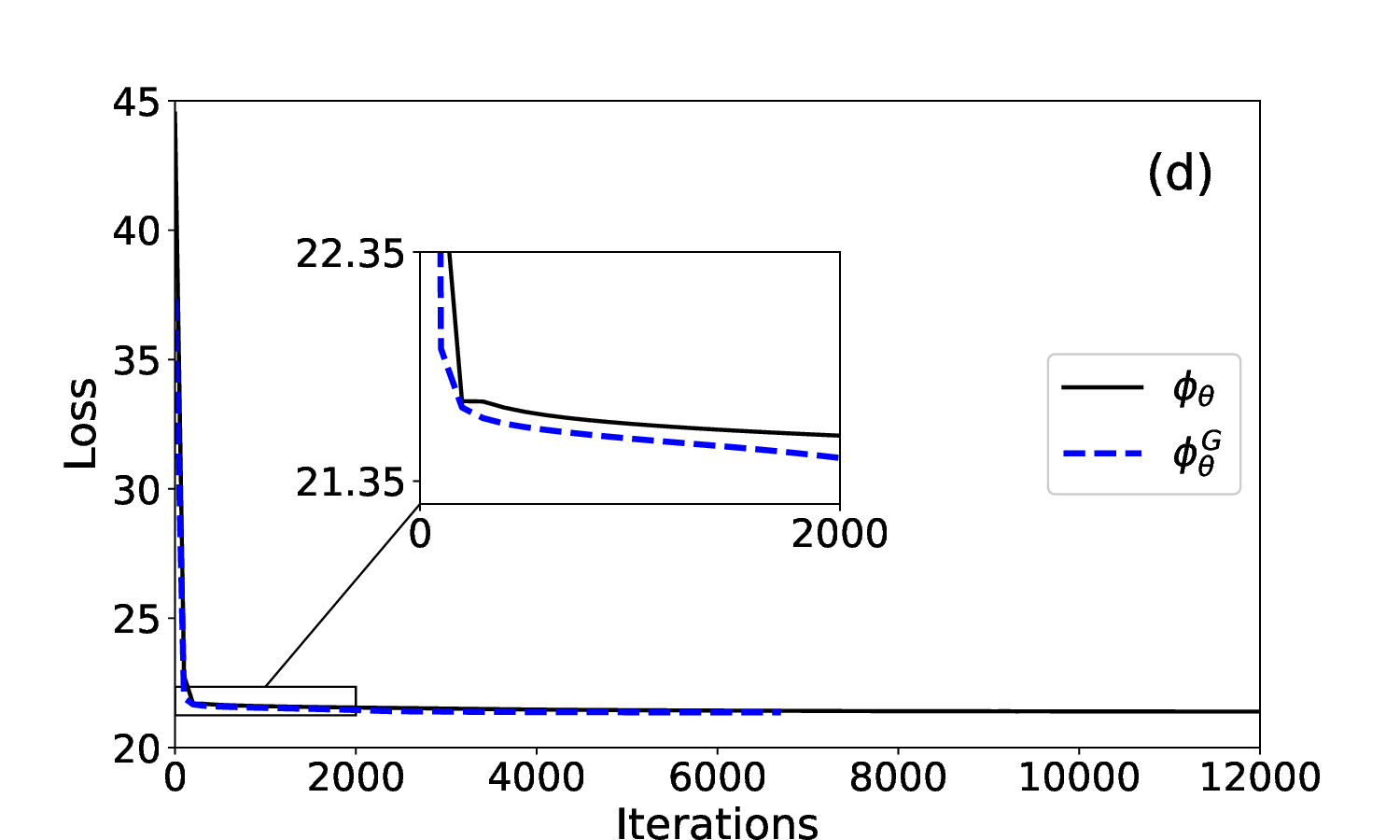}
	\end{minipage}}
    \vspace{-3mm}
	\subfigure{
		\begin{minipage}[t]{0.45\textwidth}
			\centering
			\includegraphics[width=1\textwidth]{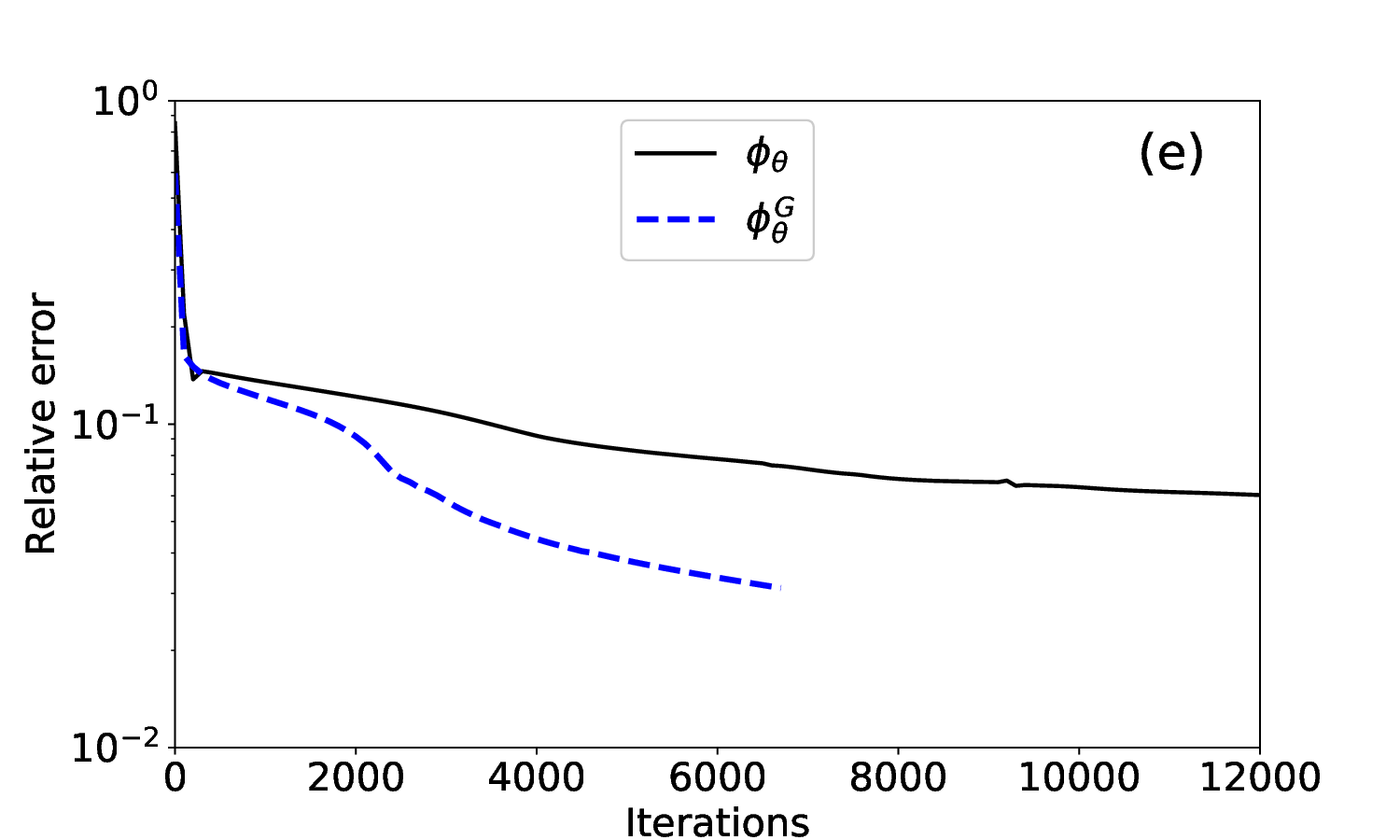}
	\end{minipage}}
	\hspace{-6mm}
	\subfigure{
		\begin{minipage}[t]{0.45\textwidth}
			\centering
			\includegraphics[width=1\textwidth]{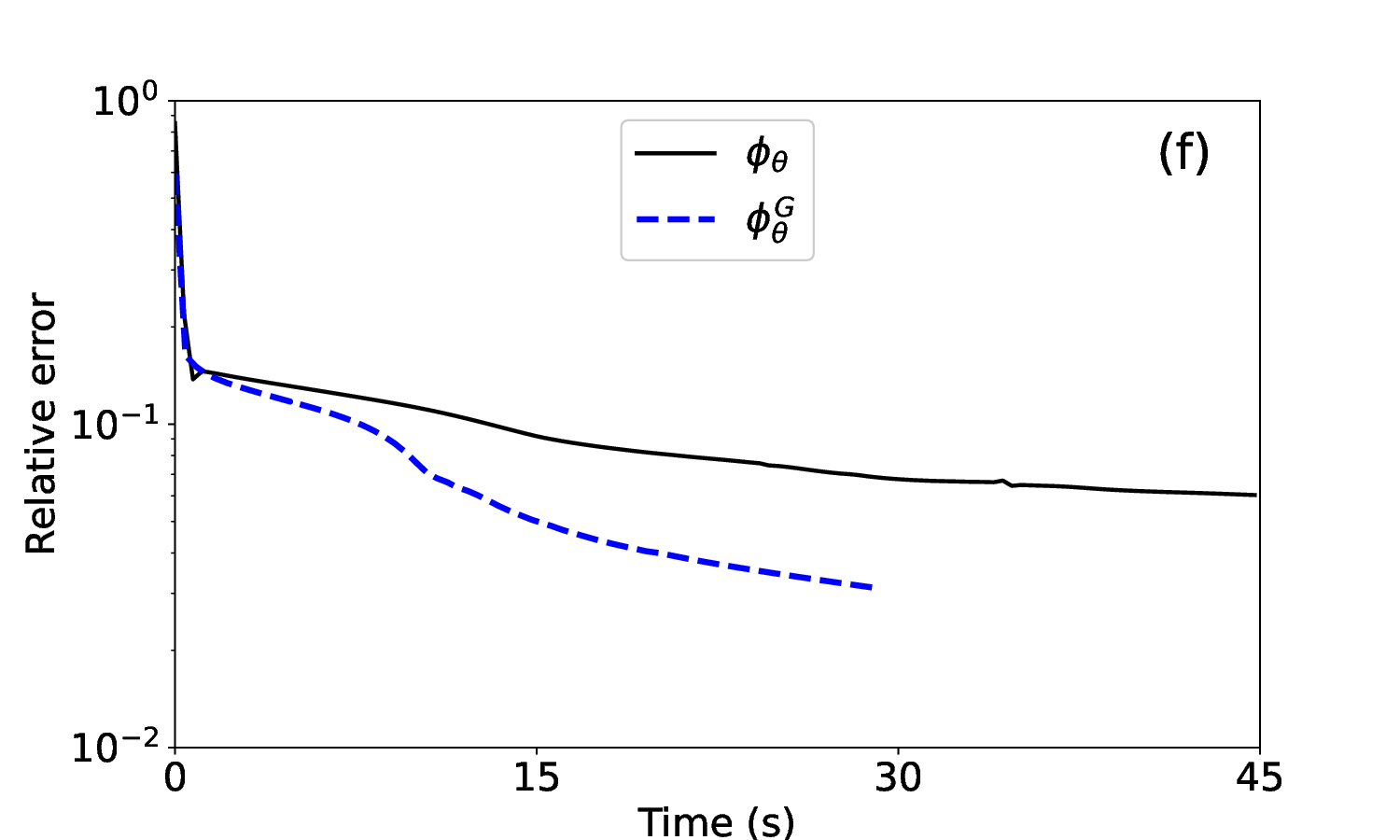}
	\end{minipage}}
    \vspace{-1mm}
    \caption{GS problem: (a) initial profile of norm-DNN from Gaussian pre-training or Xavier method \cite{glorot2010understanding}; (b) exact $\phi_g$, norm-DNN $\phi_\theta^G$ with pre-training and $\mathcal{T}$, norm-DNN $\phi_\theta$ with $\mathcal{T}$ but without  pre-training,
    norm-DNN $\phi^G$ with pre-training but without $\mathcal{T}$; (c) pointwise error; (d,e,f) loss and error (\ref{eq:relative_error}) during the iterations or computational time.}
    \label{fig:gauss}
\end{figure*}

What we propose is to utilize a Gaussian function to pre-train the norm-DNN (\ref{1d ndnn}) to get the initialization $\theta^0$ for the Adam method to solve (\ref{min dnn}). More precisely, with the same numerical example $V(x)=\frac{1}{2}x^2$ and $\beta=400$ as before, we train a norm-DNN (\ref{1d ndnn}) generated from Xavier initialization \cite{glorot2010understanding} for 1000 iterations to fit a normalized Gaussian
$\phi_0 :=\fe^{-\frac{x^2}{10}} / (5\pi)^{1/4}$. Then, we test the performance of the norm-DNN with $L=2$, $W=50$, and the other hyper-parameters are set as $tol=10^{-5}$, $N_{x}=128$.
To illustrate the improvement in efficiency, we consider two cases for comparison: $\phi_\theta^G$ from (\ref{1d ndnn}) with the described Gaussian pre-training; $\phi_\theta$ from (\ref{1d ndnn}) without using the Gaussian pre-training. The total number of iterations used in the optimization/training, the computational time (in seconds)  and the error (\ref{eq:relative_error}) are shown in Table \ref{tab:gauss}. Here and after, the computational time is counted after the pre-training. The profiles of the initial norm-DNNs, the final norm-DNNs and the training process are shown in Figure \ref{fig:gauss}. In addition, we emphasize that the Gaussian pre-training cannot replace the role of the shift layer $\mathcal{T}$. For this purpose, Figure \ref{fig:gauss}(b) shows the numerical solution $\phi^G$ from (\ref{1d ndnn}) with the Gaussian pre-training but without imposing $\mathcal{T}$.

From the numerical results, we have the following observations.
Firstly, by the red line in Figure \ref{fig:gauss}(a), we notice that the initial value generated by the Xavier method is far away from the GS. The initial function in fact does not satisfy any mathematical properties of the GS, and our norm-DNN approach can still accurately get to the GS within a certain number of iterations. This demonstrates the robustness of the proposed norm-DNN (\ref{1d ndnn}).  Secondly, as shown in Table \ref{tab:gauss},
the pre-training with a Gaussian allows norm-DNN to converge to a more accurate GS with much less iterations and  computational time. Figure \ref{fig:gauss}(e,f) illustrate that during the entire training process, norm-DNN with pre-training converges much faster in terms of both iterations and time. The pointwise accuracy of the pre-trained norm-DNN also becomes better as Figure \ref{fig:gauss}(c) indicates. Note in addition that the time needed by pre-training is very short, and this is done once for all.
Thus, we conclude that the pre-training of norm-DNN can gain significantly more efficiency. As it is already available, we will apply directly the pre-trained norm-DNN in all the following experiments.

\subsection{Influence of network architecture}\label{sec:hop}
Although there is no clear theory to guarantee that more hidden layers and more neurons can bring DNNs closer to the global minimum, the convention in deep learning agrees that DNNs with larger architecture tend to gain more accuracy  \cite{chang2022high, DRM,PINNs}.
With the pre-trained initialization fixed, here we carry out a systematical numerical study  on the influence of the network architecture for our norm-DNN (\ref{1d ndnn}).
The hyper-parameters are set as $L\in \left\{1,2,3,4\right\}$, $W\in \left\{10,50,70\right\}$, $tol=10^{-6}$, $N_{x}=128$.

\begin{table*}[h!] 	
	\centering 	
	\caption{Error (\ref{eq:relative_error}) and training time of norm-DNNs with different architecture.} 	
	\label{tab:1d} 	
	\begin{tabular}{c|ccc}
		\hline
		& $W=10$           & $W=50$           & $W=70$          \\ \hline
         $L=1$   & 5.50E-2(109s) & 5.86E-2(87s) & 6.91E-2(170s) \\
         $L=2$   & 5.27E-2(128s) & 9.26E-3(84s) & 6.79E-3(49s) \\
         $L=3$   & 5.56E-2(116s) & 7.54E-3(32s) & 4.89E-3(28s) \\
         $L=4$   & 4.23E-2(92s) & 4.89E-3(26s) & 4.57E-3(22s) \\ \hline
	\end{tabular}
\end{table*}
\begin{figure*}[h!]
	\centering
	\subfigure{
		\begin{minipage}[t]{0.4\textwidth}
			\centering
			\includegraphics[width=1\textwidth]{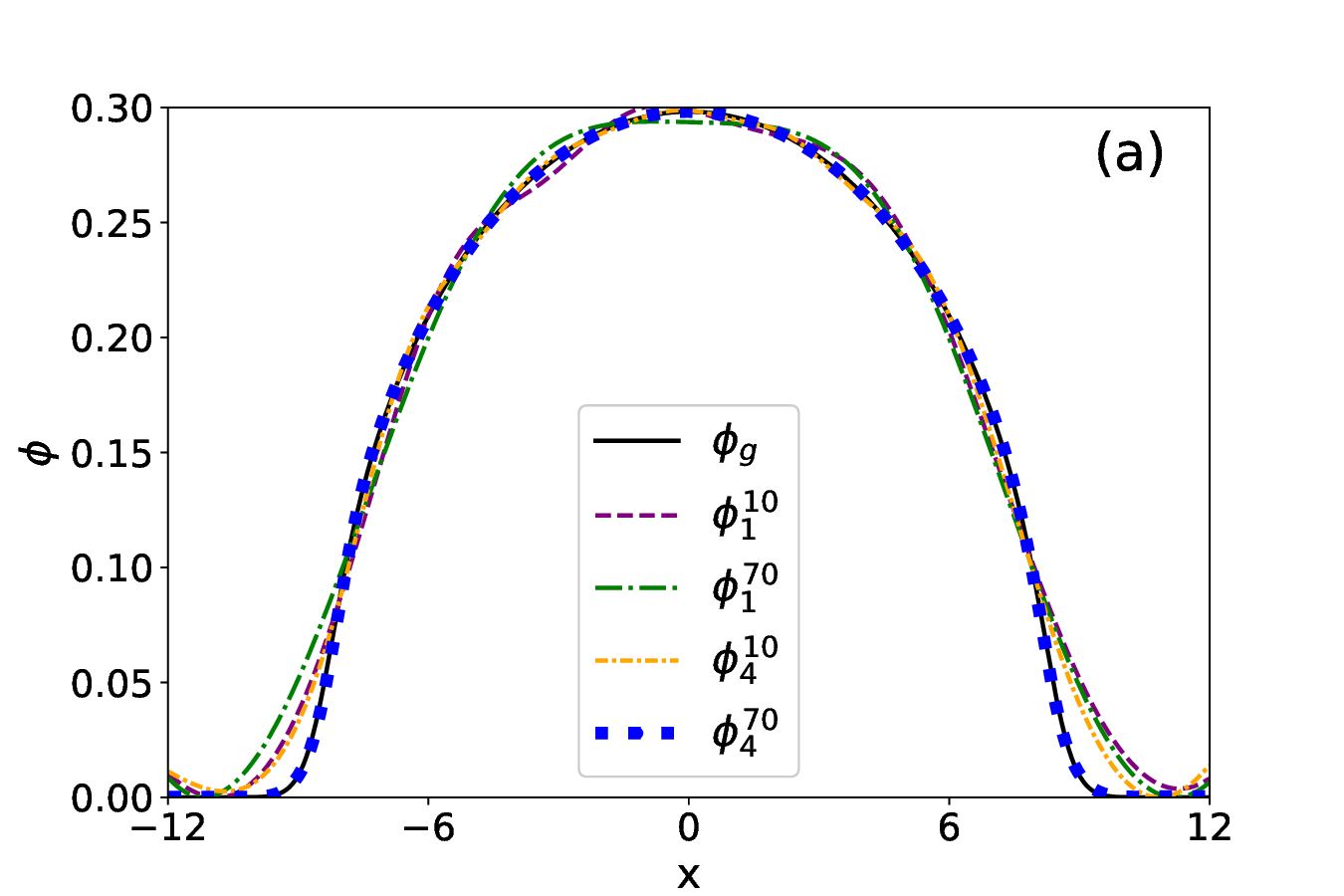}
	\end{minipage}}
	\subfigure{
	\begin{minipage}[t]{0.4\textwidth}
		\centering
		\includegraphics[width=1\textwidth]{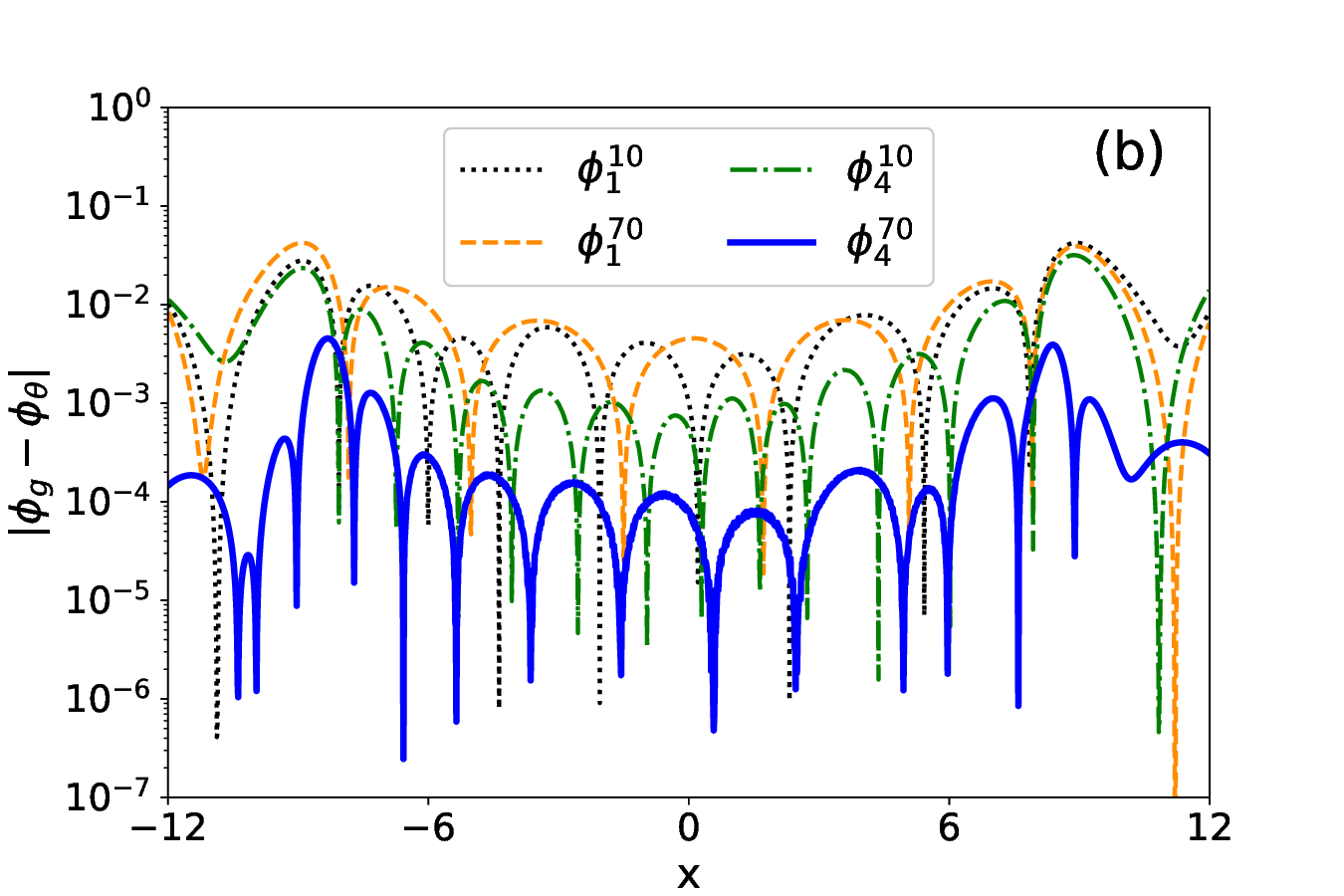}
	\end{minipage}}
	\vspace{-3mm}
	\subfigure{
		\begin{minipage}[t]{0.32\textwidth}
			\centering
			\includegraphics[width=1\textwidth]{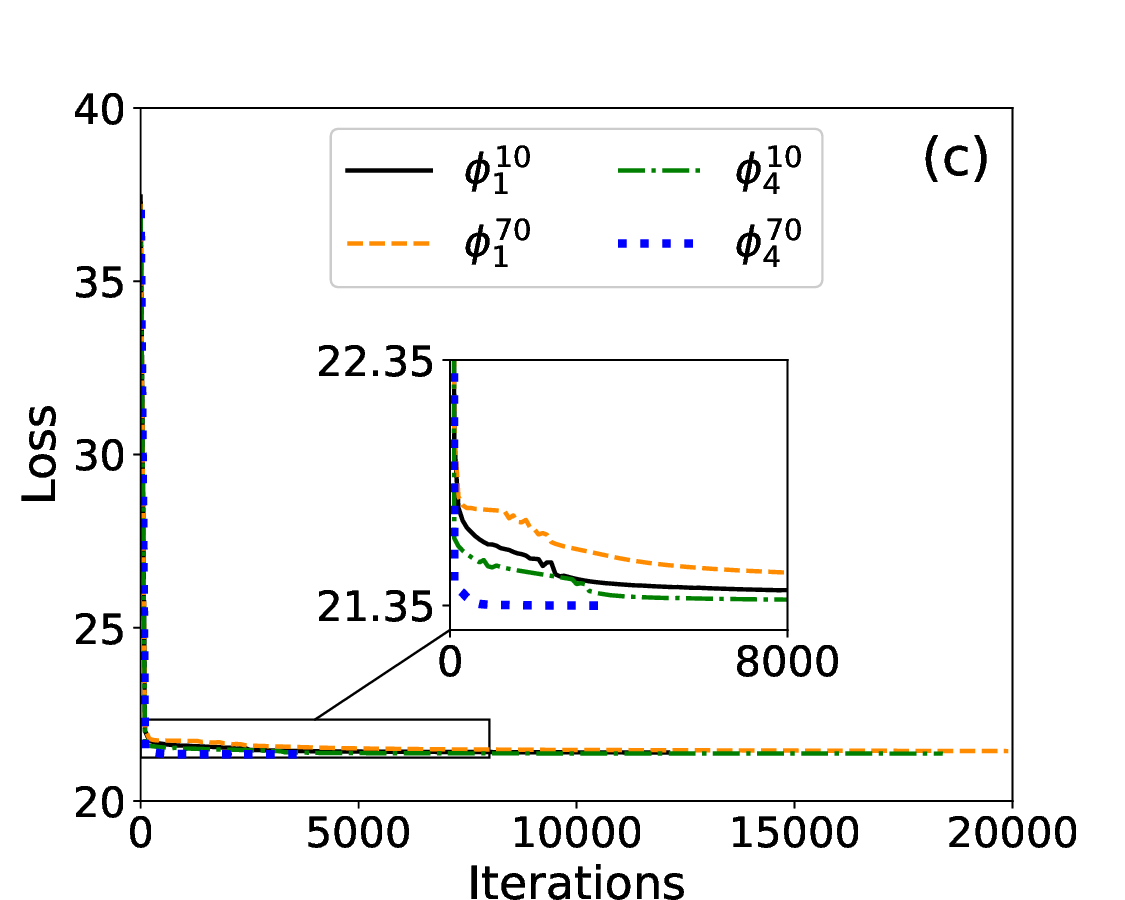}
	\end{minipage}}
	\hspace{-6mm}
	\subfigure{
		\begin{minipage}[t]{0.32\textwidth}
			\centering
			\includegraphics[width=1\textwidth]{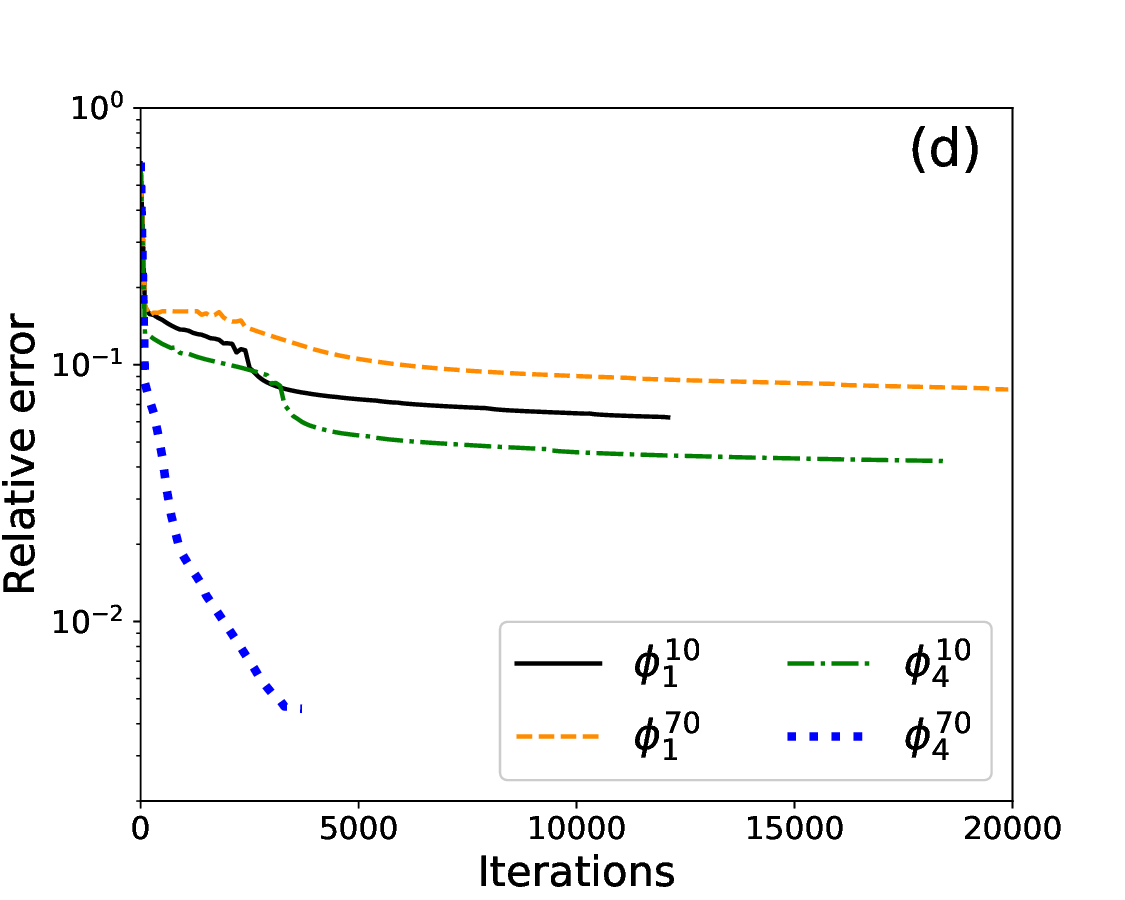}
	\end{minipage}}
	\hspace{-6mm}
	\subfigure{
	\begin{minipage}[t]{0.32\textwidth}
		\centering
		\includegraphics[width=1\textwidth]{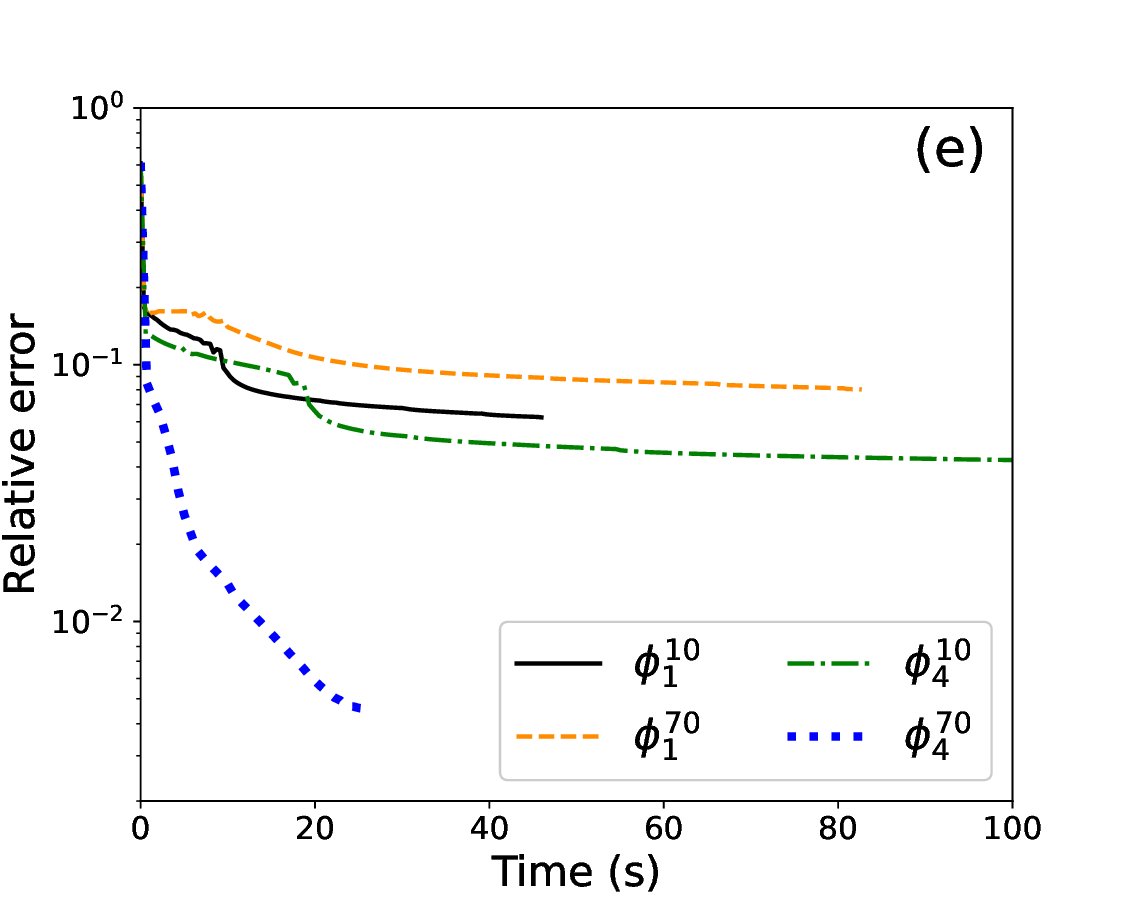}
	\end{minipage}}
        \vspace{-1mm}
	\caption{Some norm-DNNs with different architecture for GS problem: (a) exact solution and numerical solution of norm-DNN; (b) pointwise error; (c,d,e) loss and error (\ref{eq:relative_error}) during the iterations or computational time.}
    \label{fig:1d}
\end{figure*}

The errors (\ref{eq:relative_error}) and the consumed training time of norm-DNN under different $W$ and $L$ are shown in Table \ref{tab:1d}.
Clearly, we can see that both the errors and the training time are decreasing as the depth and width of the network increase.
This general trend indicates that the ability of norm-DNN to approximate GS improves with the increase of the architecture.

Denote $\phi_L^W$ for the norm-DNN (\ref{1d ndnn}) with width $W$ and depth $L$.
The fitting effects of norm-DNN with $L\in \left\{1,4\right\}$, $W\in \left\{10,70\right\}$ for the GS are shown in Figure \ref{fig:1d}(a) and their corresponding pointwise errors are shown in Figure \ref{fig:1d}(b).
It should be noted that if the width or depth is not sufficient, norm-DNN could remarkably deviate from the GS at the corners. Increasing the network architecture to $L=4$ and $W=70$ results in a very good pointwise approximation.
One concern is that the increase of the architecture will lead to the increase of computational cost per iteration in the optimization  for (\ref{eq:loss_gs}).
However, as demonstrated in Figure \ref{fig:1d}(c,d), when the network architecture is sufficiently large, norm-DNN in fact requires no more than 10000 iterations to reduce the relative error down to $4.57\times 10^{-3}$.
Consequently, the total training time is significantly reduced, though the time per iteration increases.

\subsection{Optical lattice potential}\label{sec:olp}
\begin{figure*}[h!]
	\centering
	\subfigure{
		\begin{minipage}[t]{0.4\textwidth}
			\centering
			\includegraphics[width=1\textwidth]{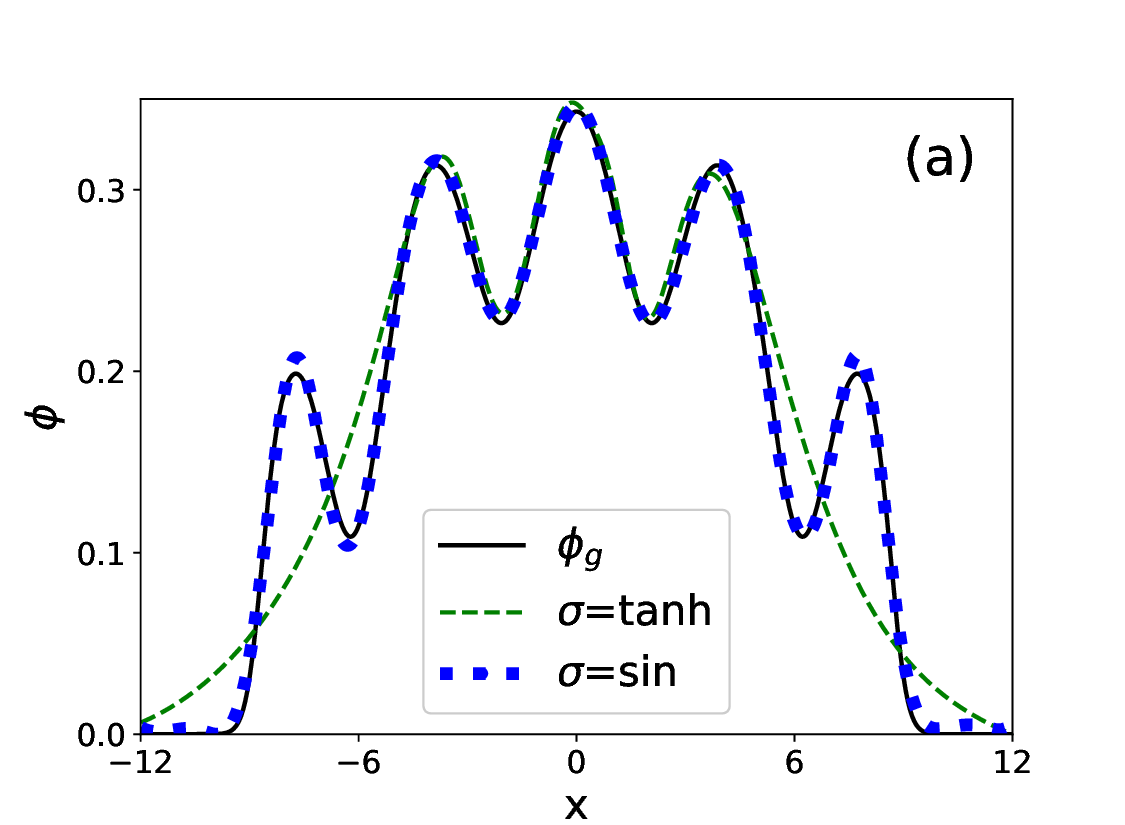}
	\end{minipage}}
	\subfigure{
		\begin{minipage}[t]{0.4\textwidth}
			\centering
			\includegraphics[width=1\textwidth]{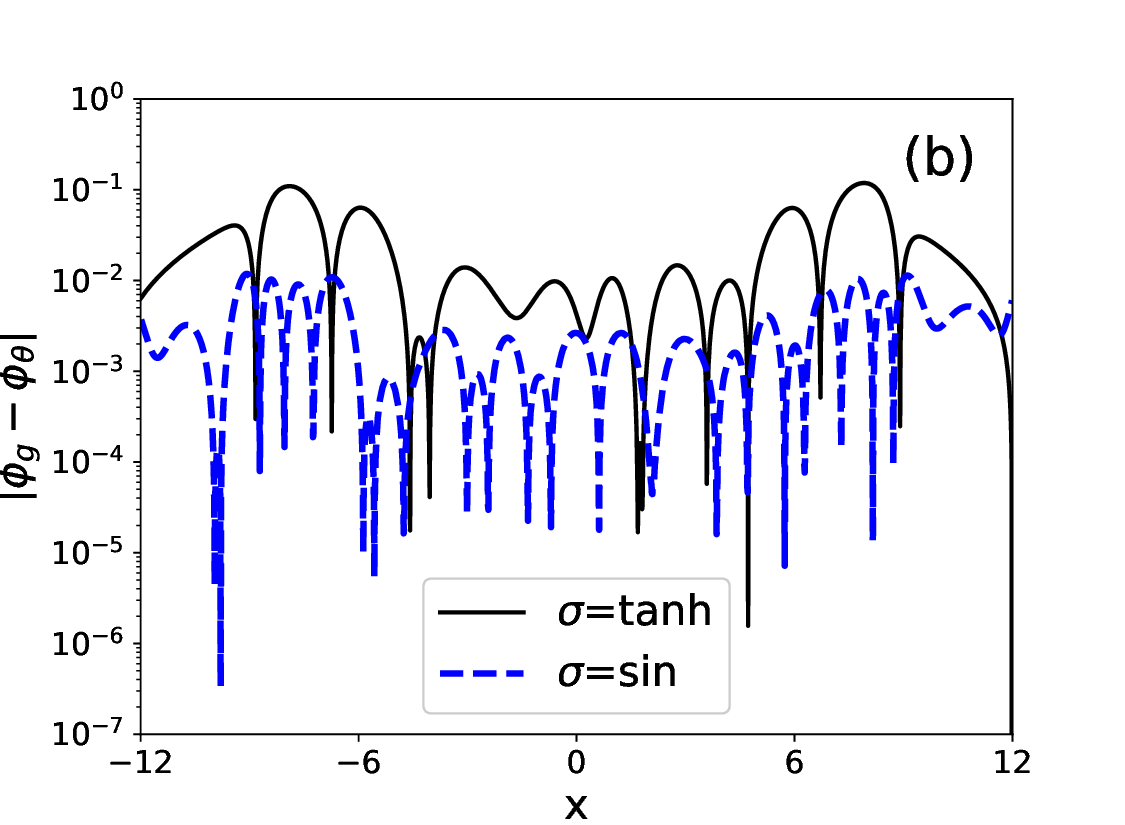}
	\end{minipage}}
        \vspace{-3mm}
	\subfigure{
		\begin{minipage}[t]{0.33\textwidth}
			\centering
			\includegraphics[width=1\textwidth]{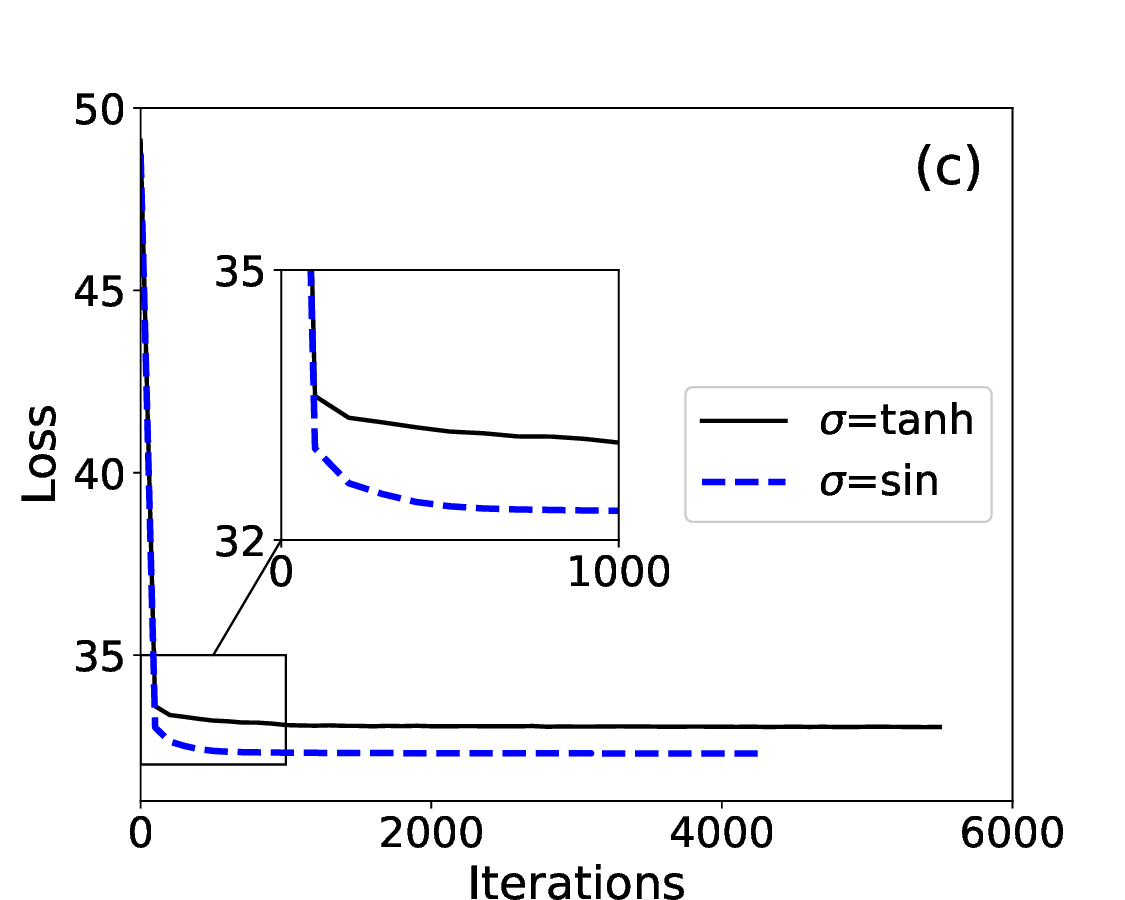}
	\end{minipage}}
	\hspace{-6mm}
	\subfigure{
		\begin{minipage}[t]{0.33\textwidth}
			\centering
			\includegraphics[width=1\textwidth]{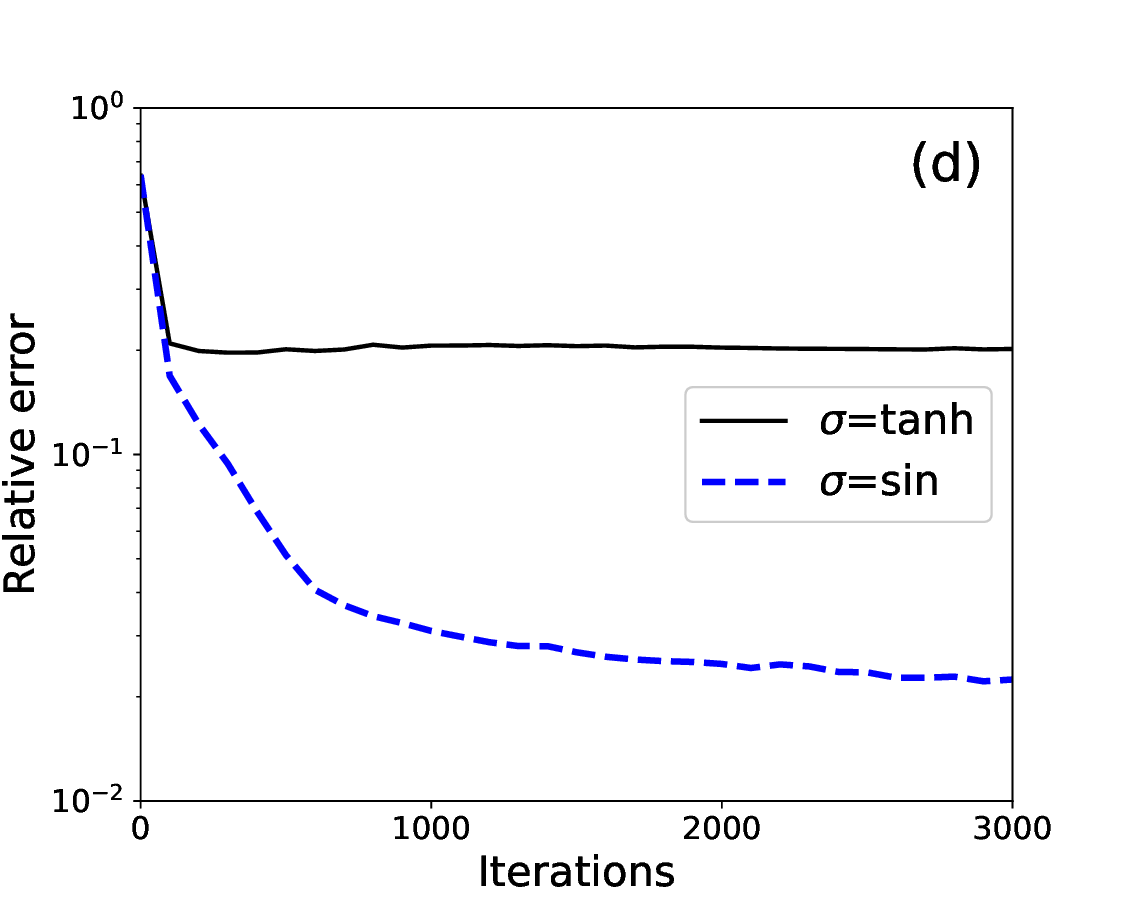}
	\end{minipage}}
	\hspace{-6mm}
	\subfigure{
		\begin{minipage}[t]{0.33\textwidth}
			\centering
			\includegraphics[width=1\textwidth]{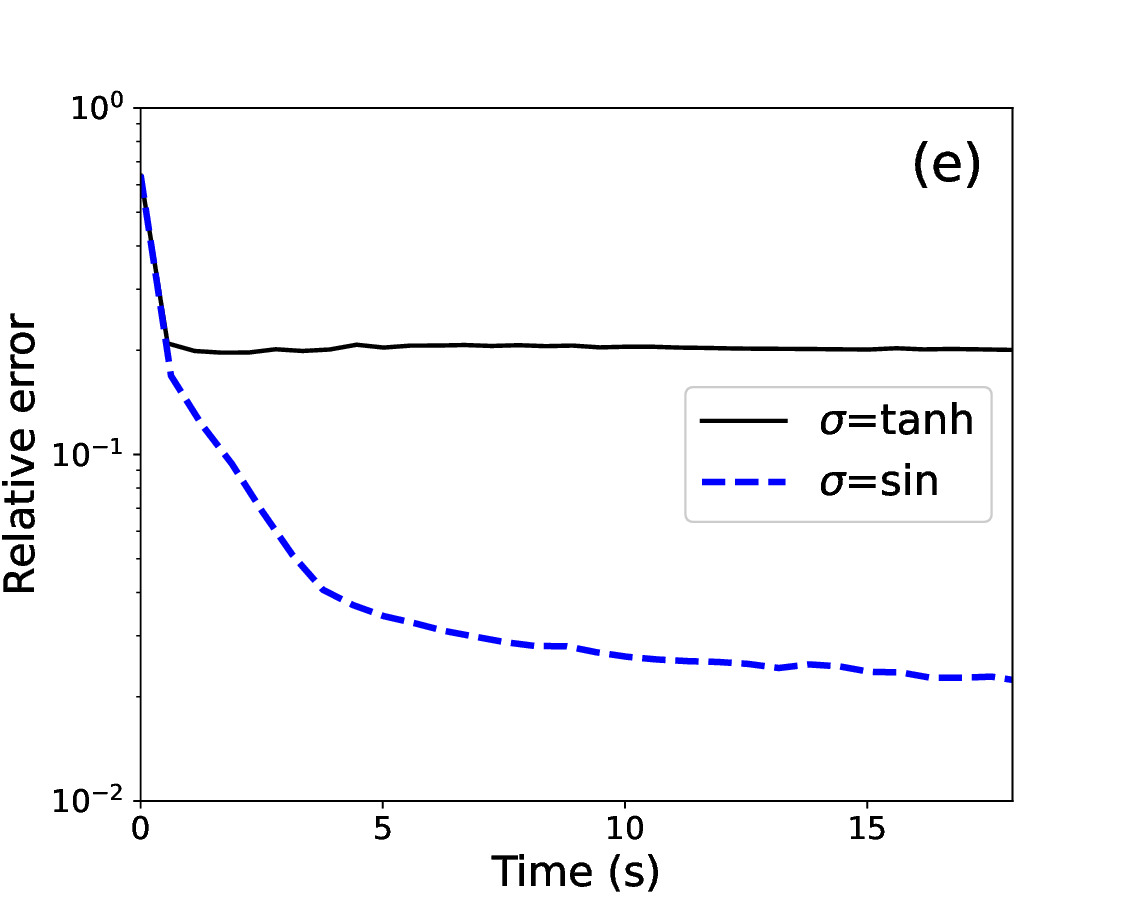}
	\end{minipage}}
        \vspace{-1mm}
	\caption{norm-DNNs with different activation functions for GS problem: (a) exact solution and numerical solution of norm-DNN; (b) pointwise error; (c,d,e) loss and error (\ref{eq:relative_error}) during the iterations or computational time.}
	\label{fig:1d_sin_com}
\end{figure*}

The previous studies are for the harmonic potential case. Now, we consider
the optical lattice potential \eqref{ho potential} case which can contain high-frequency information through the additional trigonometric terms. This will result in some local oscillations in the corresponding GS, e.g., Figure \ref{fig:1d_sin_com}(a). This kind of high-frequency information in the target function  may not be well captured by DNNs with `tanh' as the activation function \cite{chang2022high}.

To investigate such issue in norm-DNN for the GS problem, we take $V(x)= \frac{1}{2}x^{2} + 25 \sin^2(\frac{\pi x}{4})$, $\beta=400$ in  (\ref{eq:loss_gs}) for the numerical experiment. The pre-training with the normalized Gaussian is applied as before, and the hyper-parameters are set as $L\in \left\{3,4 \right\}$, $W=10$,  $tol=10^{-6}$, $N_{x}=128$.  Then, the numerical results of the proposed norm-DNN (\ref{1d ndnn}) with $\sigma=\tanh$ and $L=4$, $W=10$ are shown in Figure \ref{fig:1d_sin_com}, where the results basically tell that the approximation is invalid. To improve the performance of norm-DNN in the optical lattice potential case, we propose to consider another activation function `sin' in (\ref{1d ndnn}), i.e.,
$\sigma=\sin,$
which  brings naturally some oscillations to the network,
and the corresponding numerical results are also shown in Figure \ref{fig:1d_sin_com} as comparison.
To further illustrate the gained accuracy and efficiency, we show
the number of iterations, computational time and error (\ref{eq:relative_error}) for the norm-DNN with $\sigma=\tanh$ or $\sigma=\sin$ in Table \ref{tab:1d_sin_com}.

\begin{table*}[h!] 	
    \centering 	
    \caption{Number of iterations, computational time and error (\ref{eq:relative_error}) of norm-DNN (\ref{1d ndnn}) with `tanh' or `sin' as the activation function. } 	
    \label{tab:1d_sin_com} 	
    \begin{tabular}{c|c|ccc}
        \hline
   $\sigma$    & $W\times L$   & Iterations       & Time       & Error \eqref{eq:relative_error}      \\ \hline
      \multirow{2}{*}{tanh}   & $10\times3$ & 27800   & 131s   & 1.97E-1   \\
                                              & $10\times4$ & 5600   & 30s   & 1.98E-1   \\ \hline

      \multirow{2}{*}{sin}    & $10\times3$ & 5000   & 26s   & 2.23E-2   \\
                                              & $10\times4$ & 4400   & 27s   & 1.97E-2   \\ \hline

        \end{tabular}
\end{table*}

Clearly from the presented numerical results, we find that the norm-DNN with `sin' is able to produce the correct approximation of GS in an efficient and accurate manner. In contrast, the norm-DNN with `tanh' takes more iterations and more time for training, but it is still only able to fit three wave structures in the GS as shown in Figure \ref{fig:1d_sin_com}. The norm-DNN with `sin' can accurately capture all the local oscillatory structures.
Thus, we conclude that the `sin' activation function is the suitable choice for norm-DNN to compute GS in the optical lattice potential case.

\section{{Application to extended problems}}\label{sec:4}
{The norm-DNN method proposed for GS of {1-dimensional} BEC in the previous section, along with the conclusions drawn, is similarly applicable to high-dimensional or multi-component models.} With appropriate adjustments, it can also be applied to compute the first excited state.
In this section, we are going to present these extensions. 
We shall first study the applications of norm-DNN to the two-dimensional and the three-dimensional GS problems.
Then, we shall consider the application to the two-component GS problem and the computation of the first excited state in a sequel.

\subsection{High-dimensional {GS} problem}\label{sec:hop_2d}
We now consider the GS problem (\ref{eq:min}) for the two-dimensional (2D) BEC, i.e., $d=2$ and $\bx=(x,y)^{\top} \in\bR^2$ in (\ref{eq:GPE}). Based on our 1D study, the straightforward extension of the norm-DNN approach to 2D case then considers the optimization:
\begin{equation}
	\label{eq:loss_gs_2d}
	\min_{\theta} \left\{\text{Loss}=
	\frac{|\Omega|}{N_{x}\times N_{y}} \sum_{j=1}^{N_{x}}\sum_{k=1}^{N_{y}} \left[\frac{1}{2} |\nabla \phi_{\theta}(x_j,y_k)|^2 + V(x_j,y_k)|\phi_{\theta}(x_j,y_k)|^2 + \frac{1}{2} \beta |\phi_{\theta}(x_j,y_k)|^4 \right]\right\}
\end{equation}
with
\begin{equation}\label{2d ndnn}
    \phi_{\theta}(x,y) = \mathcal{N} \circ \mathcal{T} \circ {\bf F}_{L+1} \circ \sigma \circ {\bf F}_{L} \circ \sigma \circ\cdots \circ {\bf F}_{2} \circ \sigma \circ {\bf F}_{1}(x,y).
\end{equation}
Here $\mathcal{T}$ is the shift layer  (\ref{eq:x_min}) defined as before. Then, the 2D norm-DNN (\ref{2d ndnn}) will converge to the non-negative GS and the performance is quite similar as before. Let us illustrate this with two numerical examples.
We fix  $\beta=400$ and consider a box domain $\Omega=(-6,6)\times(-6,6)$ with equal-partitions in the following. The exact solution $\phi_g$ is obtained by traditional method. The training of norm-DNN is done by the Adam method with $tol=10^{-7}$, and a normalized Gaussian $\fe^{-(\frac{x^2}{10}+\frac{y^2}{10})} / \sqrt{5\pi}$ is used for pre-training with 1000 iterations.

\begin{figure*}[hbtp]
	\centering
	\subfigure{
		\begin{minipage}[t]{0.3\textwidth}
			\centering
			\includegraphics[width=1\textwidth]{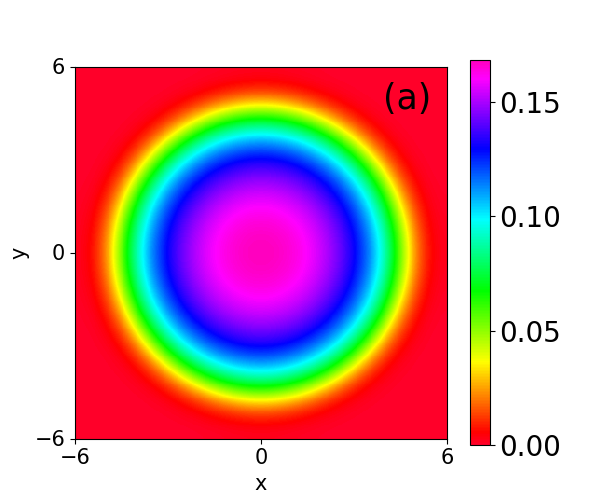}
	\end{minipage}}
	\hspace{-4mm}
	\subfigure{
		\begin{minipage}[t]{0.3\textwidth}
			\centering
			\includegraphics[width=1\textwidth]{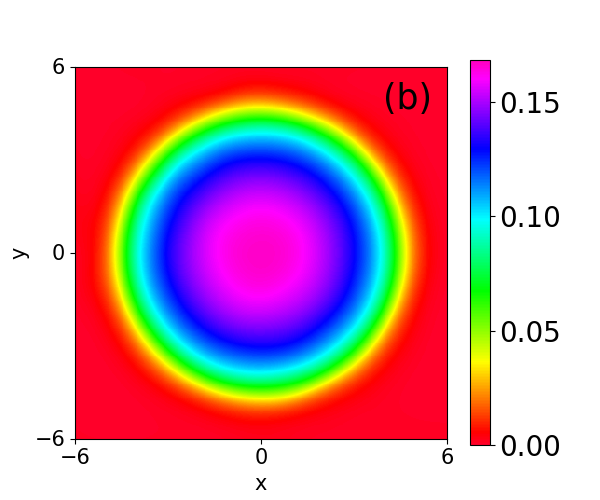}
	\end{minipage}}
	\hspace{-4mm}
	\subfigure{
		\begin{minipage}[t]{0.3\textwidth}
			\centering
			\includegraphics[width=1\textwidth]{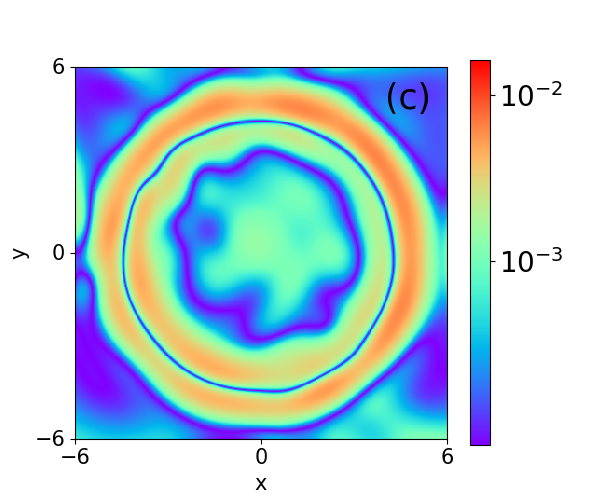}
	\end{minipage}}
	\vspace{-3mm}
	\subfigure{
		\begin{minipage}[t]{0.45\textwidth}
			\includegraphics[width=7cm,height=4cm]{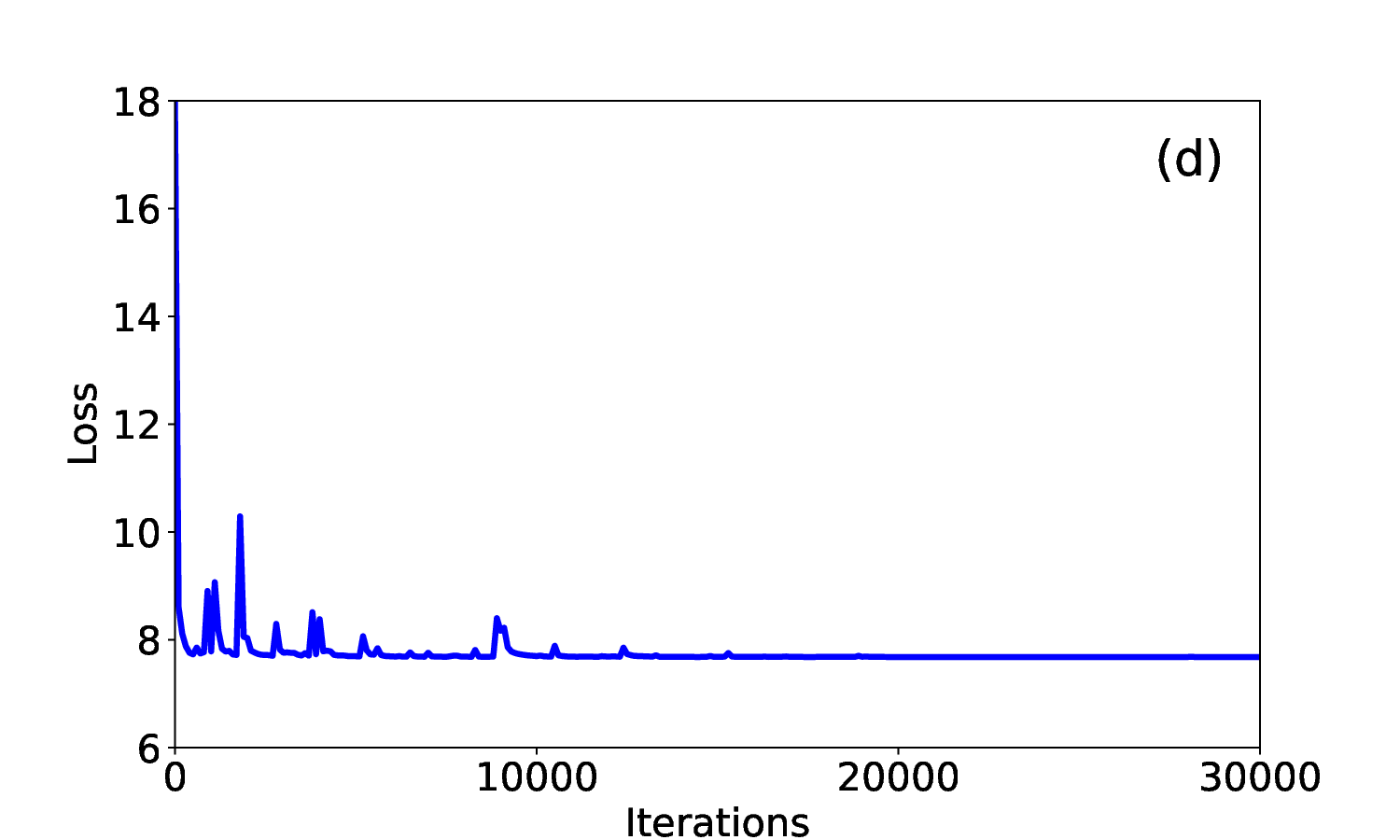}
	\end{minipage}}
	\subfigure{
		\begin{minipage}[t]{0.45\textwidth}
			\includegraphics[width=7cm,height=4cm]{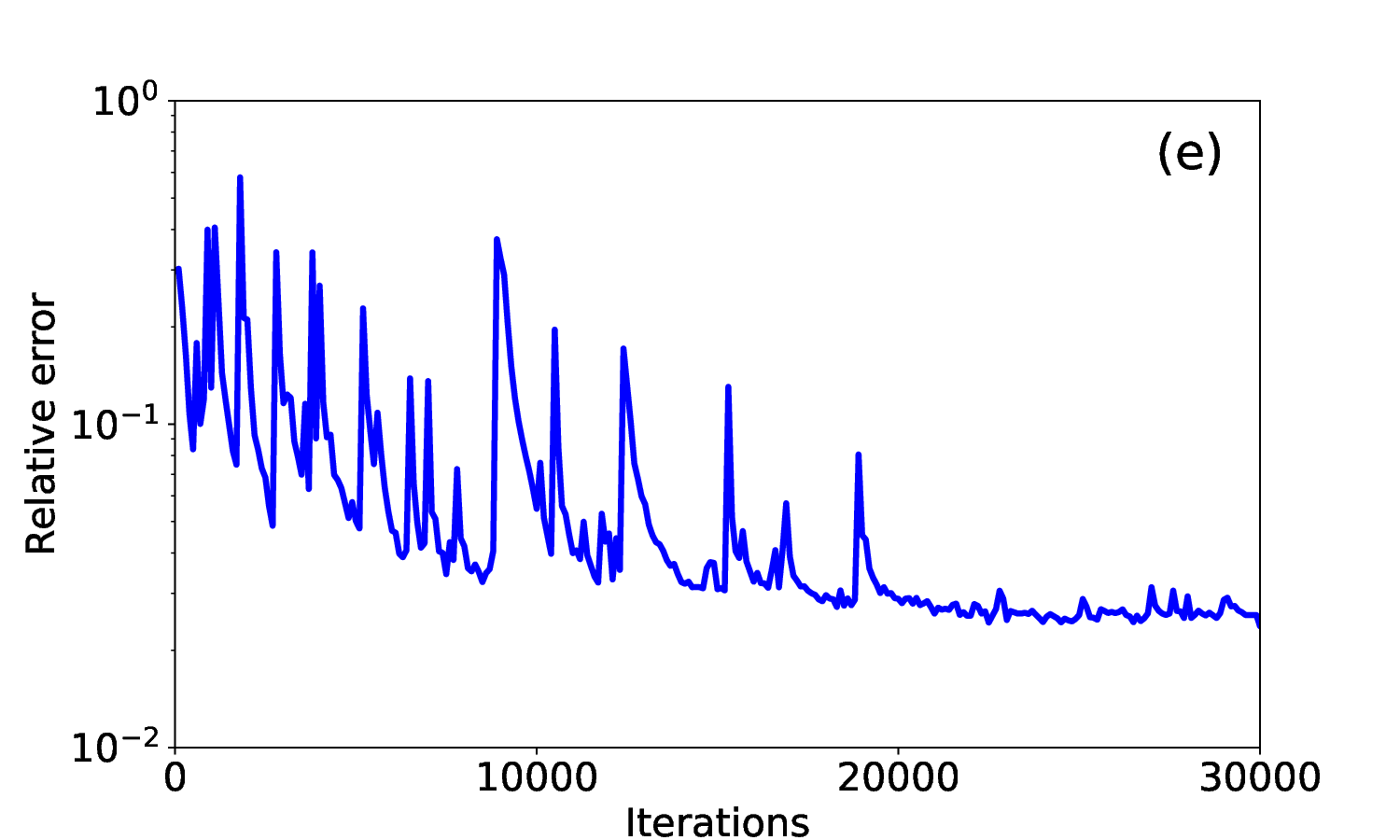}
	\end{minipage}}
        \vspace{-1mm}
	\caption{Norm-DNN for 2D GS problem Example \ref{example 1}: exact GS (a); numerical solution of norm-DNN (b); pointwise error (c); loss (d) and error (\ref{eq:relative_error}) (e) during the iterations.}
	\label{fig:2d}
\end{figure*}

\begin{example}\label{example 1}(Harmonic oscillator potential)
We first consider $V=\frac{1}{2}(x^2+y^2)$.
The hyper-parameters are set as: $L\in \left\{1,2,3,4\right\}$, $W\in \left\{10,50,70\right\}$ and $N_x=Ny=64$. As in 1D case, $\sigma=\tanh$ is used in (\ref{2d ndnn}).
Under $L=4$, $W=70$, we show the profiles of the solutions, pointwise errors and the training process of norm-DNN in Figure \ref{fig:2d}.
The errors \eqref{eq:relative_error} under different network architecture for the problem are shown in Table \ref{tab:2d}.
\end{example}

\begin{table*}[h!] 	
	\centering 	
	\caption{The relative error \eqref{eq:relative_error} of norm-DNN for the 2D GS problem in Example \ref{example 1} and Example \ref{example 2}.} 	
	\label{tab:2d} 	
	\begin{tabular}{c|ccc|c|ccc}
		\hline
		$Example$ 4.1	& $W=10$  & $W=50$ & $W=70$ & $Example$ 4.2 & $W=10$   & $W=50$   & $W=70$         \\ \hline
		$L=1$ & 1.85E-1 & 1.51E-1 & 1.40E-1 &$L=1$ & 1.59E-1 & 1.63E-1 & 1.55E-1 \\
		$L=2$ & 1.17E-1 & 4.40E-2 & 4.44E-2 &$L=2$ & 1.25E-1 & 6.41E-2 & 1.25E-1\\
		$L=3$ & 1.68E-1 & 2.36E-2 & 2.55E-2 &$L=3$ & 8.85E-2 & 3.14E-2 & 3.26E-2\\
		$L=4$ & 7.92E-2 & 2.85E-2 & 2.46E-2 &$L=4$ & 5.12E-2 & 2.86E-2 & 2.80E-2\\ \hline
	\end{tabular}
\end{table*}

\begin{figure*}[hbtp]
	\centering
	\subfigure{
		\begin{minipage}[t]{0.3\textwidth}
			\centering
			\includegraphics[width=1\textwidth]{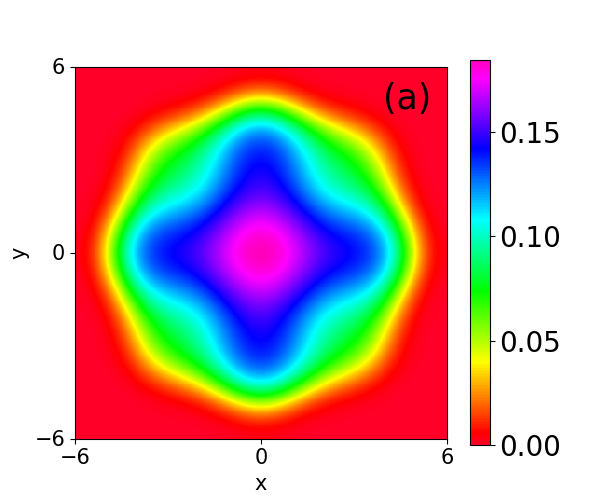}
	\end{minipage}}
	\hspace{-4mm}
	\subfigure{
		\begin{minipage}[t]{0.3\textwidth}
			\centering
			\includegraphics[width=1\textwidth]{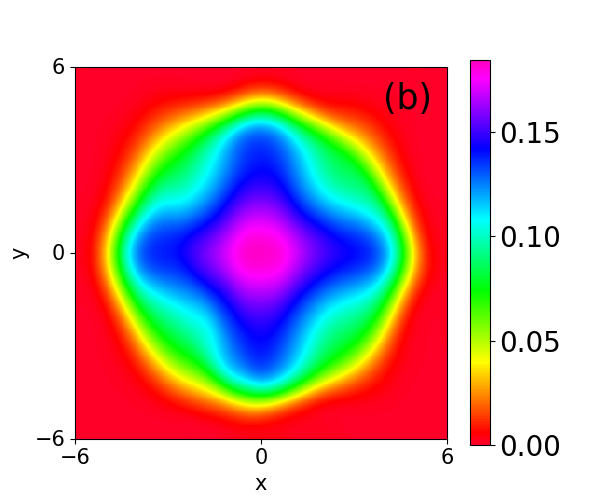}
	\end{minipage}}
	\hspace{-4mm}
	\subfigure{
		\begin{minipage}[t]{0.3\textwidth}
			\centering
			\includegraphics[width=1\textwidth]{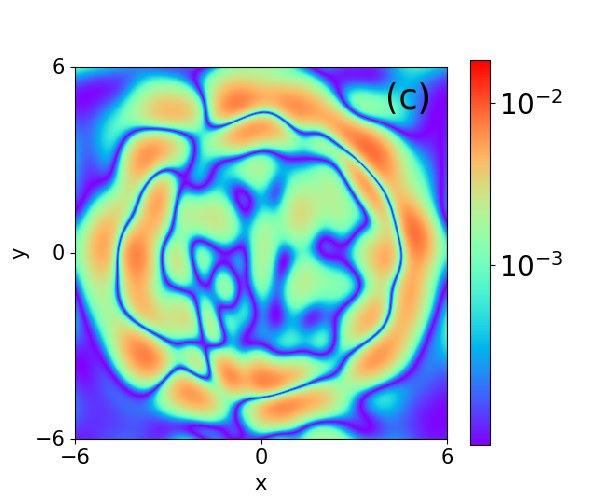}
	\end{minipage}}
	\vspace{-3mm}
	\subfigure{
		\begin{minipage}[t]{0.45\textwidth}
			\centering
			\includegraphics[width=7cm,height=4cm]{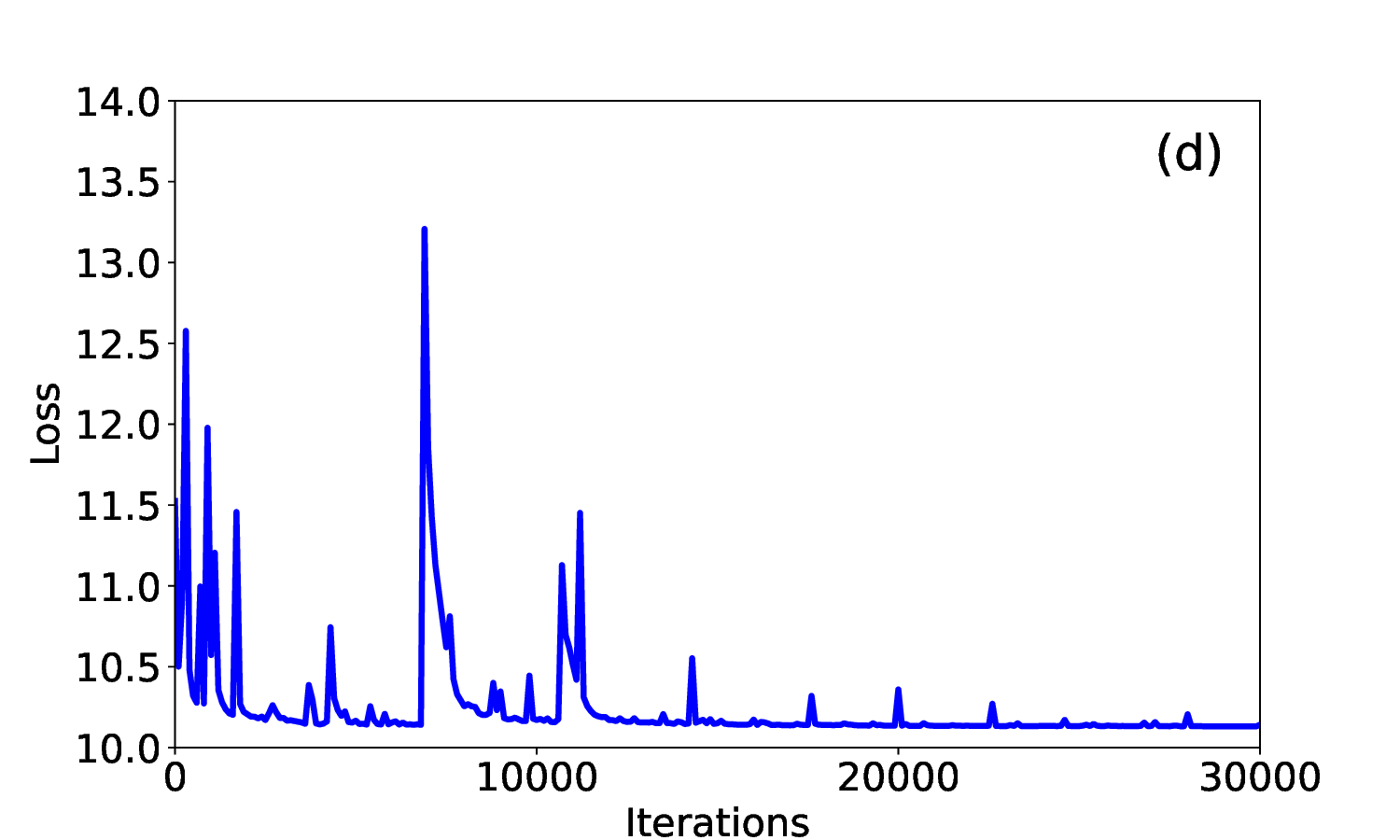}
	\end{minipage}}
	\subfigure{
		\begin{minipage}[t]{0.45\textwidth}
			\centering
			\includegraphics[width=7cm,height=4cm]{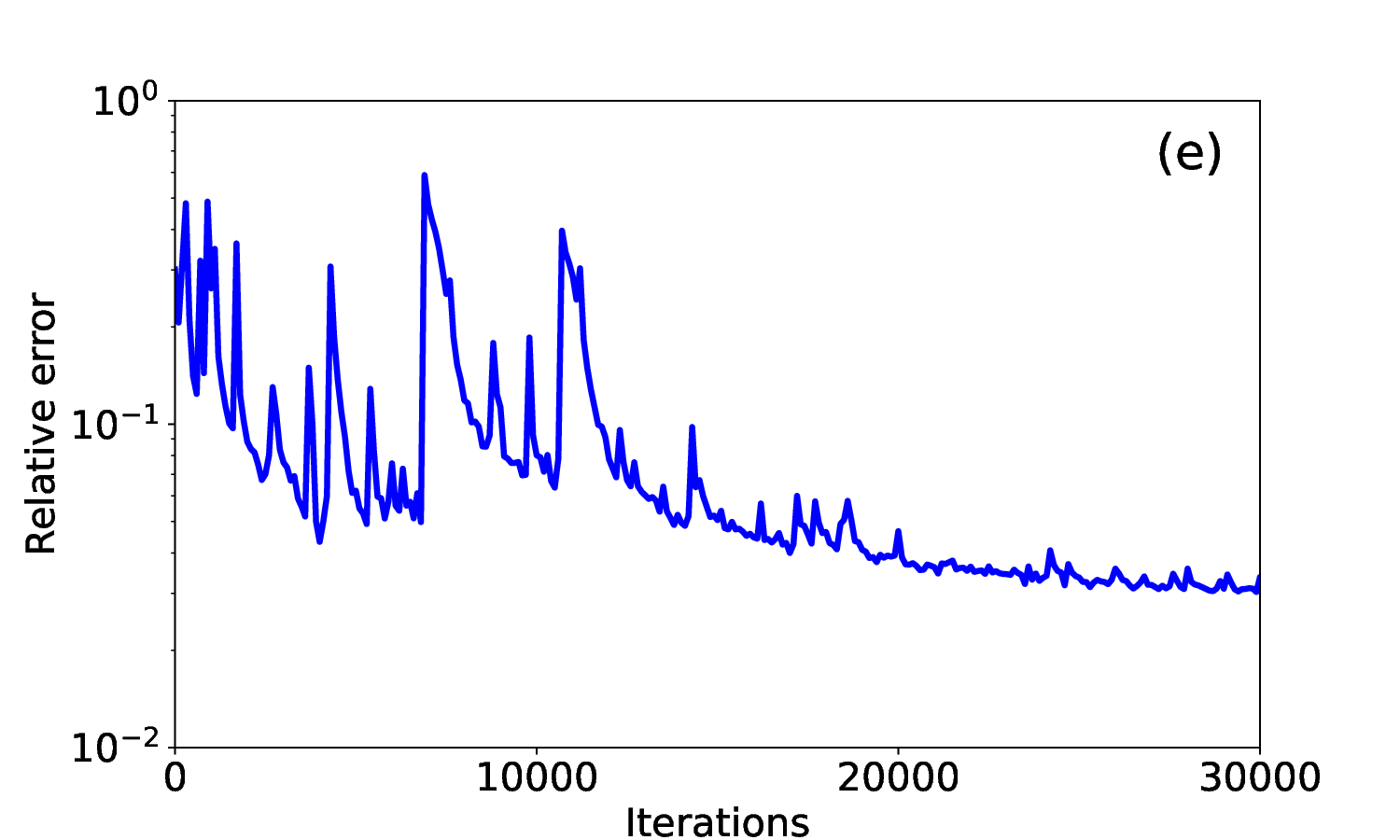}
	\end{minipage}}
        \vspace{-1mm}	
 \caption{Norm-DNN for 2D GS problem Example \ref{example 2}: exact GS (a); numerical solution of norm-DNN (b); pointwise error (c); loss (d) and error (\ref{eq:relative_error}) (e) during the iterations.}
	\label{fig:2d_sin_beta}
\end{figure*}

\begin{example}\label{example 2}(Optical lattice potential)
Next, we consider $V=\frac{1}{2}(x^2+y^2)+\frac{5}{2} (\sin^2(\frac{\pi x}{4})+\sin^2(\frac{\pi y}{4}))$.
Based on the study in Section \ref{sec:olp}, we change the activation function from `tanh' to `sin', i.e., $\sigma=\sin$ in (\ref{2d ndnn}).
The hyper-parameters are set the same as above.
The errors \eqref{eq:relative_error} are shown in Table \ref{tab:2d}.
The profiles of solutions, pointwise errors and the training process under $L=4$, $W=70$ are shown in Figure \ref{fig:2d_sin_beta}.
\end{example}

From the above two numerical examples and the presented numerical results, we can observe the following. 1) The errors \eqref{eq:relative_error} of norm-DNNs drop significantly as the network architecture increases and when $L=4$, $W=70$, the relative error is about $2\times10^{-2}$. Correspondingly, the pointwise errors are kept below 1$\%$.
2) Compared with the 1D case, we observe that the loss and error during the training process  in Figure \ref{fig:2d}(d,e) and Figure \ref{fig:2d_sin_beta}(d,e) now exhibit some oscillations which cost more efforts to train than the 1D case, but they can still converge in the end. 3) The pre-training and the adjustment of the activation function when high frequency occurs, are also very effective in 2D.
Overall, we conclude that norm-DNN can accurately capture the 2D GS, and all the strategies developed in 1D case works in high dimension.


\begin{figure*}[h]
	\centering
	\subfigure{
		\begin{minipage}[t]{0.3\textwidth}
			\centering
			\includegraphics[width=1\textwidth]{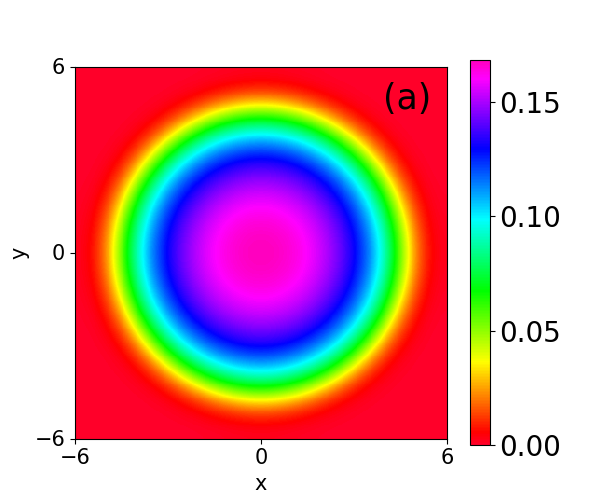}
	\end{minipage}}
	\hspace{-4mm}
 	\subfigure{
		\begin{minipage}[t]{0.3\textwidth}
			\centering
			\includegraphics[width=1\textwidth]{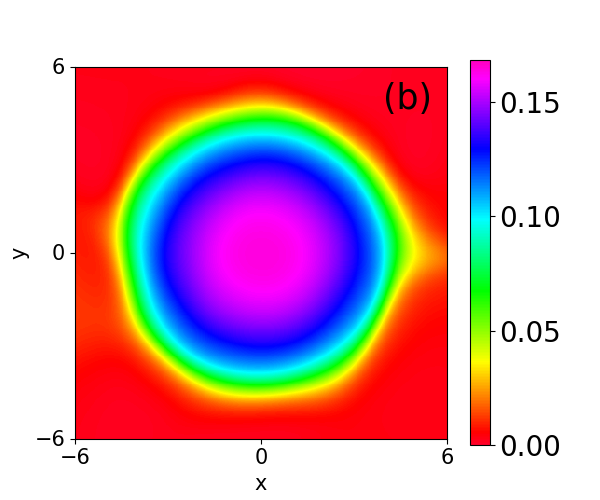}
	\end{minipage}}
 	\hspace{-4mm}
 	\subfigure{
		\begin{minipage}[t]{0.3\textwidth}
			\centering
			\includegraphics[width=1\textwidth]{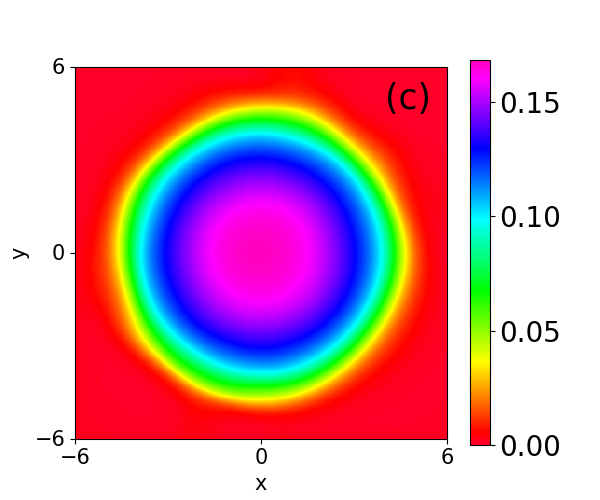}
	\end{minipage}}
        \vspace{-3mm}
	\subfigure{
		\begin{minipage}[t]{0.45\textwidth}
			\includegraphics[width=7cm,height=4cm]{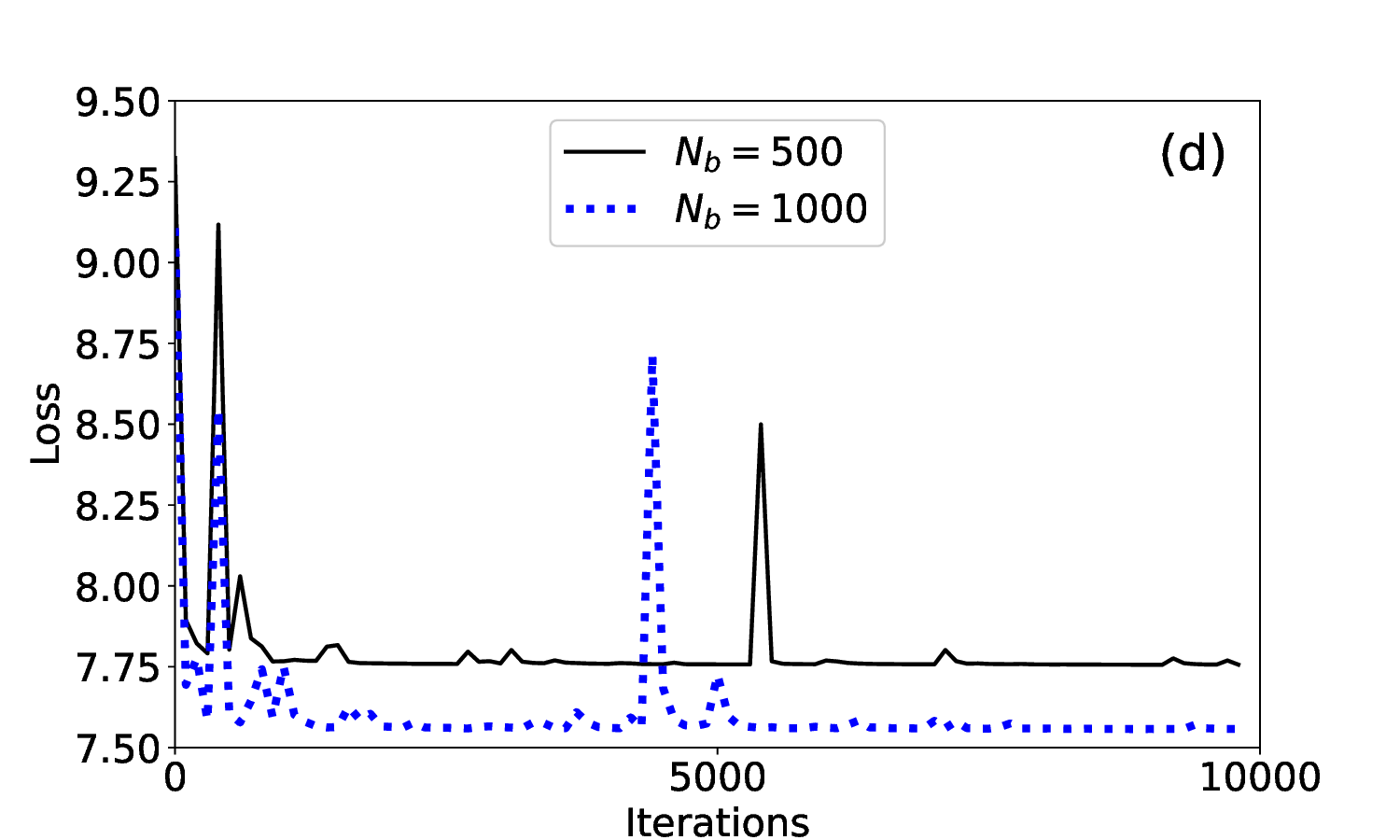}
	\end{minipage}}
	\subfigure{
		\begin{minipage}[t]{0.45\textwidth}
			\includegraphics[width=7cm,height=4cm]{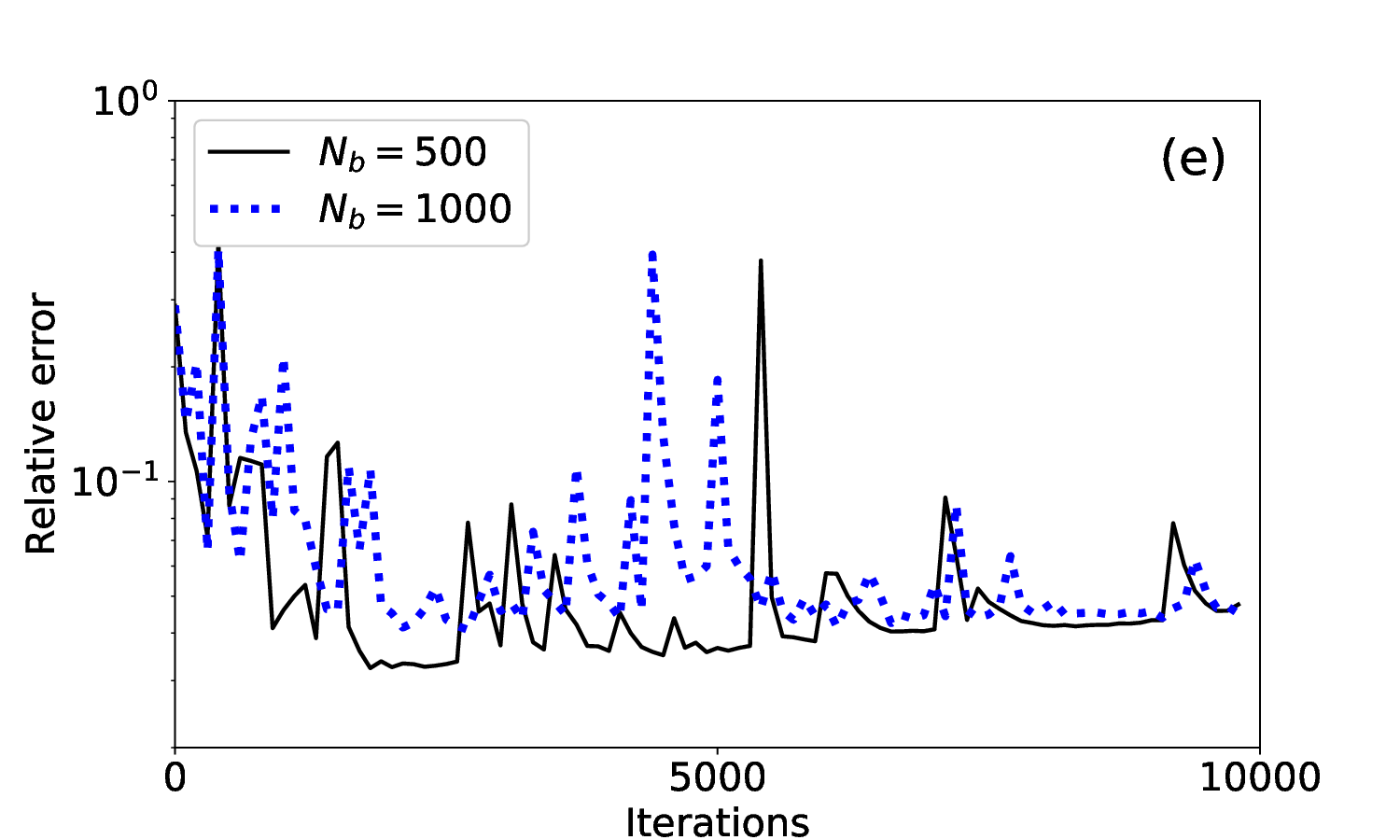}
	\end{minipage}}
        \vspace{-1mm}
	\caption{norm-DNN with randomly selected training points for 2D GS problem: exact GS (a); numerical solution of norm-DNN with $N_b=500$ (b) and $N_b=1000$ (c); loss (d) and error (\ref{eq:relative_error}) (e) during the iterations.}
	\label{fig:2d_random}
\end{figure*}

One more issue for the high-dimensional problem is the grid points  used to approximate the energy functional, e.g., the $(x_j,y_j)$ in (\ref{eq:loss_gs_2d}). The equal-partition is not possible in a very high dimension and the usual way is to consider random sampling.
To demonstrate the capacity of the proposed norm-DNN with random sampling, we consider the 2D GS problem in Example \ref{example 1}.  Now, we randomly generate some points  according to the uniform distribution in $\Omega$.
The total number of the generated points is denoted as $N_b\in \left\{500, 1000\right\}$.  Note that the number of points here is much less than that used in Example \ref{example 1} to train norm-DNN, i.e., $N_b<64\times64=4096$. The other setup remains the same as before, and we train the norm-DNN with $L=4$, $W=70$ for 10000 iterations. The solutions, pointwise errors and the changes of loss and relative error \eqref{eq:relative_error} along with the iterations are shown in Figure \ref{fig:2d_random}.

When $N_b=500$, we can clearly observe the pollution of noise in the numerical solution in Figure \ref{fig:2d_random}(b).
Nevertheless, norm-DNN in this case is still able to capture the general shape of GS. When the number of the training points is increased to $N_b=1000$, the impact of random noise is largely reduced. Thus, norm-DNN with random sampling can work for the high-dimensional problem and the numerical solution can at least be qualitatively correct. At last, we remark that
the randomness in sampling can lead to stronger oscillations in the loss and error during the training process, as shown in Figure \ref{fig:2d_random}(d,e). The  oscillations make it harder for the training to reach a stable state, and this is the reason why we set a fixed number of iterations for this test. Such issue may be relieved with a better sampling method, e.g., \cite{ADNNs,QMC}, or a better optimization method or stopping criterion, and this will be addressed in a future work.

\begin{figure*}[h!]
	\centering
	\subfigure{
		\begin{minipage}[t]{0.34\textwidth}
			\centering
			\includegraphics[width=1\textwidth]{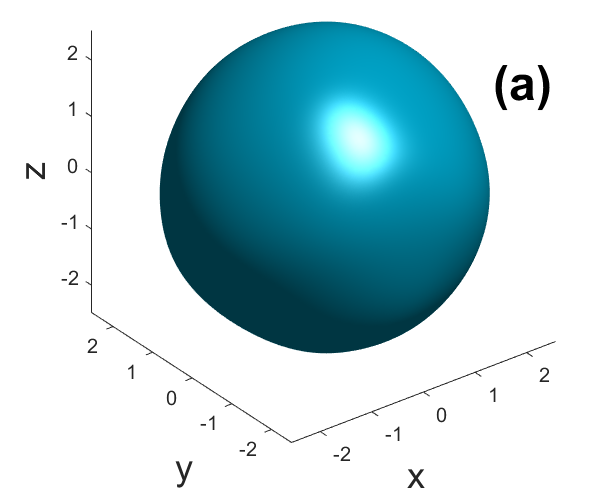}
	\end{minipage}}
	\hspace{3mm}
	\subfigure{
		\begin{minipage}[t]{0.38\textwidth}
			\centering
			\includegraphics[width=1\textwidth]{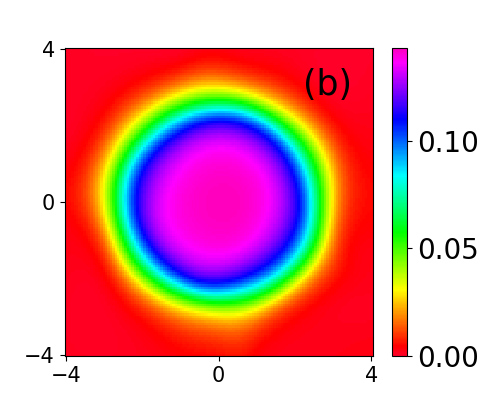}
	\end{minipage}}
        \vspace{-3mm}

        \subfigure{
		\begin{minipage}[t]{0.34\textwidth}
			\centering
			\includegraphics[width=1\textwidth]{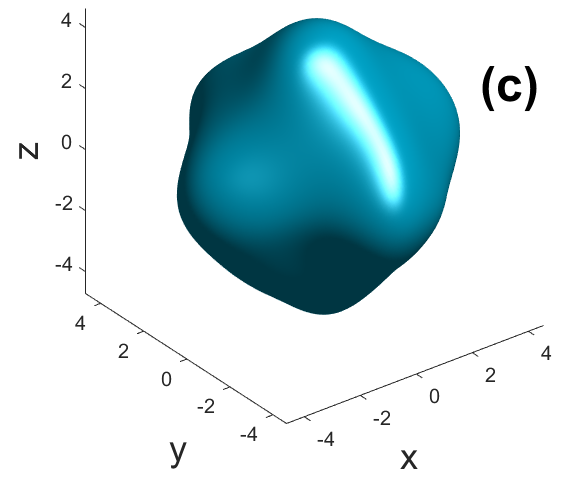}
	\end{minipage}}
	\hspace{3mm}
	\subfigure{
		\begin{minipage}[t]{0.38\textwidth}
			\centering
			\includegraphics[width=1\textwidth]{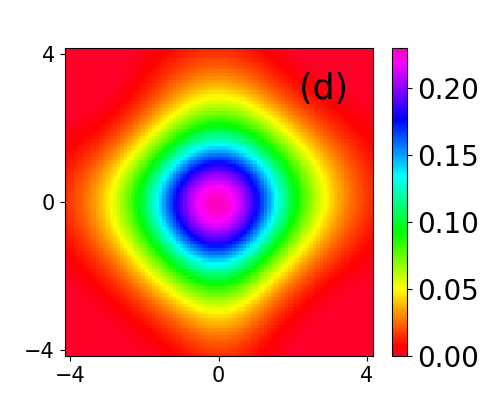}
	\end{minipage}}
 
	\caption{Numerical solutions of norm-DNN for 3D GS problem: (a) $\phi_{\theta}$ and (b) $\phi_{\theta}$ at $z=0$ for the harmonic oscillator potential $V_1$; (c) $\phi_{\theta}$ and (d) $\phi_{\theta}$ at $z=0$ for the optical lattice potential $V_2$.}
	\label{fig:3d}
\end{figure*}

Finally, we end this subsection by a three-dimensional (3D) test.
The GS problem of a 3D BEC, i.e., $d=3$ and $\bx=(x,y,z)^{\top} \in \bR^3$ in (\ref{eq:GPE}), by norm-DNN is then to consider the optimization:
\begin{equation*}
	\label{eq:loss_gs_3d}
	\min_{\theta} \left\{\text{Loss}=
	\frac{|\Omega|}{N_{3D}} \sum_{j=1}^{N_{x}}\sum_{k=1}^{N_{y}}\sum_{l=1}^{N_{z}} \left[\frac{1}{2} |\nabla \phi_{\theta}(x_j,y_k,z_l)|^2 + V(x_j,y_k,z_l)|\phi_{\theta}(x_j,y_k,z_l)|^2 + \frac{1}{2} \beta |\phi_{\theta}(x_j,y_k,z_l)|^4 \right]\right\}
\end{equation*}
with $N_{3D}=N_{x}\times N_{y}\times N_{z}$ and
    $\phi_{\theta}(x,y,z) = \mathcal{N} \circ \mathcal{T} \circ {\bf F}_{L+1} \circ \sigma \circ {\bf F}_{L} \circ \sigma \circ\cdots \circ {\bf F}_{2} \circ \sigma \circ {\bf F}_{1}(x,y,z).$
We fix $\beta=200$ and set the computation domain $\Omega=(-4,4)\times(-4,4)\times(-4,4)$ for $V=V_1=\frac{1}{2}(x^2+y^2+z^2)$ and $\Omega=(-6,6)\times(-6,6)\times(-6,6)$ for $V=V_2=\frac{1}{2}(x^2+y^2+z^2)+\frac{5}{2} (\sin^2(\frac{\pi x}{4})+\sin^2(\frac{\pi y}{4})+\sin^2(\frac{\pi z}{4})).$
We use a normalized 3D Gaussian $\fe^{-(\frac{x^2}{10}+\frac{y^2}{10}+\frac{z^2}{10})} / (5\pi)^{3/4}$ for pre-training with 1000 iterations.
A total number of the quadrature points $N_b=10000<64\times64\times64=262144$ is randomly generated via the uniform distribution in $\Omega$.
We train the norm-DNN with $L=4$, $W=100$ until the stopping condition $tol=10^{-7}$ for the Adam method is met.
In Figure \ref{fig:3d}, we display the numerical solutions of the harmonic oscillator potential $V_1$ and the optical lattice potential $V_2$ in isosurface plot and also in contour plot, where the  correct shape of 3D GS has been captured.
This further demonstrates the potential of norm-DNN to solve high-dimensional GS {problems}.
\begin{remark}
{We mention here the computational time (cputime) of GFDN  in the previous examples. 
For the 1D problem in Section \ref{sec:hop}, the cputime is  1.2s. However, for the 3D case, the cputime required by  GFDN   (912s) is almost the same as the training time of norm-DNN (980s).  It might not be fair to  compare norm-DNN with GFDN directly through the training time and the cputime.
However, this certainly indicates that norm-DNN would become much more efficient for high-dimensional problems, which is the ultimate goal of norm-DNN. Besides, for norm-DNN, once the network is trained, it  provides the solution value at any input $\textbf{x}$ effortlessly, while GFDN produces the numerical solution only on the grid points.}

\end{remark}

\subsection{Two-component BEC}
In this subsection, we consider a 1D two-component BEC with an internal atomic Josephson junction,  which in dimensionless form reads as follows \cite{Bao,BaoEast,BaoCai,Williams}:
\begin{equation}
    \label{eq:CBEC}
    \begin{split}
            i\frac{\partial \psi_1(x,t)}{\partial t} &= \Big [-\frac{1}{2} \partial^2_{x} + V(x)+ \beta_{11}|\psi_1(x,t)|^2 + \beta_{12}|\psi_2(x,t)|^2  \Big ] \psi_1(x,t) + \lambda\psi_2(x, t), \quad x\in \mathbb{R},\,t>0, \\
            i\frac{\partial \psi_2(x,t)}{\partial t} &= \Big [-\frac{1}{2} \partial^2_{x} + V(x) + \beta_{21}|\psi_1(x,t)|^2 + \beta_{22}|\psi_2(x,t)|^2 \Big ] \psi_2(x,t) + \lambda\psi_1(x, t), \quad x\in \mathbb{R},\,t>0,
    \end{split}
\end{equation}
where $\Psi(x,t)=(\psi_1(x,t), \psi_2(x,t))^{\top}$ is the unknown vector-valued {wave function}, $V$ is the real-valued potential and
$\beta_{11},\, \beta_{12}(=\beta_{21}),\, \beta_{22},\,  \lambda\in\bR$ are given constants.
The two-component BEC (\ref{eq:CBEC}) conserves the total (normalized) mass 
\begin{equation}
\label{eq:norm_cbec}
    M(\Psi(\cdot,t)):=\int_{\mathbb{R}} \left[ |\psi_1(x,t)|^2 + |\psi_2(x,t)|^2 \right] dx \equiv1,\quad t\geq0,
\end{equation}
and the energy 
\begin{equation*}
    \begin{split}
        E(\Psi(\cdot,t)):=\int_{\bR} \Big[ &\frac{1}{2}( |\partial_x \psi_1|^2 + |\partial_x \psi_2|^2 ) + V(x)(|\psi_1|^2 + |\psi_2|^2) + \frac{1}{2} \beta_{11}|\psi_1|^4 + \frac{1}{2} \beta_{22}|\psi_2|^4 \\
        &+ \beta_{12}|\psi_1|^2|\psi_2|^2 
        + 2\lambda \cdot \text{Re}(\psi_1 \overline{\psi}_2) \Big] dx\equiv E(\Psi(\cdot,0)),\quad t\geq0. 
    \end{split}
\end{equation*}

The theoretical results in \cite{Bao,BaoEast,BaoCai} indicate that the GS can be real-valued,  and thus  
the GS problem for  (\ref{eq:CBEC}) can be considered  as \cite{Bao,BaoEast}:
find $\Phi_g=(\phi_{1,g},\phi_{2,g})^{\top} \in S$ such that
\begin{equation}\label{2c_gs}
    \widetilde{E}(\Phi_g)=\mathop{\min}_{\Phi \in S} \widetilde{E}(\Phi),\ \mbox{with}\
     S=\left \{ \Phi=(\phi_1, \phi_2)^{\top}\ :\ \int_{\bR} \left[|\phi_1(x)|^2 + |\phi_2(x)|^2 \right] dx =1,\ \Phi:\bR\to\bR^2 \right \},
\end{equation}
where
\begin{equation*}
    \begin{split}
        \widetilde{E}(\Phi):=\int_{\bR} \Big[ &\frac{1}{2}( |\phi_1^{'}|^2 + |\phi_2^{'}|^2 ) + V(x)(|\phi_1|^2 + |\phi_2|^2) + \frac{1}{2} \beta_{11}|\phi_1|^4 + \frac{1}{2} \beta_{22}|\phi_2|^4 + \beta_{12}|\phi_1|^2|\phi_2|^2 + 2\lambda \phi_1 \phi_2 \Big] dx. 
    \end{split}
\end{equation*}

\begin{figure*}[htbp]
	\centering
	\subfigure{
		\begin{minipage}[t]{0.33\textwidth}
			\centering
			\includegraphics[width=1\textwidth]{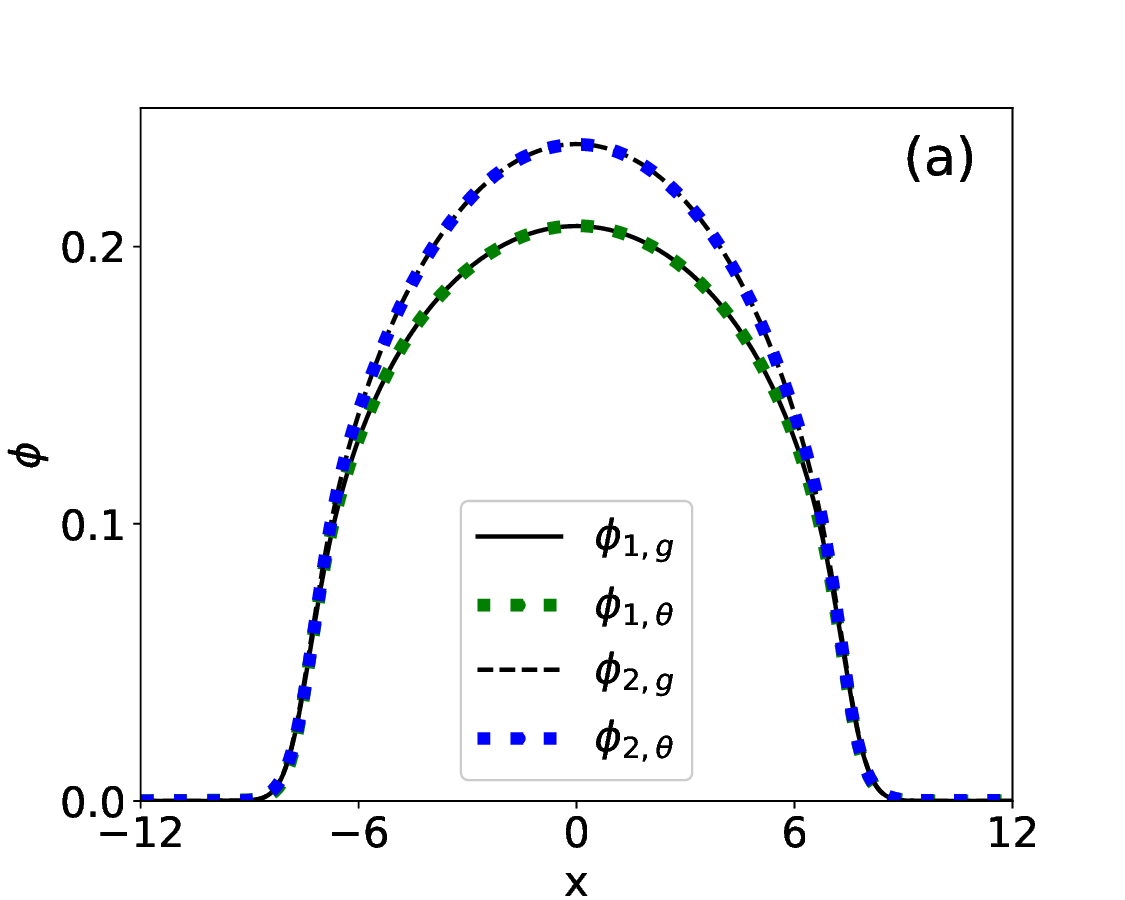}
	\end{minipage}}
	\hspace{-4mm}
	\subfigure{
		\begin{minipage}[t]{0.33\textwidth}
			\centering
			\includegraphics[width=1\textwidth]{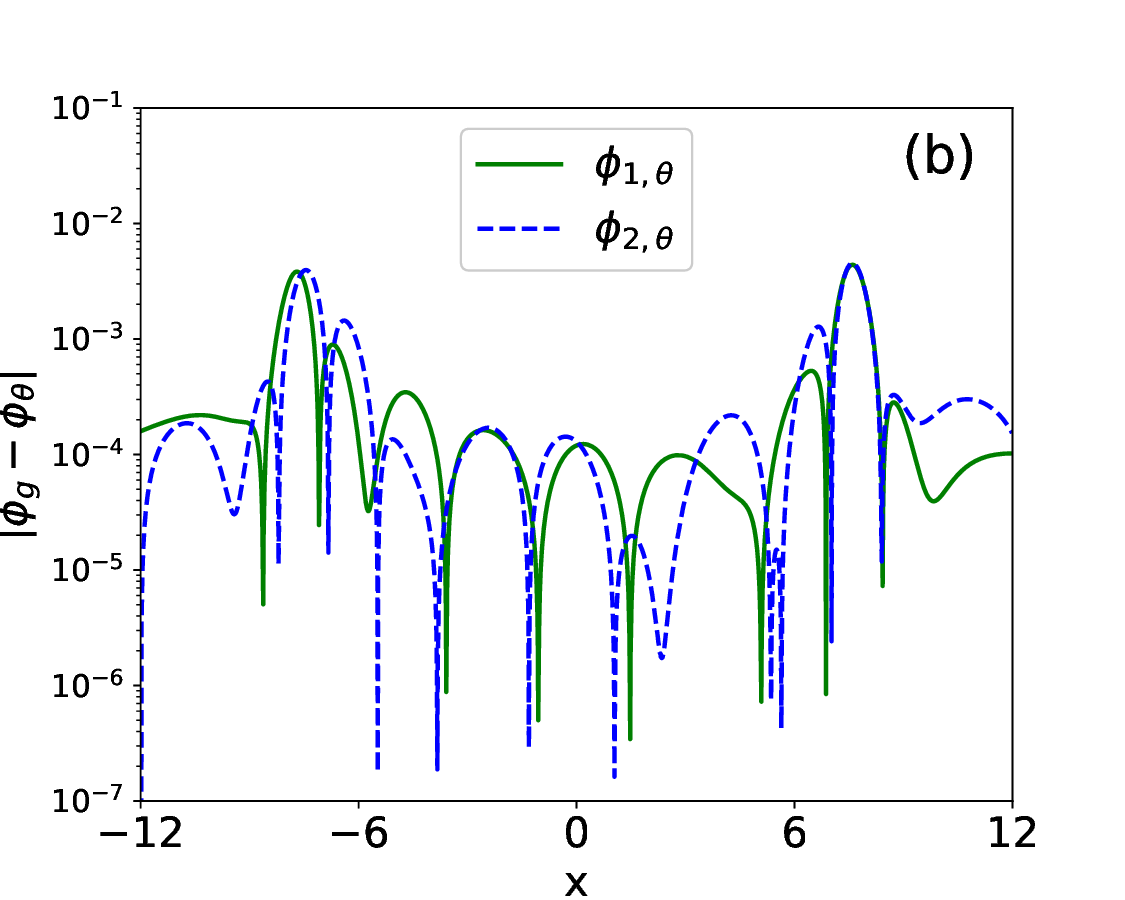}
	\end{minipage}}
	\hspace{-4mm}
	\subfigure{
		\begin{minipage}[t]{0.33\textwidth}
			\centering
			\includegraphics[width=1\textwidth]{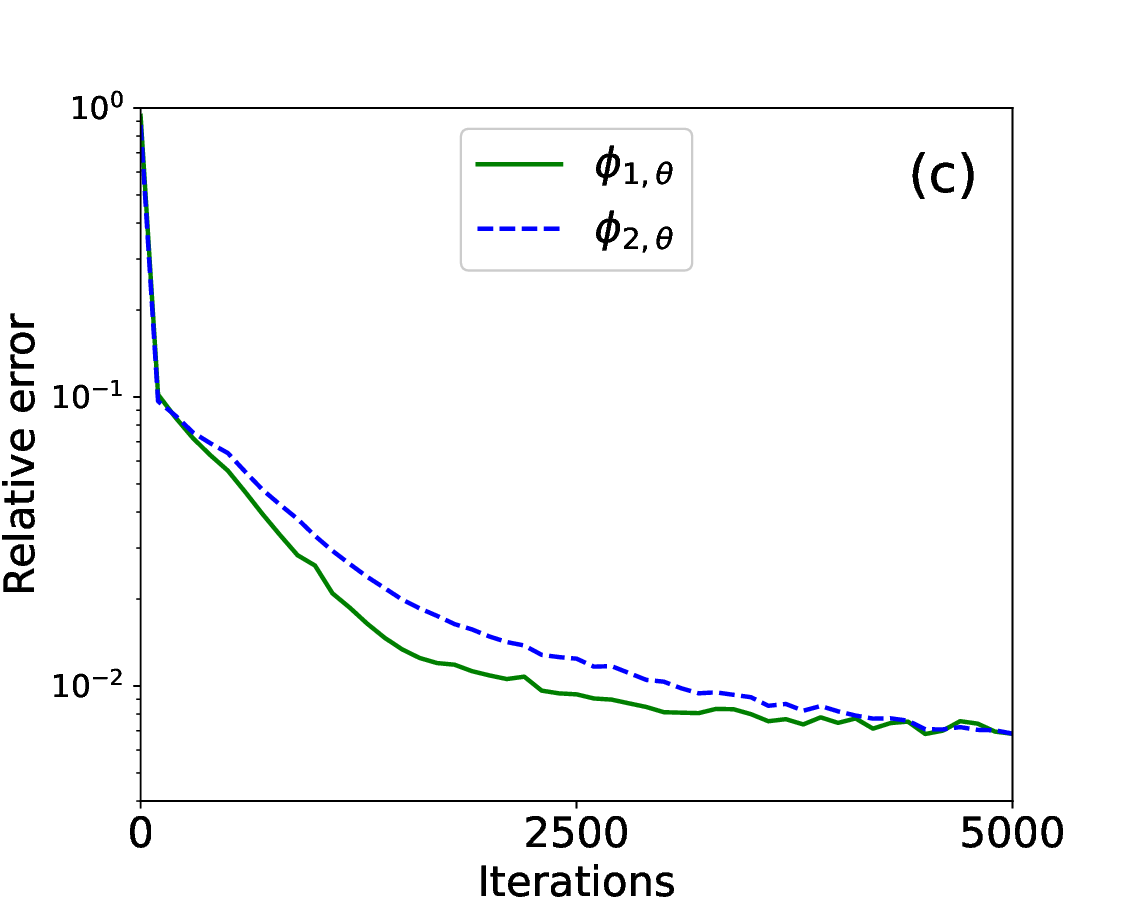}
	\end{minipage}}
 \vspace{-3mm}
	\caption{Norm-DNN for two-component GS problem with $\beta=300$. The 1st to 3rd subfigures represent in sequence: profile of solution, pointwise error,  training process of error (\ref{eq:relative_error}).}
	\label{fig:cbec1}
\end{figure*}

\begin{figure*}[htbp]
	\centering
	\subfigure{
		\begin{minipage}[t]{0.33\textwidth}
			\centering
			\includegraphics[width=1\textwidth]{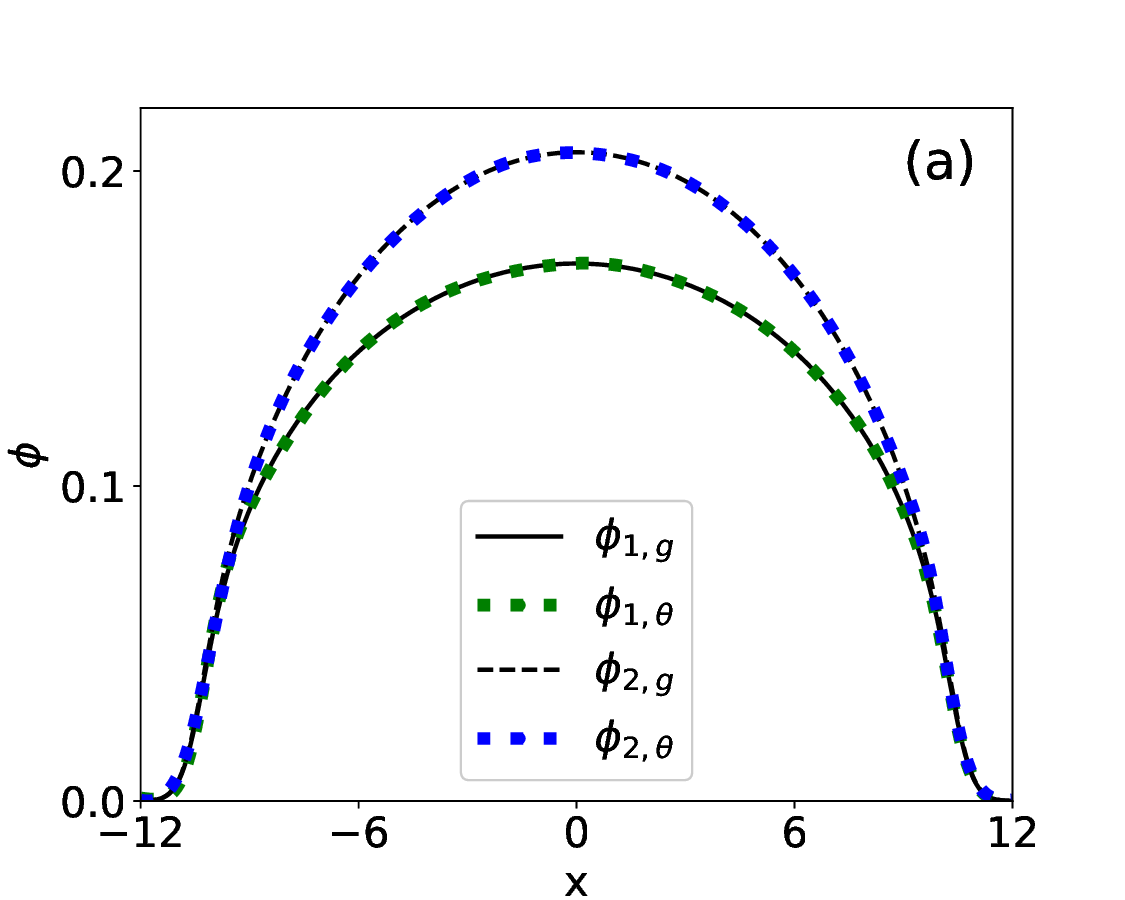}
	\end{minipage}}
	\hspace{-4mm}
	\subfigure{
		\begin{minipage}[t]{0.33\textwidth}
			\centering
			\includegraphics[width=1\textwidth]{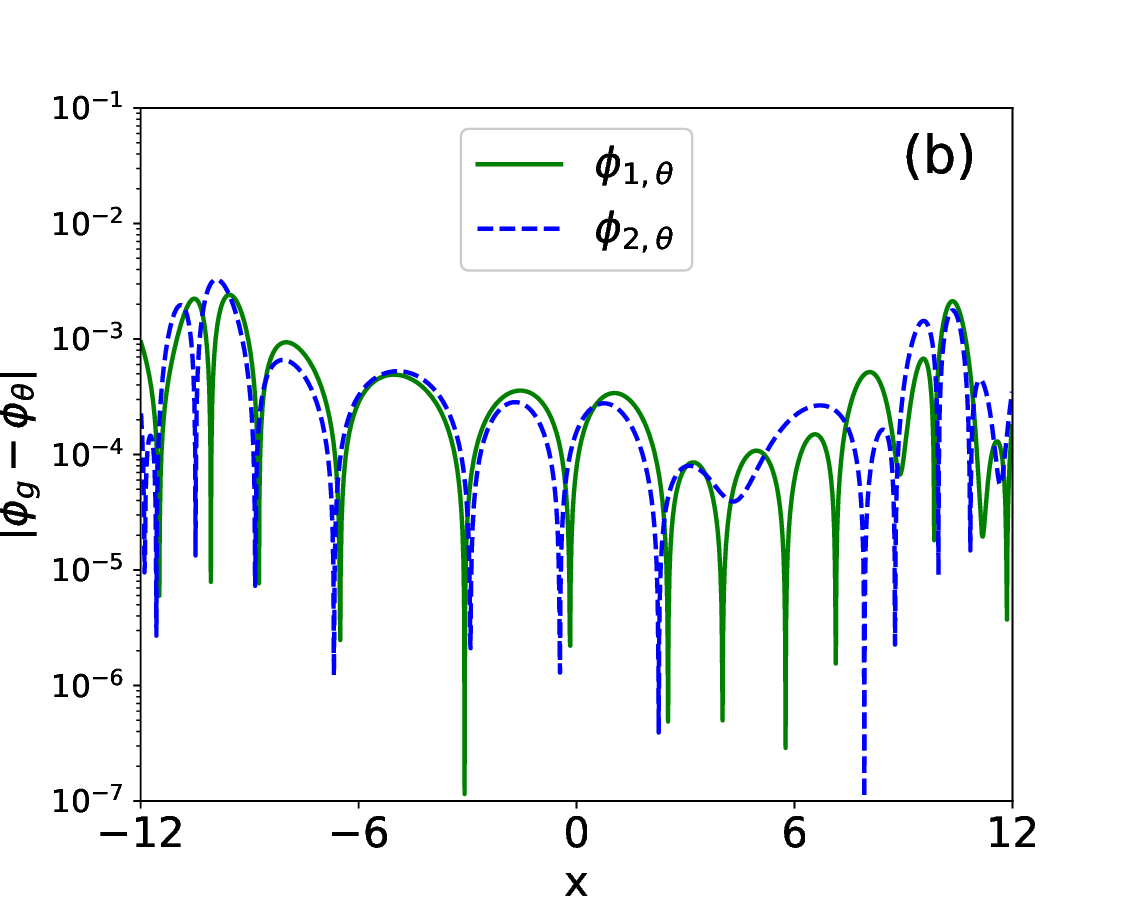}
	\end{minipage}}
	\hspace{-4mm}
	\subfigure{
		\begin{minipage}[t]{0.33\textwidth}
			\centering
			\includegraphics[width=1\textwidth]{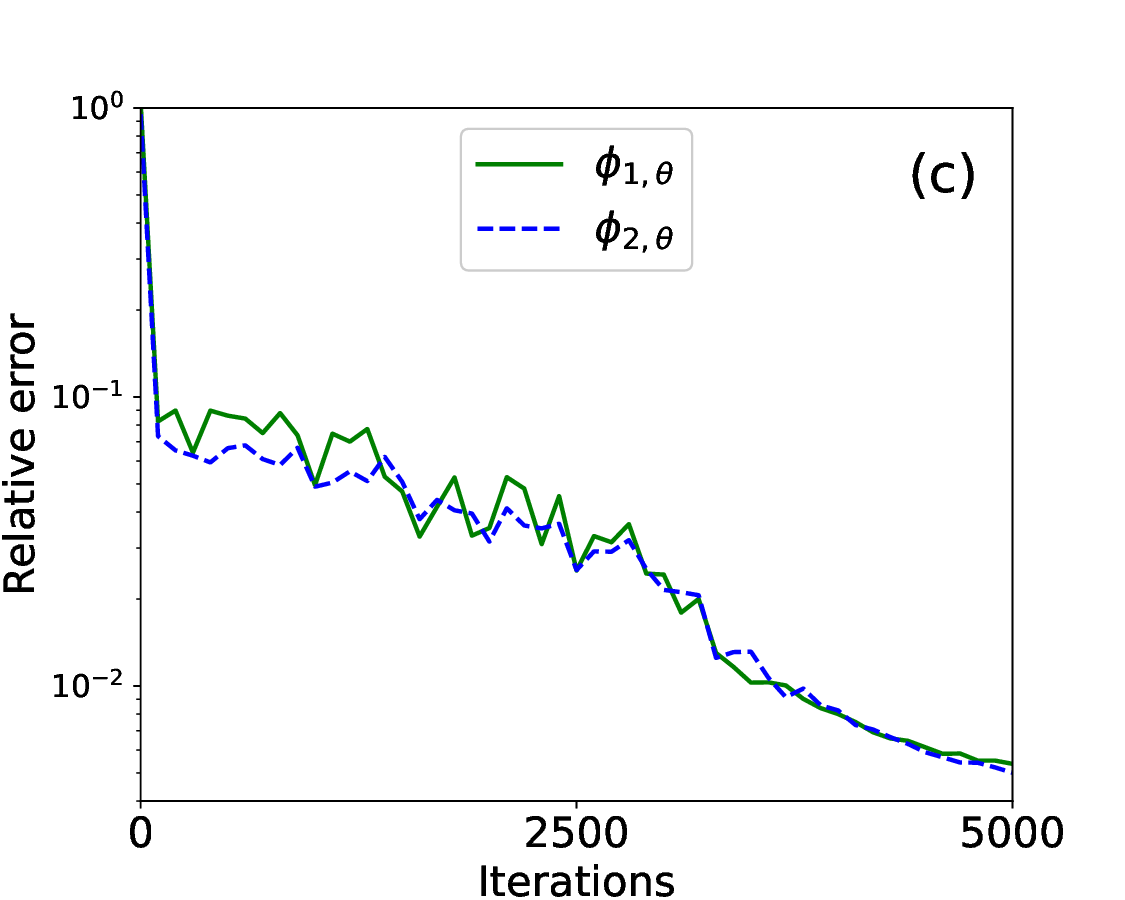}
	\end{minipage}}
 \vspace{-3mm}
	\caption{Norm-DNN for two-component GS problem with $\beta=800$. The 1st to 3rd subfigures represent in sequence: profile of solution, pointwise error,  training process of error (\ref{eq:relative_error}).}
	\label{fig:cbec2}
\end{figure*}

For the related theoretical results of (\ref{2c_gs}), we refer the readers to \cite{BaoCai}. Here let us focus on the computation of the GS $\Phi_g$ by extending the proposed norm-DNN method.    Based on our previous studies of the single component BEC,  we consider the
norm-DNN in the same form as before:
\begin{equation}
\label{cbec ndnn}
    \Phi_{\theta}(x) =\left(\phi_{1, \theta}(x), \phi_{2, \theta}(x) \right)^{\top}= \mathcal{N} \circ \mathcal{T} \circ  {\bf F}_{L+1} \circ \sigma \circ {\bf F}_{L} \circ \sigma \circ\cdots \circ {\bf F}_{2} \circ \sigma \circ {\bf F}_{1}(x).
\end{equation}
Note the only difference is now we set two neurons in the output layer, i.e.,  $n_{L+1}=2$,  to represent the approximations for $\phi_1$ and $\phi_2$. To satisfy the mass constraint (\ref{eq:norm_cbec}),
we modify the normalization layer $\mathcal{N}$ as:
\begin{equation*}
    \mathcal{N}(\mathbf{y}) = \frac{\mathbf{y}}{\sqrt{||{y_1}||^{2}_{2} + ||{y_2}||^{2}_{2}}}, \quad \mathbf{y}=\by(x)=(y_1(x), y_2(x))^{\top}:\bR\to \bR^{2},
\end{equation*}
and set the shift layer $\mathcal{T}$ as
\[
    \mathcal{T}(\by):=\mathcal{T}\left(\begin{array}{c}
    y_1 \\
    y_2
    \end{array}\right) = \left(\begin{array}{c}
    y_1 - \min_x (y_1) \\
    y_2 - \min_x (y_2)
    \end{array}\right) , \quad \mathbf{y}=\by(x)=(y_1(x), y_2(x))^{\top}:\bR\to \bR^{2}.
\]

\begin{table*}[htbp] 	
	\centering 	
	\caption{Error \eqref{eq:relative_error}, number of iterations and training time of norm-DNN for two-component GS problem. Here $\text{Error}_{1}$ and $\text{Error}_{2}$ represent the relative error \eqref{eq:relative_error} for $\phi_1$ and $\phi_2$, respectively. }
	\label{tab:cbec} 	
	\begin{tabular}{c|cccc} 		\hline 					
		& Iterations  & Time  &  $\text{Error}_{1}$  &  $\text{Error}_{2}$   \\ \hline 		
            $\beta=300$  & 6800	    & 49s  	& 6.74E-3  & 6.38E-3\\
            $\beta=800$  & 5400	    & 41s  	& 5.17E-3  & 4.95E-3\\ \hline
	\end{tabular}
\end{table*}

Now we consider a numerical example to illustrate the performance of (\ref{cbec ndnn}) for the two-component GS problem. Choose $\Omega=(-12,12)$, $V(x)=\frac{1}{2}x^2$, $\beta_{11}=\beta,\,\beta_{12}=0.94\beta,\,\beta_{22}=0.97\beta$ with $\beta=300$ or $800$, $\lambda=-1$.
The exact $\Phi_g$ can be obtained by the extended GFDN scheme from \cite{BaoCai}.
We use the same Gaussian $\phi_{0}$ as the pre-training to initialize $\phi_1$ and $\phi_2$.
The hyper-parameters are set as $L=4$, $W=70$, $tol=10^{-7}$, $N_{x}=128$. The relative error \eqref{eq:relative_error}, number of iterations and training time are shown in Table \ref{tab:cbec}.
The exact and numerical solutions of norm-DNNs for $\phi_1$ and $\phi_2$, pointwise errors, the convergence process of relative error \eqref{eq:relative_error}are shown in Figure \ref{fig:cbec1} and Figure \ref{fig:cbec2}.

From the numerical results,
we can see that norm-DNNs can accurately and quickly get the GS of the two-component BEC problem, with only the minor cost of adding an extra neuron in the output layer.
This further demonstrates the effectiveness  of the proposed norm-DNN for solving GS problems.
We remark that another common way to fit multiple target functions is to use multiple networks to fit each function individually.
For the two-component GS problem, using two different networks would cause difficulty to impose the normalization layer to meet (\ref{eq:norm_cbec}) and this will lead to additional computational costs.

\subsection{First excited state}
\begin{figure*}[h!]
	\centering
	\subfigure{
		\begin{minipage}[t]{0.45\textwidth}
			\centering
			\includegraphics[width=1\textwidth]{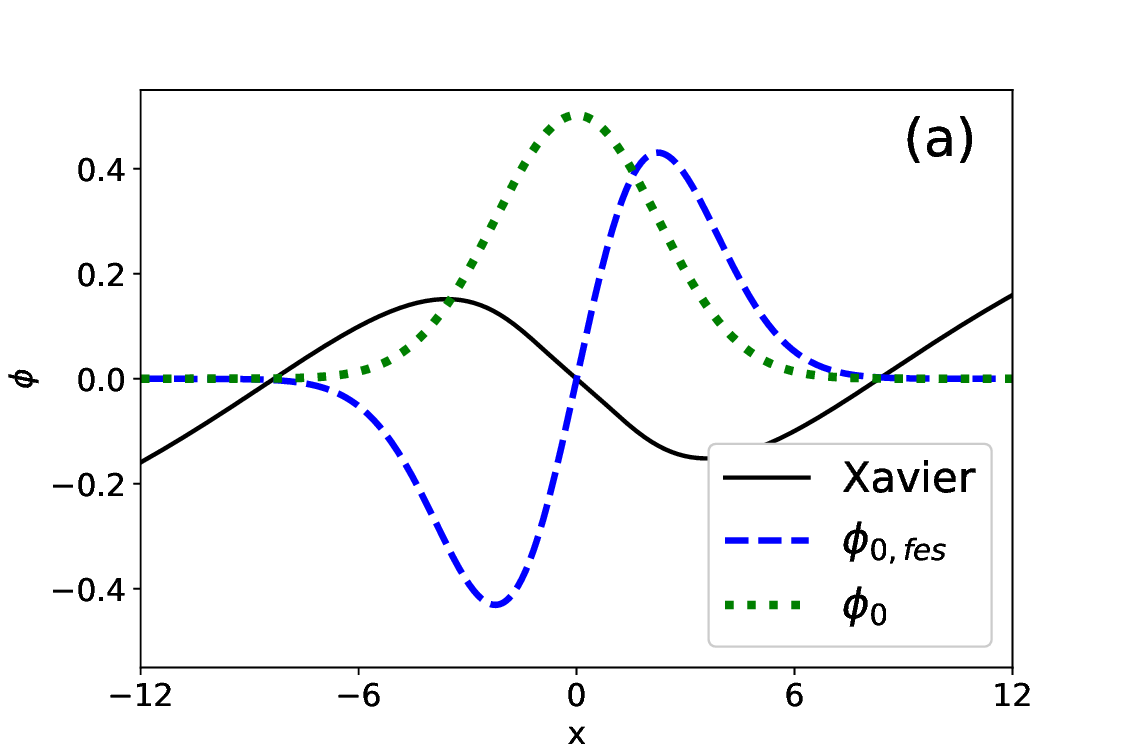}
	\end{minipage}}
	\hspace{-6mm}
	\subfigure{
		\begin{minipage}[t]{0.45\textwidth}
			\centering
			\includegraphics[width=1\textwidth]{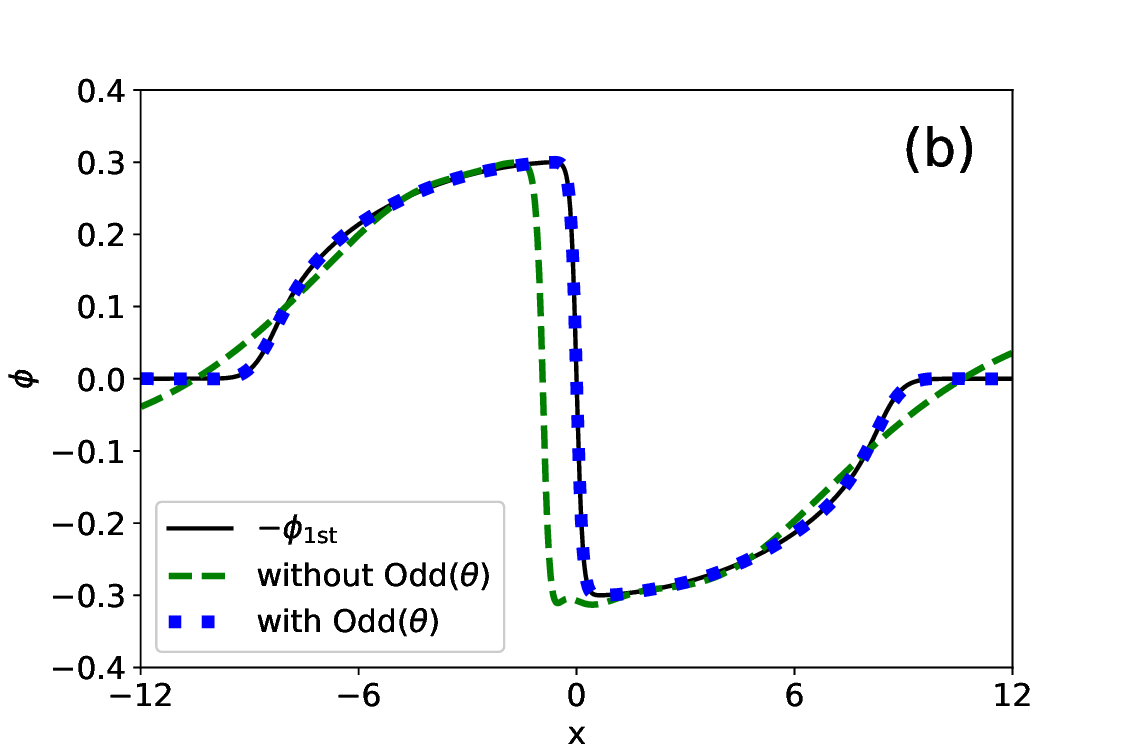}
	\end{minipage}}
        \vspace{-3mm}
	\subfigure{
		\begin{minipage}[t]{0.45\textwidth}
			\centering
			\includegraphics[width=1\textwidth]{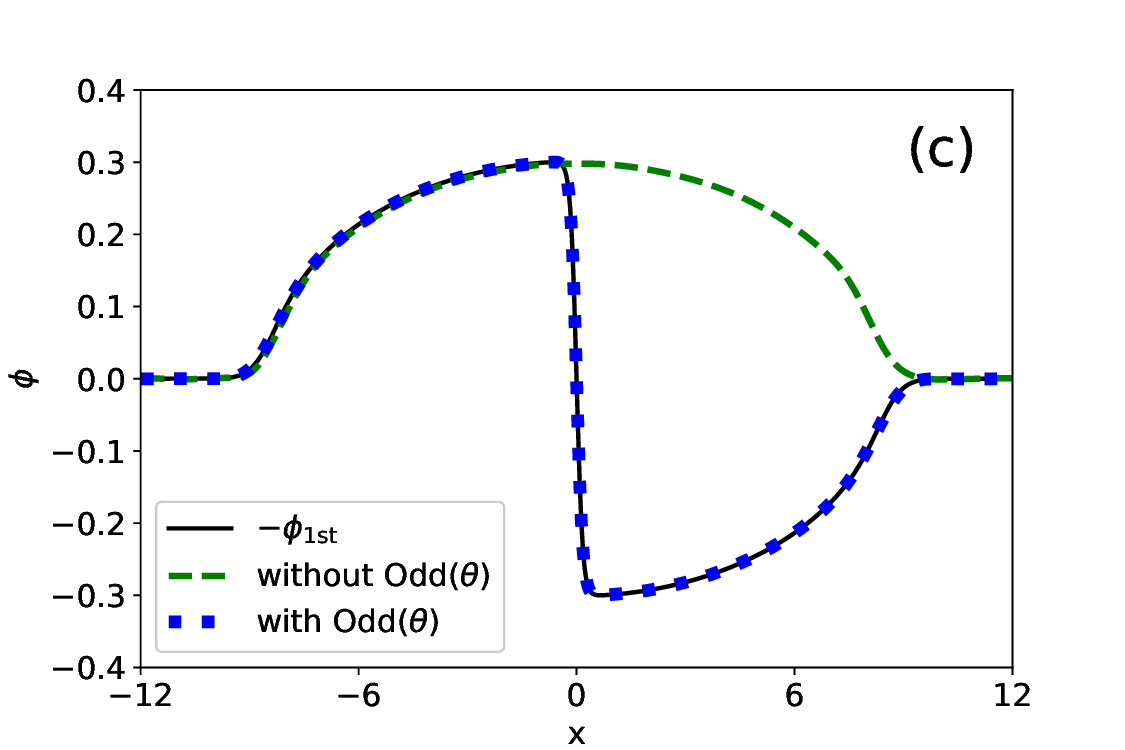}
	\end{minipage}}
	\hspace{-6mm}
	\subfigure{
		\begin{minipage}[t]{0.45\textwidth}
			\centering
			\includegraphics[width=1\textwidth]{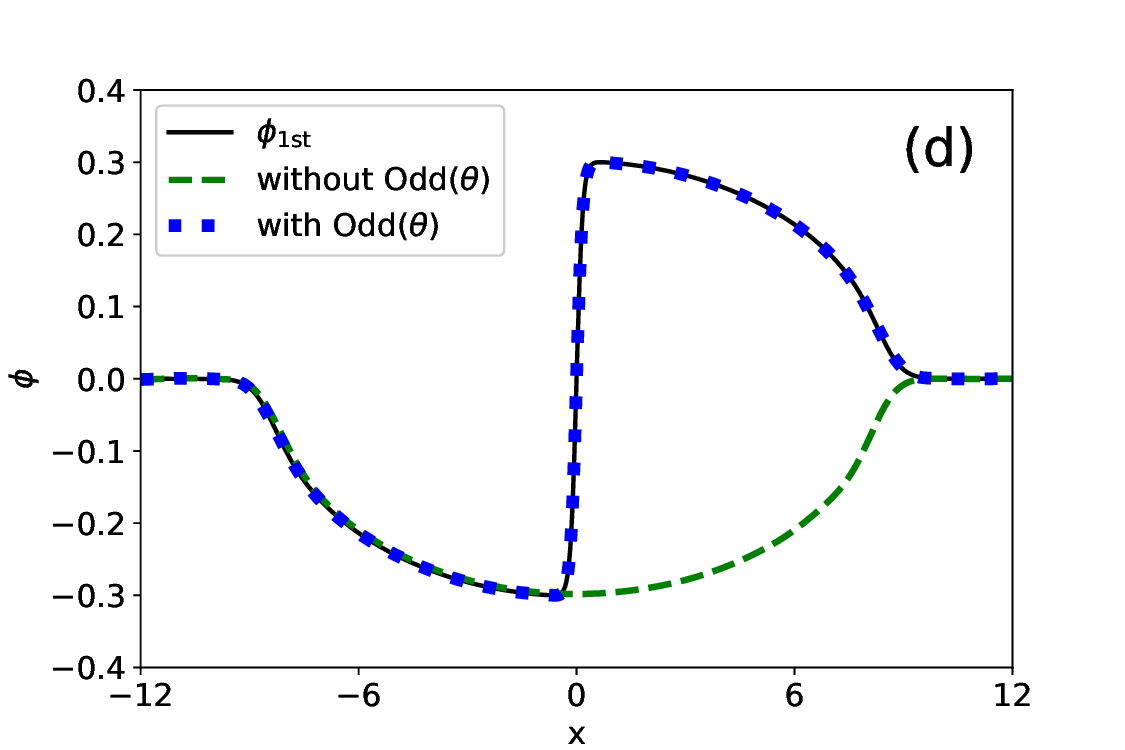}
	\end{minipage}}
        \vspace{-1mm}	
 \caption{FES problem: (a) initial profile of norm-DNN from pre-training or Xavier method \cite{glorot2010understanding}; (b) norm-DNN $\phi_{\theta}$ with Xavier method; (c) norm-DNN $\phi_{\theta}$ pre-trained from $\phi_{0}$; and (d) norm-DNN $\phi_{\theta}$ pre-trained from  $\phi_{0,fes}$. }
	\label{fig:fes_1d}
\end{figure*}

The other eigenfunctions of \eqref{eq:eigen} with energy larger than  $E_g$ are referred as excited states.
In addition to GS, the first excited state (FES) of the BEC is also widely concerned \cite{nmfes}.
It can be characterized as the  minimizer of  the energy functional in the orthogonal space of GS under the mass constraint, i.e.,
	\begin{equation}
		\label{eq:excite}
		\phi_{\mathrm{1st}}\in\argmin\{E(\phi)\,:\, \Vert \phi \Vert_2 = 1, \ <\phi,\phi_g>=0\},
	\end{equation}
	with $<\cdot,\cdot>$ denotes the inner product in $L^2(\bR^d)$.
Now we present the special strategy for norm-DNN to compute it.
As given in Section \eqref{sec:pre_gs}, the FES $\phi_{\mathrm{1st}}$ is orthogonal to $\phi_g$ in  {$L^2$}. When the potential $V(x)$ is an even function, $\phi_{\mathrm{1st}}$ is in fact known to be an odd function \cite{BaoCai} that minimizes the energy on the unit sphere of  {$L^2$}.
The traditional GFDN method therefore will choose an odd initial guess, e.g., $\phi_{0,fes}(x):= \sqrt{2}x\fe^{-\frac{x^2}{10}} / (5^{3} \pi)^{1/4}$, and then the flow will stay in the correct functional space to reach $\phi_{\mathrm{1st}}$.
Unfortunately, our norm-DNN cannot preserve the parity of the initial data to the end of training.
This fact has already been shown in Figure \ref{fig:gauss}(b), where training the norm-DNN from an even function $\phi_0$ may yield an odd numerical solution. On the other hand, the training from a normalized odd function $\phi_{0,fes}$ may yield an even solution, as will be shown in Figure \ref{fig:fes_1d}(b,c,d).

To efficiently and accurately fit the FES, we propose the following two modifications for the norm-DNN approach:
\begin{itemize}
	\item  {Unlike} GS, FES does not satisfy the non-negative  property, and so we do not need the shift layer $\mathcal{T}$ (\ref{eq:x_min}) in (\ref{1d ndnn}). Therefore for the 1D FES problem, the norm-DNN simply reads:
	\begin{equation}\label{fes ndnn}
		\phi_{\theta}(x) = \mathcal{N} \circ {\bf F}_{L+1} \circ \sigma \circ {\bf F}_{L} \circ \sigma \circ\cdots \circ {\bf F}_{2} \circ \sigma \circ {\bf F}_{1}(x).
	\end{equation}
	
	\item We add a regularization term in the loss to restrict $\phi_{\theta}(x)$ being odd in space, and so we solve the following minimization problem:
	\begin{equation}
		\label{eq:loss_odd}
		\min_{\theta}\left\{
		\text{Loss}(\theta) + \text{Odd}(\theta)\right\},\qquad \text{Odd}(\theta):=\frac{\delta}{N_o} \sum_{j=1}^{N_o} |\phi_\theta(x_j^o) + \phi_\theta(-x_j^o)|^2,
	\end{equation}
	with a wight $\delta>0$ and the training points $\big \{x^{o}_{j}\big \}_{j=1}^{N_{o}} \subset \Omega\cap\bR^+$ for the {odd}-function restriction.
\end{itemize}

\begin{table*}[h!] 	
	\centering 	
	\caption{Number of iterations, computational time, relative error \eqref{eq:relative_error}, the error of numerical energy $E(\phi_\theta)$  of norm-DNN for FES problem (exact FES energy $E(\phi_{\mathrm{1st}})=22.0777$). }	
	\label{tab:fes_1d} 	
	\begin{tabular}{c|c|cccc} 		\hline 					
		& Initialization & Iterations & Time & Error \eqref{eq:relative_error} & $|E(\phi_{\mathrm{1st}}) - E(\phi_{\theta})|$   \\ \hline
		\multirow{3}{*}{with Odd}  &$\phi_{0, fes}$  & 2400   &  19s    & 2.68E-3 & 1.00E-4 \\
		&$\phi_{0}$ & 2100    &  17s     & 2.07E-3 & 1.00E-4 \\
		& Xavier \cite{glorot2010understanding}     & 15500   &  118s    & 2.90E-3 & 1.00E-4 \\ \hline
		\multirow{3}{*}{without Odd} &$\phi_{0, fes}$  & 2300   &  26s    & 1.41E+0  & 7.28E-1\\
		&$\phi_{0}$ & 4600    &  34s    & 1.41E+0  & 7.28E-1 \\
		&Xavier \cite{glorot2010understanding}      & 2400   &  16s    & 1.93E+0  & 1.15E-1\\ \hline
	\end{tabular}
\end{table*}

Now let us fix $V(x)=\frac{1}{2}x^2$, $\beta=400$ for numerical investigations.
The hyper-parameters are set as $L=4$, $W=70$, $tol=10^{-6}$, $N_{x}=512$. The weight in (\ref{eq:loss_odd}) is taken as $\delta= 100$ and the points $x_o^j$ are taken as the grid points in $[0,12]$.  In Table \ref{tab:fes_1d},
we present the number of iterations, computational time, the relative function error \eqref{eq:relative_error} and the error of the numerical energy $E(\phi_{\theta})$ of norm-DNN for the FES problem. To address the necessity of the regularization in (\ref{eq:loss_odd}), we have presented the numerical results of norm-DNN with or without using $\text{Odd}(\theta)$ for comparison. Moreover, the Xavier method \cite{glorot2010understanding} or the pre-training with the Gaussian $\phi_0$ or the odd function $\phi_{0,fes}$ has been used for the initialization. The profiles of the solutions are shown in Figure \ref{fig:fes_1d}.

From the numerical results, we have the following findings.
1) Norm-DNN with the regularization $\text{Odd}(\theta)$ provides accurate and efficient approximation to the FES, and the energy error $|E(\phi_{\mathrm{1st}}) - E(\phi_{\theta})|$ can reach $10^{-4}$.
2) The random initialization for (\ref{eq:loss_odd}) which at the beginning is far away from the FES, can still converge to the FES $\phi_{\mathrm{1st}}$ and approximate it accurately. This indicates the robustness of the proposed approach for FES problem.
3) Pre-training with the Gaussian $\phi_{0}$ or the odd function $\phi_{0,fes}$ can significantly accelerate the convergence. The difference between the two is subtle.
4) Without using $\text{Odd}(\theta)$, norm-DNN could produce the wrong solution despite of the used initialization.
It turns to GS  {or some other states} instead of FES.

\begin{figure*}[h!]
	\centering
	\subfigure{
		\begin{minipage}[t]{0.4\textwidth}
			\centering
			\includegraphics[width=1\textwidth]{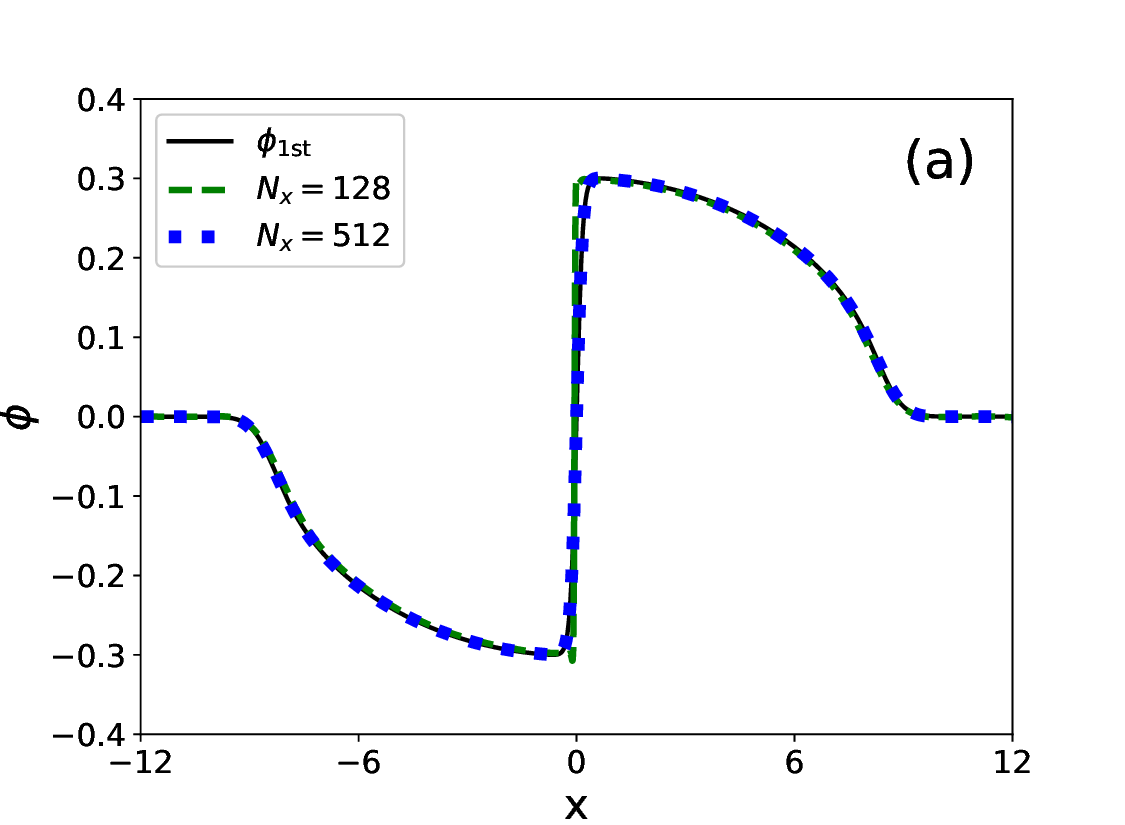}
	\end{minipage}}
	\subfigure{
		\begin{minipage}[t]{0.4\textwidth}
			\centering
			\includegraphics[width=1\textwidth]{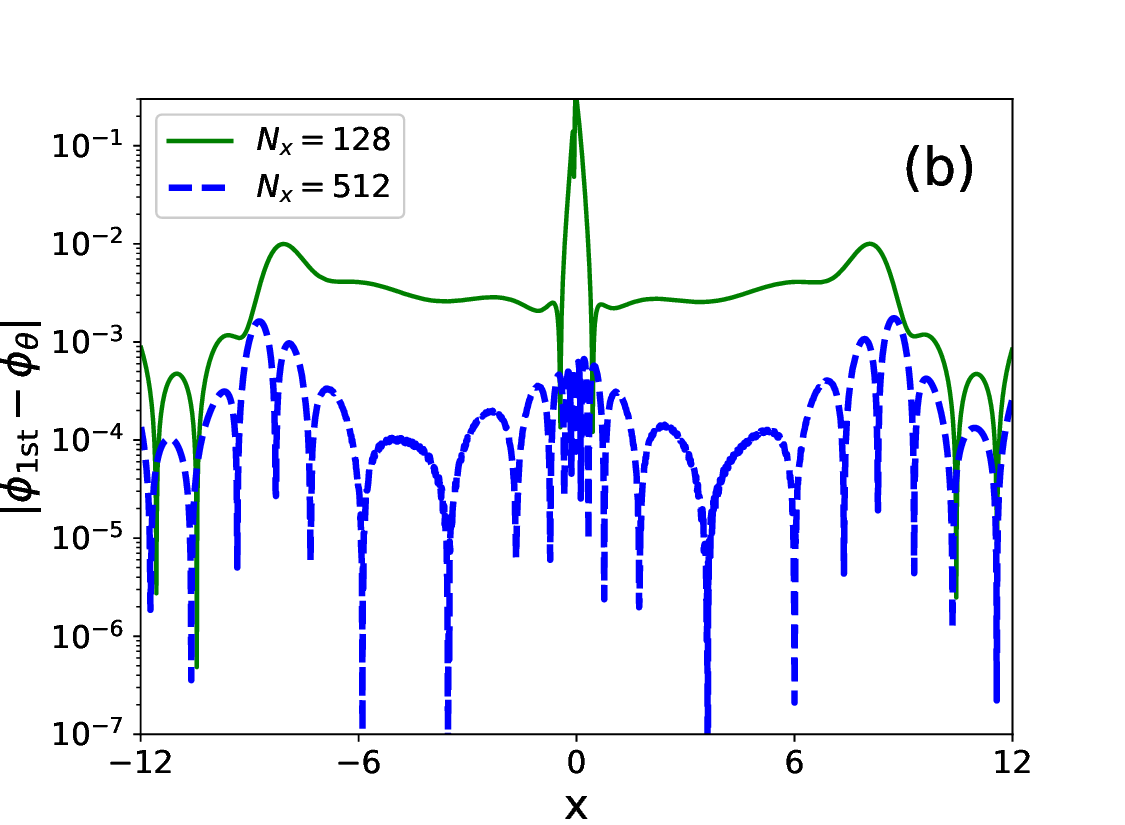}
	\end{minipage}}
	\vspace{-3mm}
	\subfigure{
		\begin{minipage}[t]{0.33\textwidth}
			\centering
			\includegraphics[width=1\textwidth]{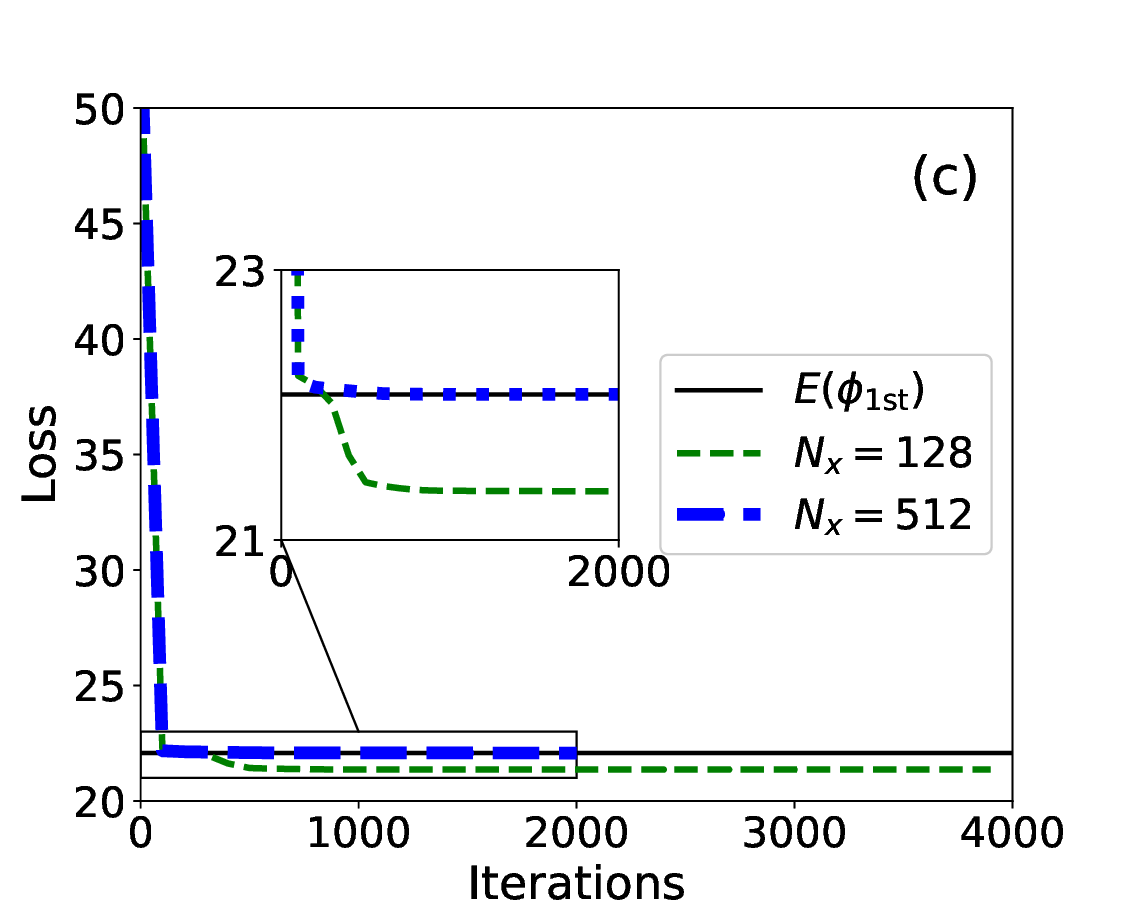}
	\end{minipage}}
	\hspace{-6mm}
	\subfigure{
		\begin{minipage}[t]{0.33\textwidth}
			\centering
			\includegraphics[width=1\textwidth]{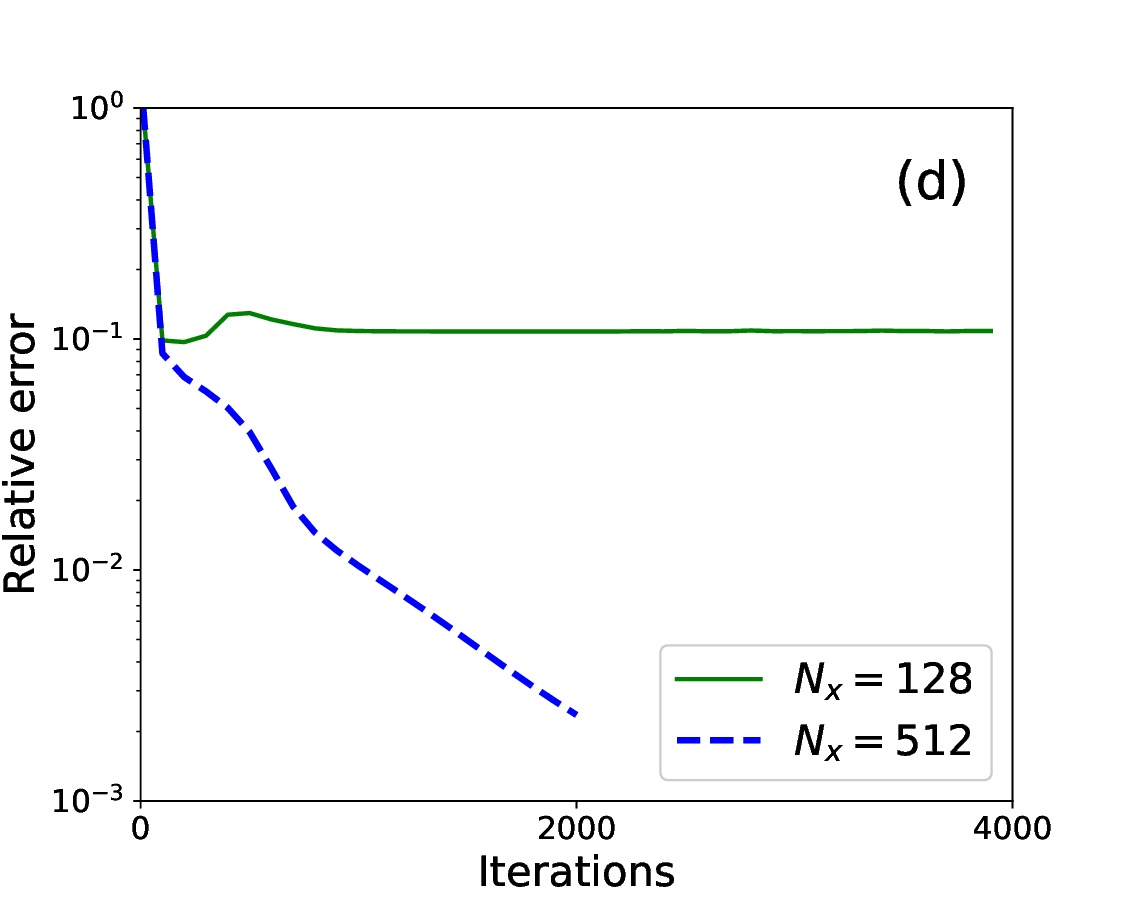}
	\end{minipage}}
	\hspace{-6mm}
	\subfigure{
		\begin{minipage}[t]{0.33\textwidth}
			\centering
			\includegraphics[width=1\textwidth]{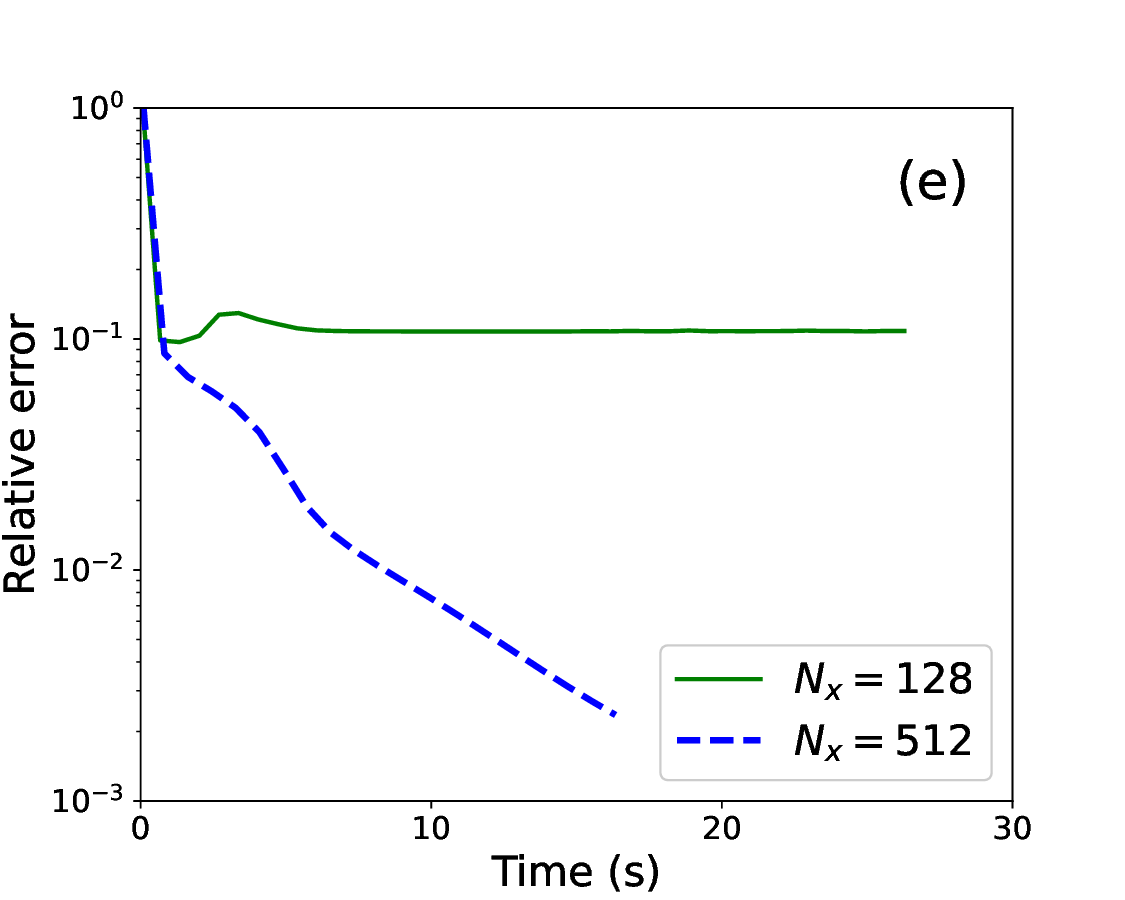}
	\end{minipage}}
 \vspace{-1mm}
	\caption{Norm-DNN with different $N_x$ for FES problem: (a) exact solution and numerical solution of norm-DNN; (b) pointwise error; (c,d,e) loss and error (\ref{eq:relative_error}) during the iterations or computational time.}
	\label{fig:fes_1d_nx}
\end{figure*}

\begin{table*}[h!] 	
	\centering 	
	\caption{The relative error \eqref{eq:relative_error} and training time of norm-DNN with different $N_x$ for FES problem. } 	
	\label{tab:fes_1d_nx} 	
	\begin{tabular}{c|cccc}
		\hline
		$N_x$  & $64$ & $128$ & $256$ & $512$         \\ \hline
		Error(Time) & 1.34E-1(15s) & 1.08E-1(26s) & 1.03E-1(60s) & 2.07E-3(17s)    \\ \hline
	\end{tabular}
\end{table*}

One more thing that we would like to address here is the number of spatial grid points used for FES.
The previous experiments for the GS problem \eqref{sec:hop} use $N_x=128$, which is enough for the training to converge quickly and the error can reach the order of $10^{-3}$. Here for the FES problem, we need to take more grid points to accurately capture the sudden jump in $\phi_{1\text{st}}$ at the origin.  To illustrate this issue, we test the performance of the training for $N_{x}\in \left\{64, 128,256,512\right\}$. The other parameters are fixed as before and the Gaussian $\phi_0$ for pre-training is used.
The relative error \eqref{eq:relative_error} and training time of norm-DNNs are given in Table \ref{tab:fes_1d_nx}. The piecewise errors and the  changes of loss and relative error \eqref{eq:relative_error} along with the iterations and time are shown in Figure \ref{fig:fes_1d_nx}.
The numerical results show that norm-DNN with $N_x=512$ gives the fastest convergence rate and best accuracy for computing FES. When there is insufficient training data, the approximation could be poor especially at the jump, and overfittings may occur as shown by the green line in Figure \ref{fig:fes_1d_nx}(d,e).

\section{Generalization and comparison}\label{sec:5}
The previous experiments have investigated the norm-DNN for approximating a single GS of (\ref{eq:GPE}) under a fixed set of physical parameters, e.g., $\beta$ and $\gamma_j$ in $V$. Mathematically, the GS $\phi_g$ can be viewed as a function depending on these  physical parameters.
Thus for applications that concern parameters from a certain range, it would be more interesting and useful to have a fully trained norm-DNN in parameters space. Such trained norm-DNN will then be able to directly provide a valid numerical approximation of GS for any  parameters within the range, which makes it very simple and efficient for applications. This is considered as the parameter generalization of norm-DNN, which we shall study in this section.

\subsection{Generalization}\label{sec:parameters}
The traditional approaches for computing GS  are all designed under fixed parameters. It means that one need to specify the values of all the physical parameters in order to run the numerical scheme to get the specific solution. Whenever the value of one parameter is  changed, one will have to run the scheme once again for the new solution. This could certainly be very costly and tedious for applications.
The DNNs in classical machine learning are known to have strong parameter generalization ability.
We now propose such parameter generalization for norm-DNN to compute GS within a range of parameters and investigate its performance.
Let us consider the 1D GS problem as the illustrative example for simplicity. We take the harmonic oscillator potential $V(x)=\gamma x^2$.  Then, the concerned physical parameters in the model (\ref{eq:loss_gs}) are $\gamma$ and $\beta$. The parameter generalization procedure is presented in Algorithm \ref{alg:1}.

\begin{algorithm}
    \renewcommand{\algorithmicrequire}{\textbf{Input:}}
    \renewcommand{\algorithmicensure}{\textbf{Output:}}
    \caption{Parameter generalization for norm-DNN}
    \label{alg:1}
    \begin{algorithmic}[1]
        \REQUIRE Domain for parameters $P=\{(\gamma, \beta)\}$ and physical domain $\Omega$
        \ENSURE Fully trained norm-DNN $f_{\Theta}(\gamma,\beta)\approx \phi_g$ for any $(\gamma, \beta)\in P$
        \STATE Sample points for training: $P_s=\{(\gamma_i, \beta_j)\}\subset P$, $\left\{x_{k} \right\}_{k=1}^{N_x}\subset\Omega$
        \STATE For each $(\gamma_i, \beta_j)\in P_s$, compute the norm-DNN from (\ref{eq:loss_gs}) and denote it as $\phi_{\theta}^{i,j}(x)$
        \STATE Give spatial points $\left\{x_{l} \right\}_{l=1}^{N_t}\subset\Omega$ where the value of $\phi_g$ is of interest, then
         evaluate $\phi_{\theta}^{i,j}$ at $\left\{x_{l} \right\}_{l=1}^{N_t}$
        \STATE Set $\phi_{\theta}^{i,j}(x_l)$ as the label and do supervised training: generate a FCNN $f_{\Theta}(\gamma, \beta)=(f_{\Theta}^1,\ldots,f_{\Theta}^{N_t})^\top$, i.e., \eqref{eq:DNN} with $n_{L+1}=N_t,\,n_0=2$; train it by considering
        \begin{equation}
            \label{eq:loss_params}
                    \min_{\Theta} \sum_{P_s} \sum_{l=1}^{N_t} |\phi_{\theta}^{i,j}(x_l)-f_\Theta^l(\gamma_i, \beta_j)|^2
        \end{equation}

    \end{algorithmic}
\end{algorithm}

\begin{remark}
The output of Algorithm \ref{alg:1} takes discrete values in the physical domain $\Omega$, i.e., $f_{\Theta}\in\bR^{N_t}$ for every $(\gamma,\beta)$. It approximates $\phi_g$ at the specified points $\left\{x_{l} \right\}_{l=1}^{N_t}$.  These points can be chosen or changed based on the practical need. If changed, one can just repeat the steps 3-4 of Algorithm \ref{alg:1} with the new $x_l$ to get the FCNN. 
{Besides, combining the parameter $\theta$ and spatial variable $x$ optimization, an object would be training $f_{\Theta}$ as a function of the trio $(\gamma,\beta,x)$, which via the framework of PINO \cite{PINO} or DeepONet \cite{DeepONet} is possible and may  enhance the performance, and this will be considered in a future work.}
\end{remark}

\begin{table*}[h!] 	
	\centering 	
	\caption{Error \eqref{eq:relative_error} of norm-DNN on grid points of parameter space.} 	
	\label{tab:params_error} 	
	\begin{tabular}{|c|c|c|c|c|c|c|}
        \hline
        \diagbox{$\beta$}{$\gamma$} & 0.5 &0.6  &0.7  &0.8  &0.9  &1.0  \\ \hline
        300 & 7.46E-3 & 8.18E-3 & 7.69E-3 & 9.13E-3 & 1.01E-2 & 8.59E-3 \\ \hline
        310 & 9.02E-3 & 7.89E-3 & 7.45E-3 & 8.83E-3 & 8.81E-3 & 8.99E-3 \\ \hline
        320 & 9.52E-3 & 9.00E-3 & 6.84E-3 & 9.43E-3 & 9.91E-3 & 1.07E-2 \\ \hline
        330 & 1.14E-2 & 8.62E-3 & 8.90E-3 & 6.85E-3 & 1.08E-2 & 9.56E-3 \\ \hline
        340 & 1.05E-2 & 9.00E-3 & 8.40E-3 & 6.98E-3 & 9.02E-3 & 9.87E-3 \\ \hline
        350 & 1.20E-2 & 8.32E-3 & 9.01E-3 & 8.15E-3 & 9.03E-3 & 9.01E-3 \\ \hline
        360 & 1.29E-2 & 9.04E-3 & 8.12E-3 & 6.67E-3 & 9.15E-3 & 1.07E-2 \\ \hline
        370 & 1.34E-2 & 9.37E-3 & 8.20E-3 & 8.08E-3 & 8.33E-3 & 9.87E-3 \\ \hline
        380 & 1.28E-2 & 9.43E-3 & 8.68E-3 & 9.74E-3 & 7.37E-3 & 9.72E-3 \\ \hline
        390 & 1.25E-2 & 1.10E-2 & 8.31E-3 & 9.36E-3 & 7.92E-3 & 9.78E-3 \\ \hline
        400 & 1.20E-2 & 1.11E-2 & 7.76E-3 & 9.67E-3 & 7.71E-3 & 7.64E-3 \\ \hline
	\end{tabular}
\end{table*}

\begin{table*}[h!] 	
	\centering 	
	\caption{Error \eqref{eq:relative_error} of FCNN on the four test points in parameter space.} 	
	\label{tab:params} 	
	\begin{tabular}{c|cc}
		\hline
	Error \eqref{eq:relative_error}    & $\beta=355$   & $\beta=365$        \\ \hline
		$\gamma=0.725$ & 4.19E-2  & 3.26E-2  \\
		$\gamma=0.825$ & 1.84E-2  & 1.51E-2  \\ \hline
	\end{tabular}
\end{table*}

\begin{figure*}[h!]
	\centering
	\subfigure{
		\begin{minipage}[t]{0.45\textwidth}
			\centering
			\includegraphics[width=1\textwidth]{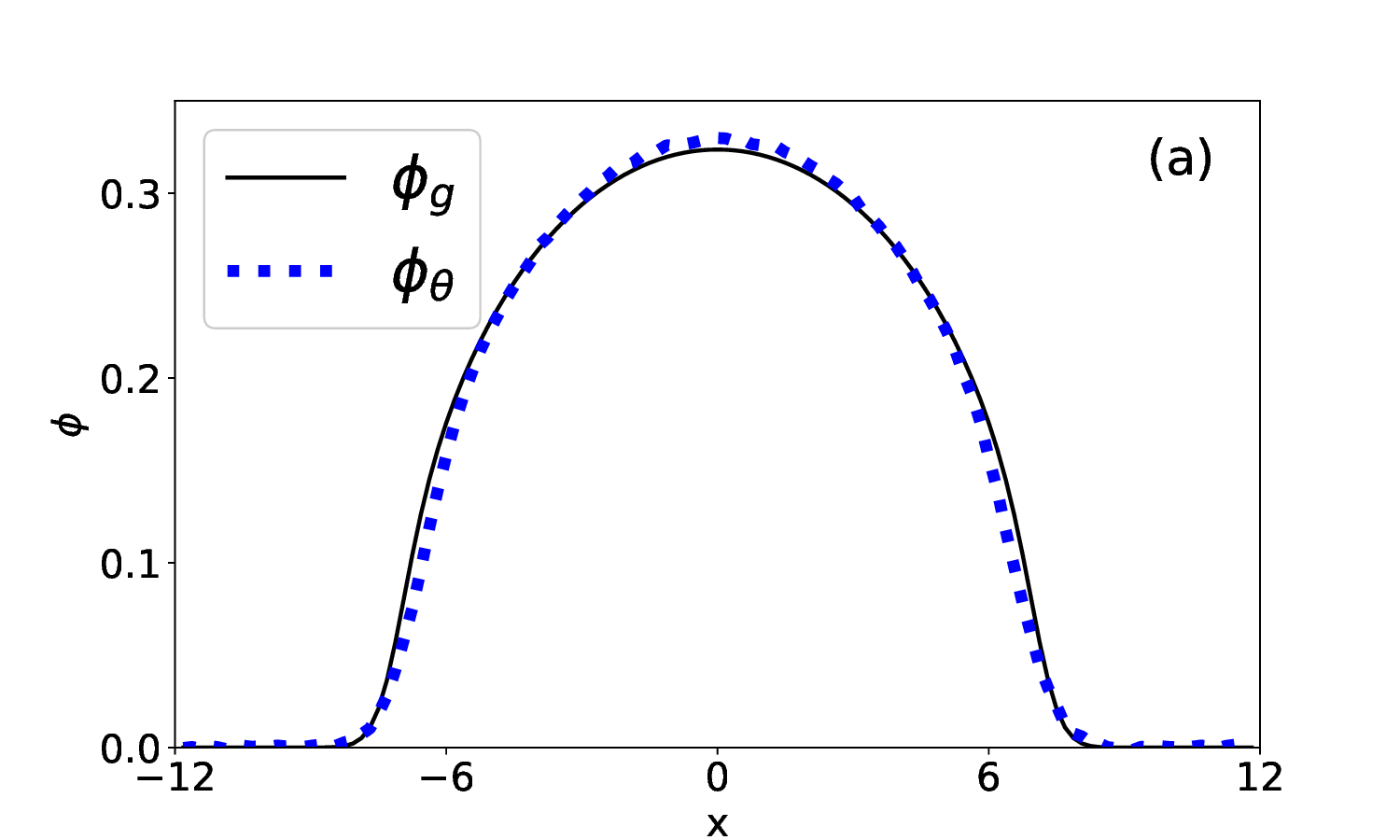}
	\end{minipage}}
	\hspace{-5mm}
	\subfigure{
		\begin{minipage}[t]{0.45\textwidth}
			\centering
			\includegraphics[width=1\textwidth]{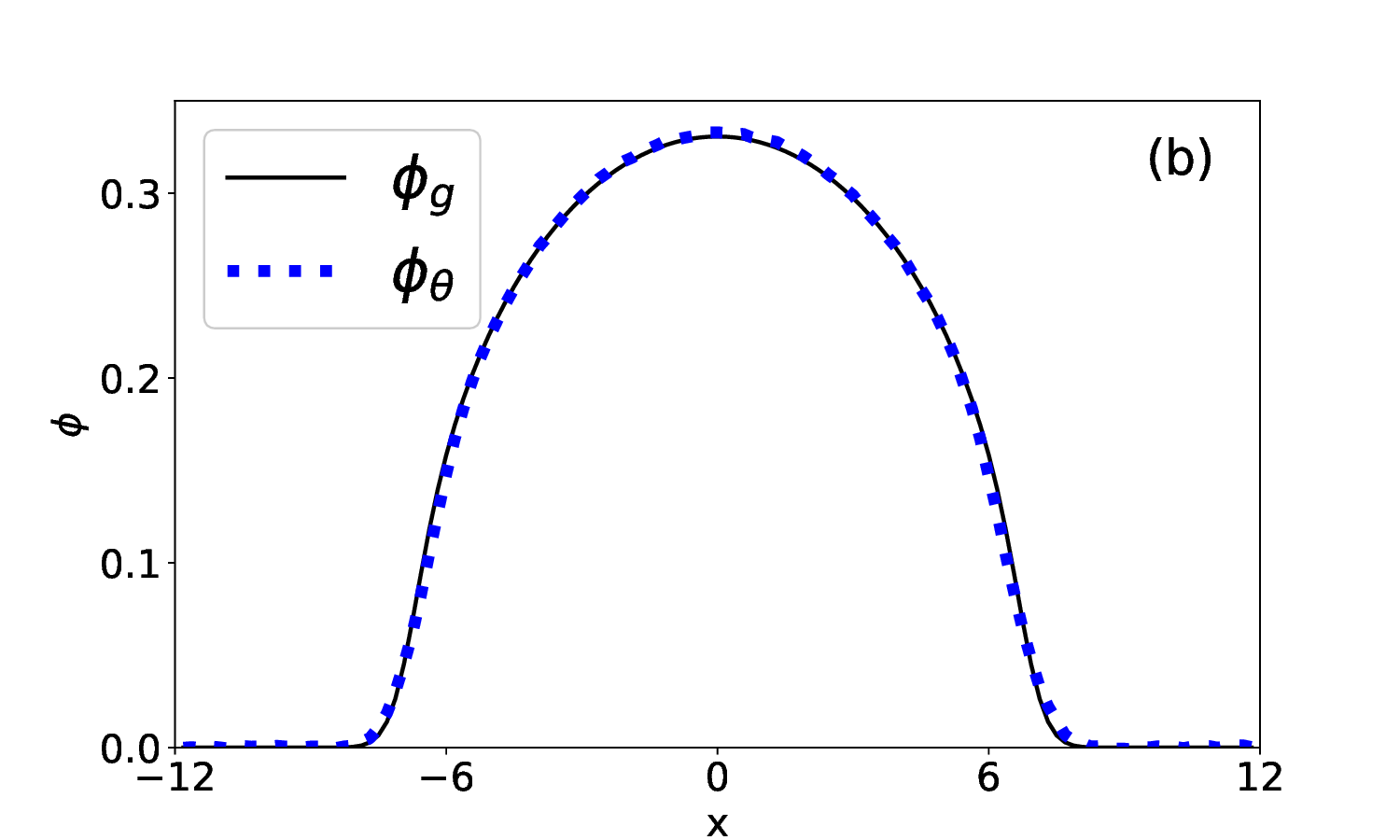}
	\end{minipage}}
        \vspace{-3mm}
	\subfigure{
		\begin{minipage}[t]{0.45\textwidth}
			\centering
			\includegraphics[width=1\textwidth]{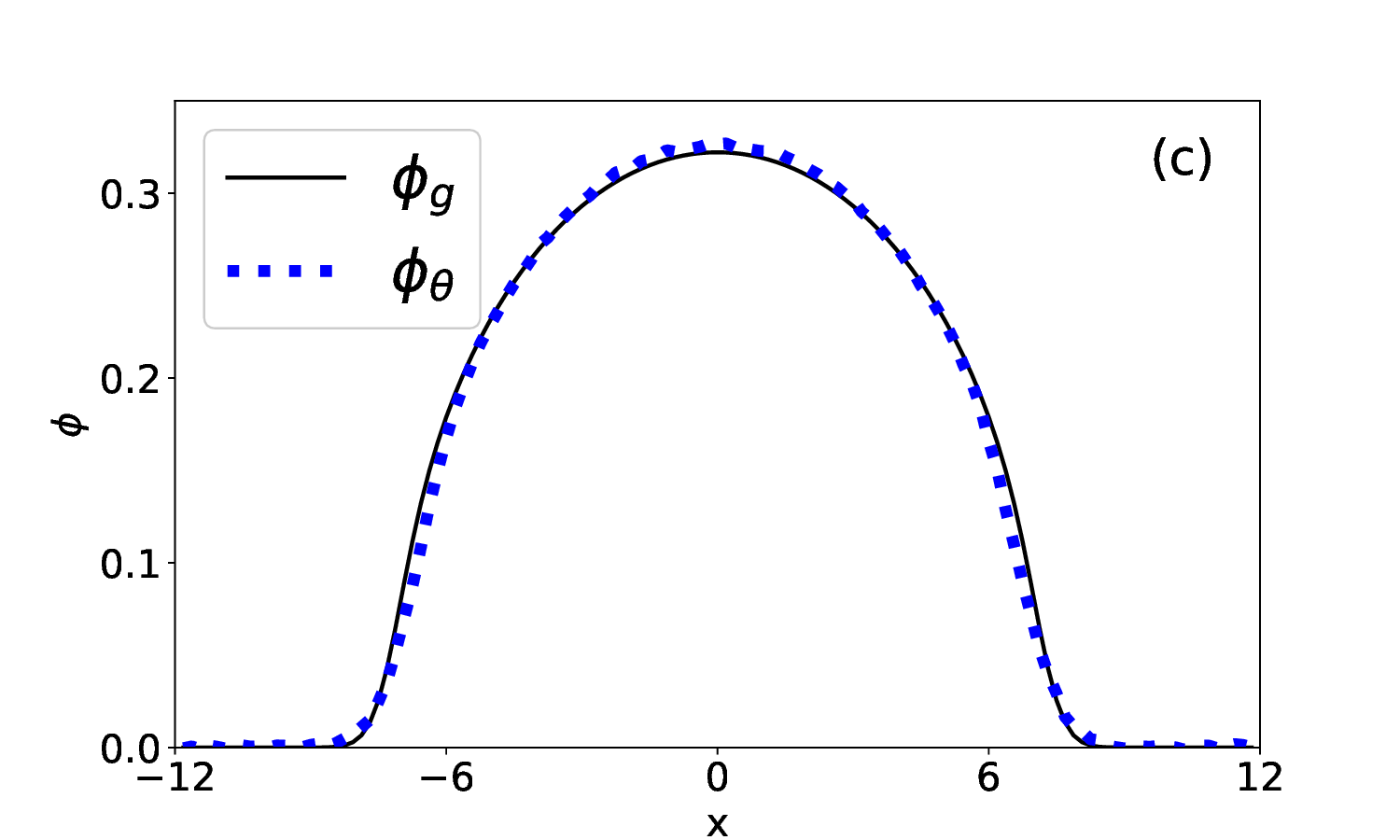}
	\end{minipage}}
	\hspace{-5mm}
	\subfigure{
		\begin{minipage}[t]{0.45\textwidth}
			\centering
			\includegraphics[width=1\textwidth]{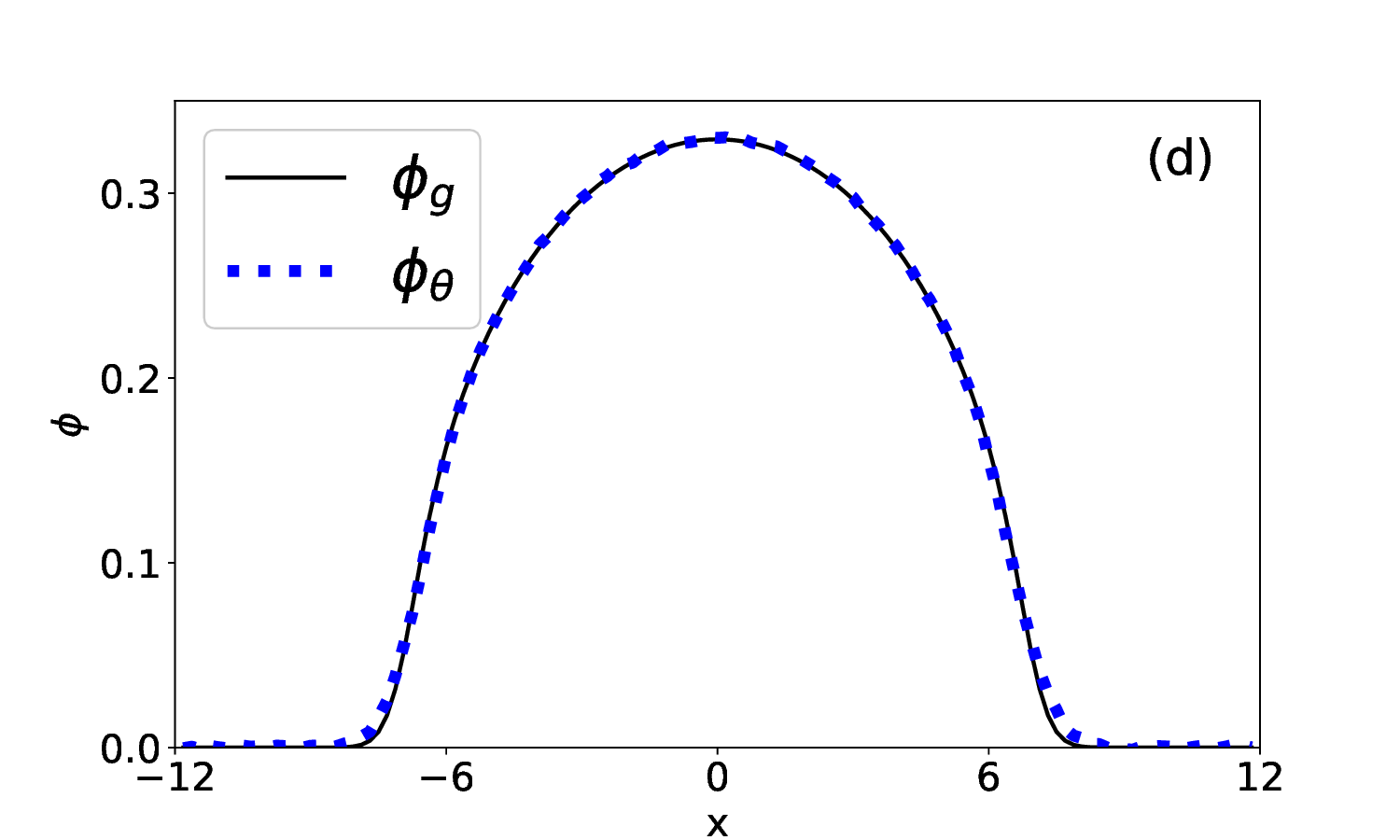}
	\end{minipage}}
 \vspace{-1mm}
	\caption{Profiles of exact solution and numerical solution from norm-DNN/FCNN at the four test points in parameter space: (a) $\beta=355$, $\gamma=0.725$; (b) $\beta=355$, $\gamma=0.825$; (c) $\beta=365$, $\gamma=0.725$; (d) $\beta=365$, $\gamma=0.825$.}
	\label{fig:params}
\end{figure*}

To investigate the performance of Algorithm \ref{alg:1}, we consider a precise experiment.
Fix the physical domain $\Omega=(-12,12)$ and choose the parameter domain as $P = \left\lbrace (\gamma, \beta)\, :\, 0.5\leq\gamma\leq1,\, 300\leq\beta\leq 400 \right\rbrace $.
A {$11\times 11$} equally spaced grid of $P$ is used to train the norm-DNN in the parameter space. Four test points $\left\lbrace 0.725,0.825\right\rbrace\times\left\lbrace 355,365\right\rbrace$ outside the grid points of $P$ are used to  compute the error and measure the generalization ability. We take $L=4$, $W=70$ for both of the networks, i.e., norm-DNN and FCNN in Algorithm \ref{alg:1}. The other hyper-parameters are set as: $N_x=N_t=128$ and $tol=10^{-5}$. The minimization at the supervised learning step in Algorithm \ref{alg:1} is done by Adam method with fixed $10000$ iterations.
In Table \ref{tab:params_error},  we first show the error of norm-DNN for computing the GS with fixed parameters at some training grid points in $P$. On the four test parameter points, the relative error \eqref{eq:relative_error} (norm computed over the grid with $N_x$) are shown in Table \ref{tab:params}. The profiles of the solutions at the test points are shown in Figure \ref{fig:params}.

From the numerical results,
it can be observed that for different parameters, norm-DNNs can get all the GS with relative error at about 1$\%$,
which to some extent indicates the robustness of norm-DNN in parameters.
For the parameters outside the grid points for training, the FCNN is also able to offer the approximation with the same magnitude of accuracy. This shows the validity of the fully trained norm-DNN and justifies the generalization ability of norm-DNN.

\subsection{Comparison with traditional DNN}
To further clarify the efficiency, we now carry out some comparisons between the proposed norm-DNN approach and the traditional DNN with  supervised learning approach (SL-DNN) for the GS problem.

\begin{figure*}[t!]
 	\centering
 	\subfigure{
 		\begin{minipage}[t]{0.45\textwidth}
 			\centering
 			\includegraphics[width=1\textwidth]{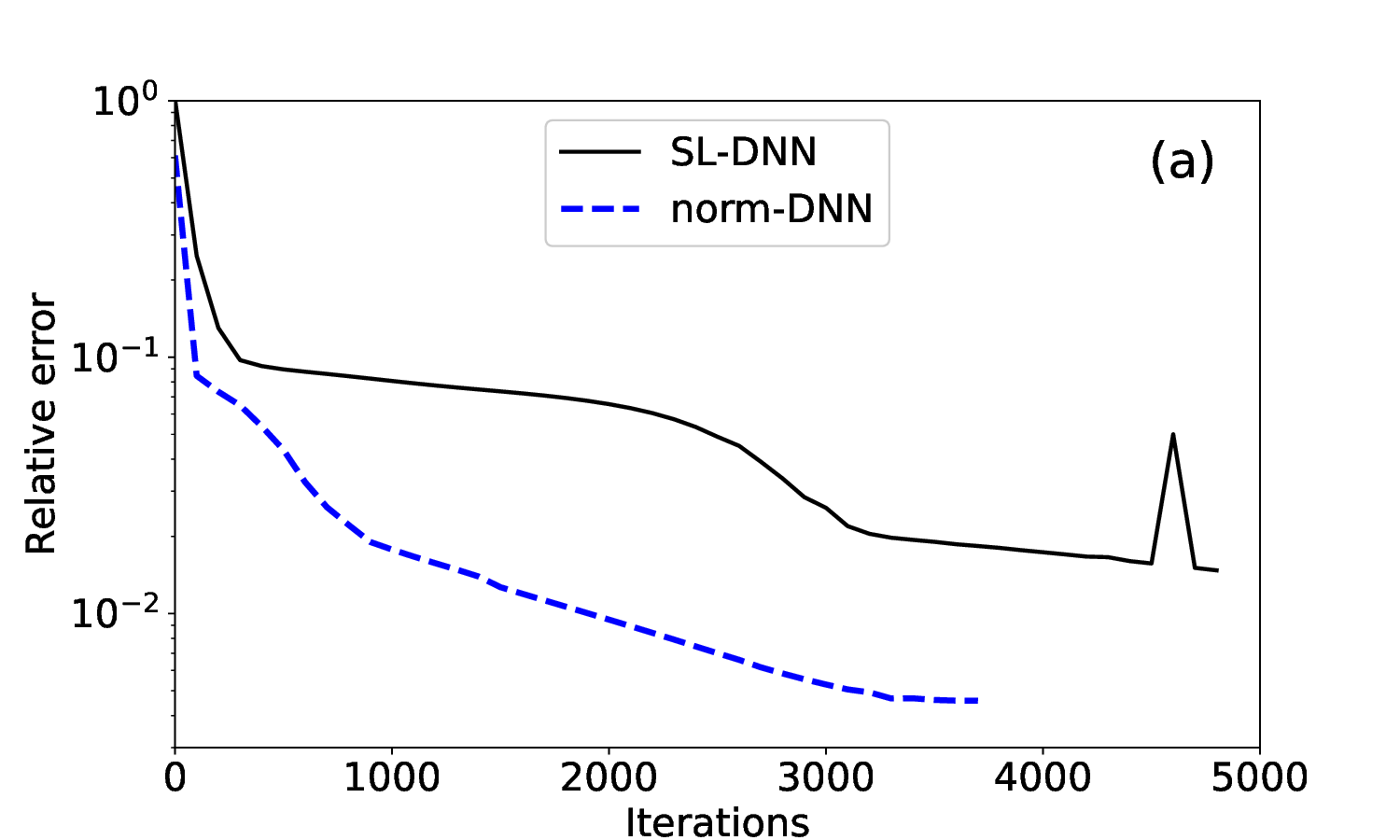}
 	\end{minipage}}
 	\hspace{-4mm}
 	\subfigure{
 		\begin{minipage}[t]{0.45\textwidth}
 			\centering
 			\includegraphics[width=1\textwidth]{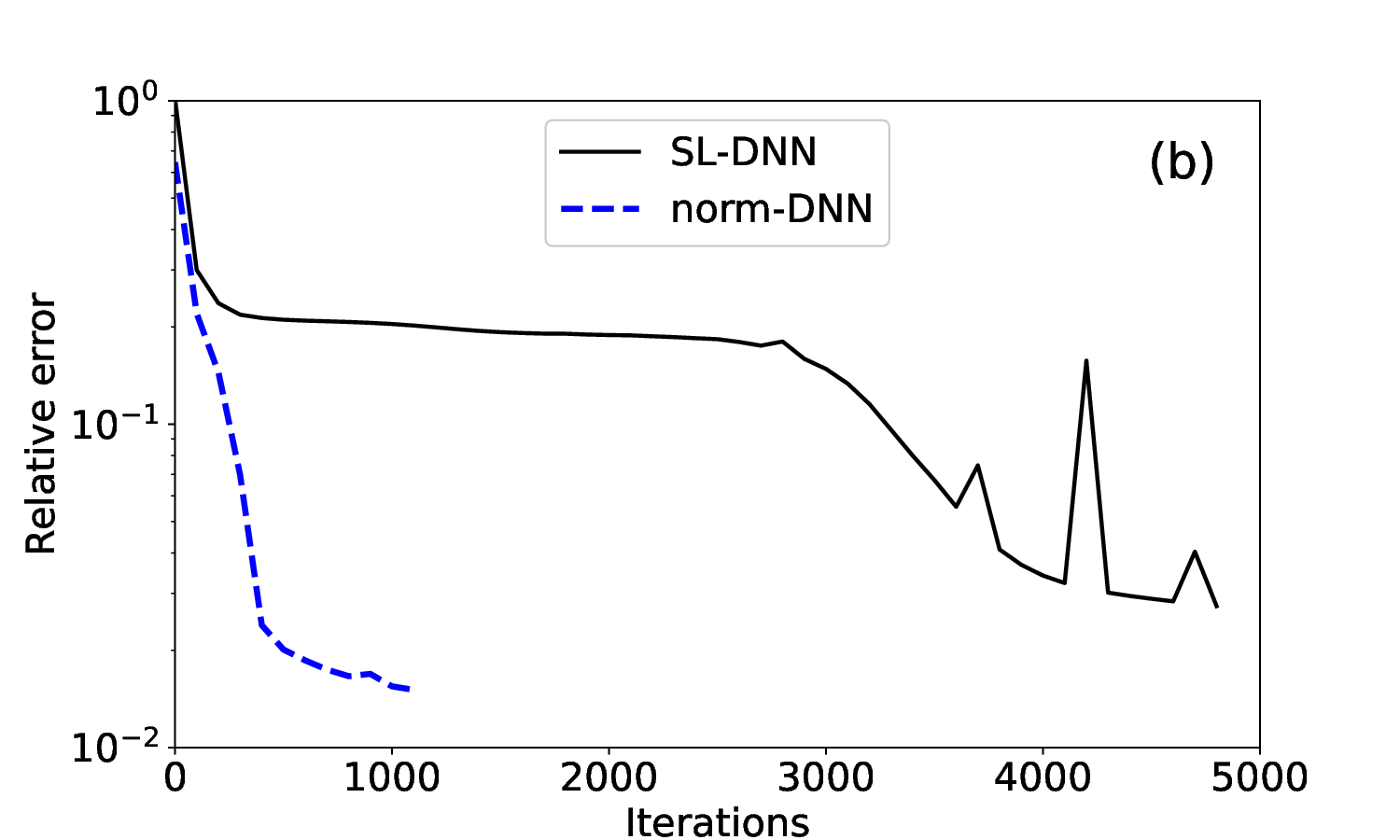}
 	\end{minipage}}
        \vspace{-3mm}
 	\subfigure{
 		\begin{minipage}[t]{0.45\textwidth}
 			\centering
 			\includegraphics[width=1\textwidth]{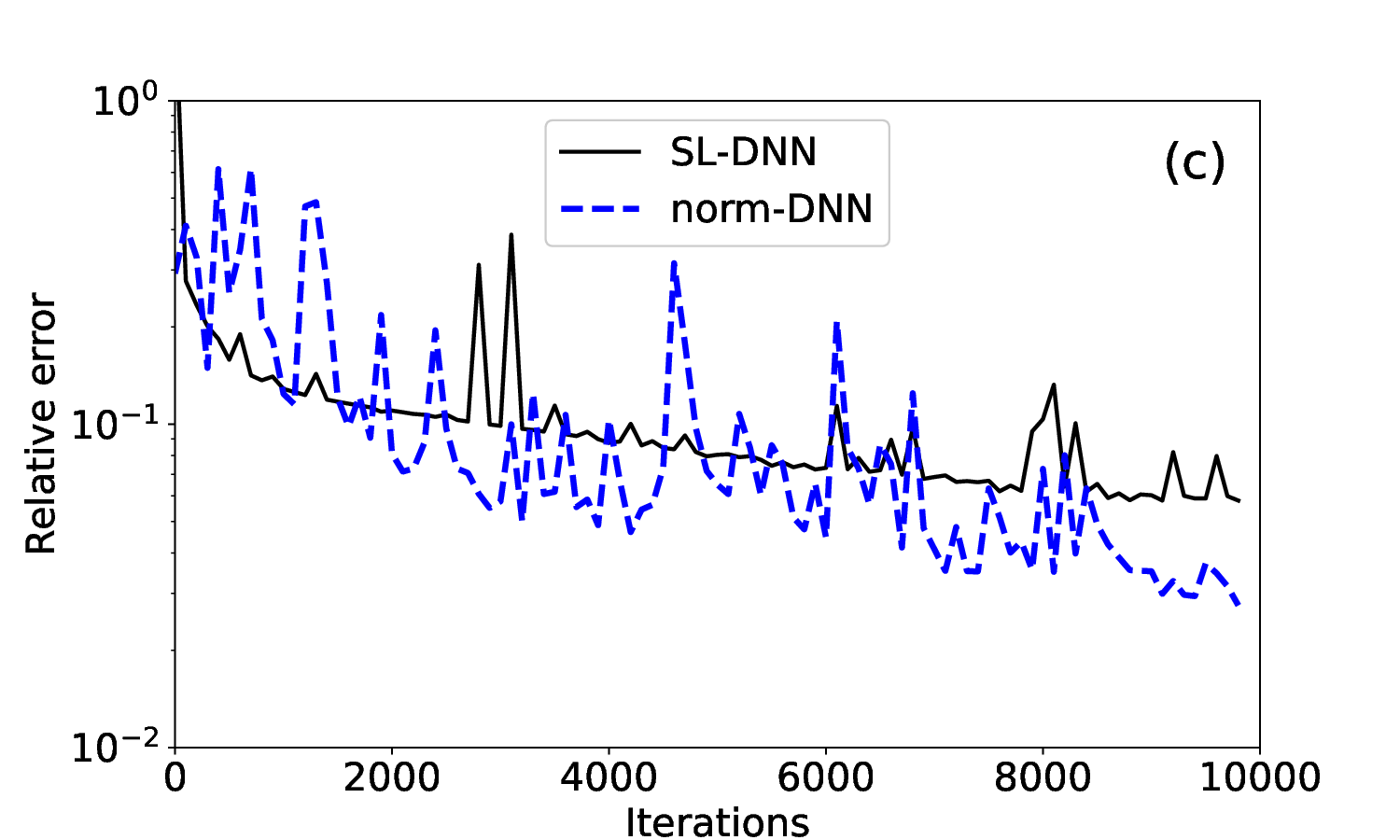}
 	\end{minipage}}
 	\hspace{-4mm}
 	\subfigure{
 		\begin{minipage}[t]{0.45\textwidth}
 			\centering
 			\includegraphics[width=1\textwidth]{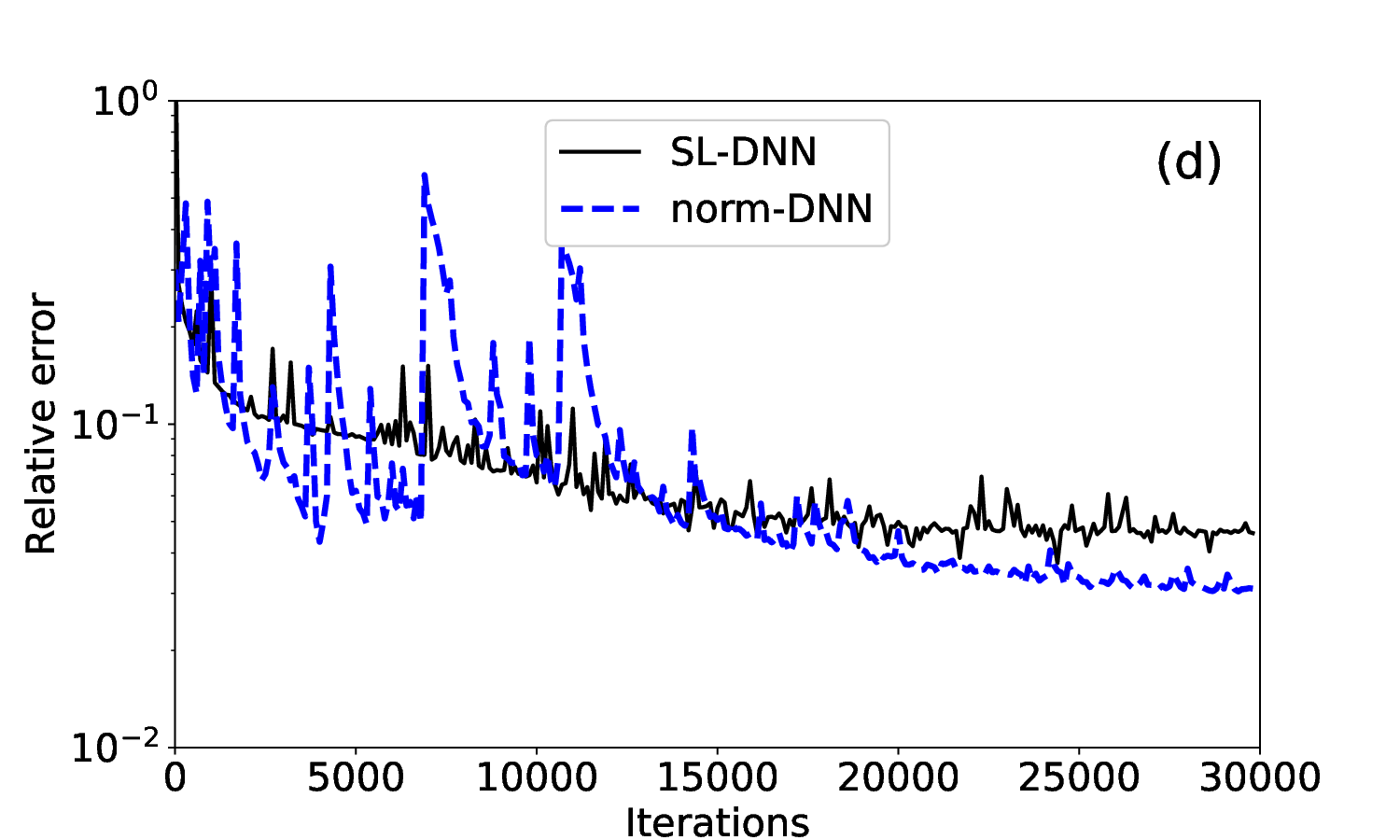}
 	\end{minipage}}
        \vspace{-1mm}
 	\caption{Error (\ref{eq:relative_error}) of the SL-DNN and norm-DNN during the iterations for the fixed-parameter experiments (a)-(d).}
 	\label{fig:com}
 \end{figure*}

Firstly, we consider the approximation of GS under fixed parameters. For the traditional SL-DNN, we first obtain the GS $\phi_g$ by the GFDN method (\ref{eq:gfdn}), and then we use its function values at the spatial grid points as the labeled data for supervised learning: generate a FCNN $f_\theta(x)$ as in (\ref{eq:DNN}) to solve $$\min_{\theta} \sum_{j=1}^{N_x} |\phi_{g}(x_j)-f_\theta(x_j)|^2.$$ With the obtained SL-DNN $f_\theta(x)$, we compute its relative error (\ref{eq:relative_error}) as before, and we compare it with the result from norm-DNN. The following four examples from previous studies will be used for numerical experiments:
\begin{itemize}
	\item[(a)] 1D problem with $\Omega=(-12,12)$, $\beta=400$, $V(x)=\frac{1}{2}x^2$;
	\item[(b)] 1D problem with $\Omega=(-12,12)$, $\beta=400$, $V(x)=\frac{1}{2} x^2 + 25 \sin^{2}(\frac{\pi x}{4})$;
	\item[(c)] 2D problem with $\Omega=(-6,6)\times(-6,6)$, $\beta=400$, $V(x,y)=\frac{1}{2}(x^2+y^2)$;
	\item[(d)] 2D problem with $\Omega=(-6,6)\times(-6,6)$, $\beta=400$, $V(x,y)=\frac{1}{2}(x^2+y^2)+\frac{5}{2}(\sin^2(\frac{\pi x}{4})+\sin^2(\frac{\pi y}{4}))$.
\end{itemize}
The hyper-parameters are set the same as in Section \ref{sec:parameters}. Then for the experiments (a)-(d), the corresponding relative error \eqref{eq:relative_error} and the training process are shown in Table \ref{tab:com} and Figure \ref{fig:com}. It can be seen from the numerical results that norm-DNN is more accurate than the supervised learning DNN. The training process of norm-DNN is at least as fast as that of the SL-DNN.
Note that the SL-DNN requires the exact solution for training data, while norm-DNN is completely unsupervised and does not require any data points from the unknown. This shows the significant advantage of {norm-DNN}.

\begin{table*}[h!] 	
 	\centering 	
 	\caption{Error \eqref{eq:relative_error} of supervised learning DNN and norm-DNN for fixed-parameter experiments (a)-(d).} 	
 	\label{tab:com} 	
	\begin{tabular}{c|cc}
		\hline
		Experiments& SL-DNN& norm-DNN     \\ \hline
		(a)           & 1.47E-2 & 4.57E-3 \\
		(b)          & 2.74E-2 & 1.51E-2 \\
		(c)          & 5.81E-2 & 2.75E-2 \\
		(d)           & 4.61E-2 & 3.10E-2 \\ \hline
	\end{tabular}
 \end{table*}

 \begin{figure*}[h!]
	\centering
	\subfigure{
		\begin{minipage}[t]{0.45\textwidth}
			\centering
			\includegraphics[width=1\textwidth]{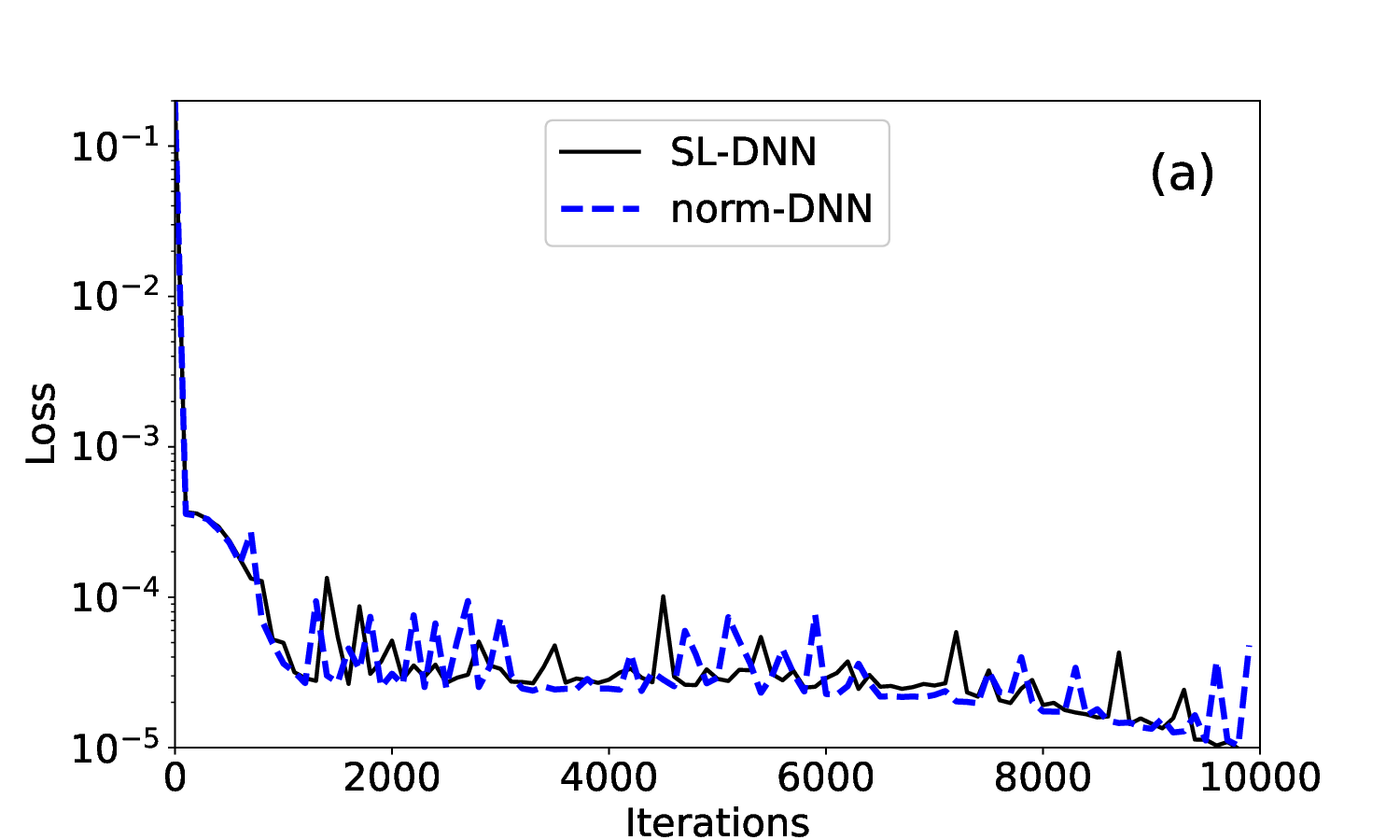}
	\end{minipage}}
	\hspace{-4mm}
	\subfigure{
		\begin{minipage}[t]{0.45\textwidth}
			\centering
			\includegraphics[width=1\textwidth]{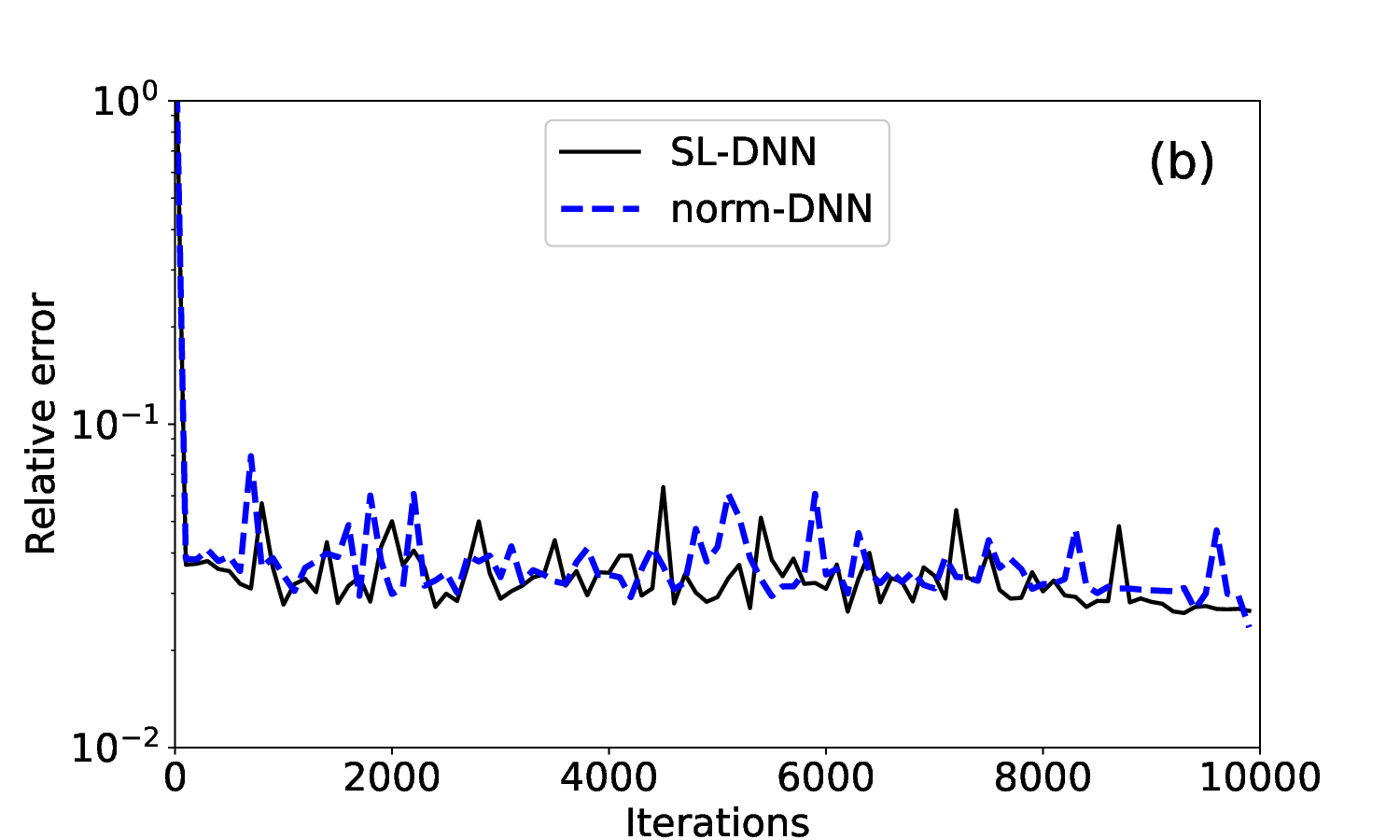}
	\end{minipage}}
	\vspace{-3mm}
	\caption{Loss \eqref{eq:loss_params} and error \eqref{eq:relative_error_F} during the iterations for parameter generalization problem.}
	\label{fig:params_com}
\end{figure*}

We also compare the parameter generalization of norm-DNN with the SL-DNN. Consider the numerical example from Section \ref{sec:parameters} with the same computational setup.
The procedure of the SL-DNN method in this case goes similarly as (\ref{eq:loss_params}) but with the labeled data  now obtained from GFDN for each parameter $(\gamma_{i}, \beta_{j})$.
With the four test points: $\left\{0.725,0.825 \right\}\times\left\{ 355, 365\right\}$ in the parameter space, we quantify the accuracy by computing
\begin{equation}
	\label{eq:relative_error_F}
	\text{error}= \left\|\phi^{\text{test}}_\theta-\phi^{\text{test}}_g\right\|_F/\left\|\phi^{\text{test}}_g\right\|_F,
\end{equation}
where $\phi^{\text{test}}_g, \phi^{\text{test}}_\theta \in \mathbb{R}^{4\times N_x}$ are the augmented exact and numerical GS solutions on the four test points and $\|\cdot\|_{F}$ denotes the Frobenius norm.
The loss \eqref{eq:loss_params} and relative error \eqref{eq:relative_error_F} are shown in Figure \ref{fig:params_com}.
Clearly the results show that for the parameter generalization problem, norm-DNN without data points can achieve the same accuracy and efficiency as the SL-DNN.

\section{Conclusion}\label{sec:6}
The ground state (GS) problem of the Gross-Pitaevskii model for Bose-Einstein condensate is a constraint optimization in functional space. In this paper, we applied the deep neural network (DNN) technique to numerically solve the problem with the intention to overcome possible curse-of-dimensionality and parameter generalization issues. In order to satisfy the mass constraint and guide the learning to GS, we proposed a normalized DNN (norm-DNN) by introducing a normalization layer and a shift layer {into} the traditional DNN. The norm-DNN converts the problem into a {unconstrained} optimization in finite dimensional parameter space, where GS as well as the  excited states can be efficiently computed via unsupervised learning. Experiments from  simple one-dimensional tests to the {high-dimensional} or  multi-component cases have been done. Practical strategies for the choice of activation function, initialization and regularization of loss have been suggested under specific physical setups.
The extensive numerical results illustrated the effectiveness and efficiency of the proposed norm-DNN approach.
The generation of norm-DNN in the domain for physical parameters  were also investigated, and comparisons were made with existing supervised learning. {Of course, the truly meaningful and challenging problem is the solution of the many-body Schr\"odinger equation. Nevertheless, the findings in this paper shows a promising perspective of norm-DNN in this direction, and this will be our future task.}

\appendix
\section*{Acknowledgements}
W. Bao is supported by the Ministry of Education of Singapore under its AcRF Tier 2 funding MOE-T2EP20122-0002 (A-8000962-00-00).
Z. Chang and X. Zhao are partially supported by the National Natural Science Foundation of China 12271413 and the Natural Science Foundation of Hubei Province No. 2019CFA007.

\bibliographystyle{model1-num-names}

\end{document}